\documentclass[preprint,nofootinbib]{revtex4}
\usepackage{framed}
\usepackage{bbold}
\usepackage{slashed}
\usepackage{graphicx,hyperref}
\usepackage{amsmath}
\usepackage{amssymb}
\usepackage{epsfig}
\usepackage{hhline}
\usepackage{soul}
\usepackage{tabularx,pifont,multirow}
\usepackage{enumitem}
\usepackage{colordvi}
\usepackage{soul}
\usepackage{xcolor}
\usepackage{yfonts}

\newcommand{\be}{\begin{equation}}
\newcommand{\ee}{\end{equation}}

\newcommand{\newtext}[1]{{\textcolor{black}{#1}}}
\newcommand{\newnewtext}[1]{{\textcolor{black}{#1}}}
\newcommand{\noveltext}[1]{{\textcolor{black}{#1}}}
\newcommand{\joantext}[1]{{\textcolor{black}{#1}}}
\newcommand{\nickcorr}[1]{{\textcolor{black}{#1}}}

\newcommand{\nftext}[1]{{\textcolor{black}{#1}}}
\newcommand{\nprooftext}[1]{{\textcolor{black}{#1}}}

\newcommand{\CC}{\Lambda}
\newcommand{\rL}{\rho_{\Lambda}}
\newcommand{\rvm}{\rho^\CC_{\rm RVM}}
\newcommand{\MPl}{M_{\rm Pl}}

\definecolor{darkgreen}{rgb}{0,0.3,0.05}
\makeatletter                                                         %
\newcommand*\rel@kern[1]{\kern#1\dimexpr\macc@kerna}                  %
\newcommand*\widebar[1]{                                              %
  \begingroup                                                         %
  \def\mathaccent##1##2{                                              %
    \rel@kern{0.8}                                                    %
    \overline{\rel@kern{-0.8}\macc@nucleus\rel@kern{0.2}}             %
    \rel@kern{-0.2}                                                   %
  }                                                                   %
  \macc@depth\@ne                                                     %
  \let\math@bgroup\@empty \let\math@egroup\macc@set@skewchar          %
  \mathsurround\z@ \frozen@everymath{\mathgroup\macc@group\relax}     %
  \macc@set@skewchar\relax                                            %
  \let\mathaccentV\macc@nested@a                                      %
  \macc@nested@a\relax111{#1}                                         %
  \endgroup                                                           %
}                                                                     %
\makeatother                                                          %
\begin{document}

\preprint[\leftline{KCL-PH-TH/2019-{\bf 44}}

% This is needed to format the full author list
%\long\def\inst#1{\par\nobreak\kern 4pt\nobreak
%{\it #1}\par\vskip 10pt plus 3pt minus 3pt}
%

% Title of the paper
\title{\Large {\bf Gravitational and Chiral Anomalies in the Running
Vacuum Universe and Matter-Antimatter Asymmetry } \vspace{0.0cm}}

\author{\large \bf Spyros Basilakos$^{a,b}$, Nick E. Mavromatos$^{c}$ and Joan Sol\`a Peracaula$^d$ \vspace{0.5cm}}

\affiliation{$^a$Academy of Athens, Research Center for Astronomy and Applied Mathematics, Soranou Efessiou 4, 115 27 Athens, Greece. \\
$^b$ National Observatory of Athens, Lofos Nymfon,
11852, Athens, Greece. \vspace{0.5cm}\\
$^c$Theoretical Particle Physics and Cosmology Group, Physics Department, King's College London, Strand, London WC2R 2LS.
\vspace{0.5cm}\\
 $^{d}$\nprooftext{Departament de F\'\i sica Qu\`antica i Astrof\'\i sica, \\ and \\ Institute of Cosmos Sciences (ICCUB), 
 Universitat de Barcelona, Avinguda Diagonal 647 E-08028 Barcelona, Catalonia, Spain.}}

%\date{\today}

\begin{abstract}
\vspace{0.05cm}
%\centerline{\bf Abstract }
%\noindent\\[-2mm]
We present a model for the Universe in which quantum anomalies are argued to play an important dual r\^ole: they are responsible for generating matter-antimatter asymmetry in the Cosmos, but also provide time-dependent contributions to the vacuum energy density of ``running-vacuum'' type, which drive the Universe evolution.
According to this scenario, during the inflationary phase of a string-inspired Universe, and its subsequent exit,
the existence of primordial gravitational waves induce gravitational anomalies, which couple to the (Kalb-Ramond (KR)) axion field emerging from the antisymmetric tensor field of the massless gravitational multiplet of the string. Such anomalous CP violating interactions
have two important effects: first, they lead to contributions to the vacuum energy density of the form appearing in the ``{\it running vacuum  model''} (RVM) framework, which are proportional to \noveltext{both, the square and the fourth power} of the effective Hubble parameter, $H^2$ \noveltext{and $H^4$ respectively. The $H^4$ terms may lead to inflation, in a dynamical scenario whereby the r\^ole of the inflaton is played by the effective scalar-field (``vacuumon'') representation of the RVM.} Second, there is an {\it undiluted} KR axion at the end of inflation, which plays an important r\^ole in generating matter-antimatter asymmetry in the Cosmos, through baryogenesis via leptogenesis in models involving heavy right handed neutrinos. As the Universe exits inflation and enters a radiation dominated era, the generation of chiral fermionic matter is responsible for the
{\it cancellation} of gravitational anomalies, thus restoring diffeomorphism invariance for the matter/radiation (quantum) theory, as required for consistency. \newnewtext{Chiral U(1) anomalies} may remain uncompensated, though, during matter/radiation dominance, \joantext{providing RVM-like $H^2$ and $H^4$ contributions to the Universe energy density}.
Finally, in the current era, when the Universe enters a de Sitter phase again, and matter is no longer dominant, gravitational anomalies resurface, leading to
\joantext{RVM-like  $H^2$ contributions to the vacuum energy density}, which are however much more suppressed, as compared to their counterparts during inflation, due to the smallness of the present era's Hubble parameter $H_0$.  In turn, this feature endows the observed dark energy  with a dynamical character that follows the RVM pattern, a fact which has been shown to improve the global fits to the current cosmological observations as compared to the  concordance $\Lambda$CDM with its rigid cosmological constant, \newnewtext{  $\Lambda > 0$}.  Our model favours axionic Dark Matter, the source of which can be the KR axion. The uncompensated chiral anomalies in late epochs of the Universe are argued to play an important r\^ole in this, in the context of cosmological models characterised by the presence of large-scale cosmic magnetic fields at late eras.
\end{abstract}
\maketitle
%\newpage

%\tableofcontents

\section{Introduction and Motivation: Running Vacuum Model for the Universe}\label{Introduction}

Over the last two decades, a plethora of cosmological observations~\cite{planck} have changed drastically our perception of the Universe. Strong evidence points towards the fact that the energy budget of the Cosmos in the current epoch consists mostly ($\sim 69\%$) of an unknown form of energy (termed ``dark energy'' (DE)), whose equation of state is close to that of a cosmological constant, $w\simeq -1$. In addition, $\sim 26\%$ consists of `` dark matter'' (DM), and thus only about $\sim 5\%$ of the Universe energy budget \newtext{corresponds to the known form of matter which we call baryonic matter}. The dominance of the DE component results in the observed acceleration of the Universe at late eras, while its equation of state $w \simeq -1$ points towards the fact that the Universe enters again, for a second time (the first being during inflation), a de-Sitter-type phase.

\newtext{Let us remark that most of the phenomenological description of the cosmological data has been obtained in the context of the Cosmological-Constant-Cold-Dark-Matter ($\CC$CDM), the standard or   ``concordance''  model of cosmology,  which is characterized by a \newnewtext{positive} cosmological constant $\CC$ and its associated vacuum energy density,   $\rho_{\CC} = \CC/ 8\pi {\rm G}$ (${\rm G}$ being Newton's gravitational constant).   The latter  plays  the role of DE, and in fact it is the  canonical DE candidate.  In most cases the data is fitted to the  spatially-flat $6$-parameter canonical version of the $\CC$CDM,  the so-called  ``base $\CC$CDM''~\cite{planck}.  The simplicity of the $\CC$CDM, however, may be \newnewtext{to the} detriment of its ability to provide a better description of the \newnewtext{cosmological observations as a whole}. In fact, this could be at the root of the observed discrepancies  or ``tensions'' which are being persistently observed in some observables, as we shall discuss later on. }

An important question is \newnewtext{therefore} whether the current de Sitter phase of the Universe is due to  the dominance of a purely cosmological-constant type DE, with $w=-1$ exactly, or there is a time-dependent vacuum energy density that resembles to a good approximation the de Sitter phase. \newtext{At a more fundamental level,  the vacuum energy is probably the result of quantum gravity effects, and in this sense, understanding its microscopic nature might have to wait for some time, until a satisfactory theory of quantum gravity, supported by observations, becomes available. \newnewtext{This will also lead to a resolution of  the longstanding cosmological constant problem}\,\cite{Weinberg}.}
Nonetheless, like with all other fundamental interactions in Nature, there might be an effective field theory description that captures the essential features and is in agreement with observations, even providing further insights for them.
\newtext{Such an attempt has been made by the development of the so-called
``Running Vacuum Model'' (RVM)
~\cite{rvm1,rvm2,rvm3} -- see  also
\cite{JSPrevs,SolGo2015} and references therein for a
detailed review.  Numerous studies of that model on its cosmological evolution from the early universe to the present  day can be found in~\cite{bls1,bls2,bls3,bls4,GRF2015,bls5}. Furthermore,  detailed confrontations  with the recent cosmological data has been presented in  \cite{JJA,AdriaJoansigma8,JAJ}, which extend the analyses of \cite{BPS09,Grande2011,GoSolBas2015} and of older works\,\cite{Cristina2003}.}

An important feature of RVM is the existence of a `de-Sitter--like' vacuum energy term in the total stress tensor, with an equation of state $w_{\rm RVM}=-1$, which however is
time dependent, $\rvm(t) = \Lambda (t)/ 8\pi {\rm G}$.\footnote{We note at this point that such a model for the Universe vacuum energy has also been advocated within the context of string brane Universes  in the presence of space-time brane defects~\cite{ncstrings}; quantum fluctuations of the latter induce  a {\it non-criticality} of the string Universe~\cite{aben}, manifested through the generation of a target-space vacuum energy dependent on the Liouville mode, which is identified with the cosmic time~\cite{time}. Given the connection of the Liouville mode with a world-sheet local renormalisation group (RG) scale, this picture provides an interpretation of the cosmic time as some sort of RG scale. Such a RG-like picture also lies at the heart of the RVM evolution, but from a rather different perspective~\cite{rvm1,rvm2,rvm3}, not associated with specific string models, as we shall review below.}. \newtext{Let us emphasize, however,  that the time dependence of the vacuum energy density in the RVM is only through the Hubble rate (and its time derivatives), i.e.  $\rvm(t)=\rvm(H(t), \dot{H}(t),...)$, in contrast to the old phenomenological time-evolving \newnewtext{models~\cite{old}}. This feature is connected to the renormalization group (RG) in curved spacetime, as we shall see below.}
Ordinary matter and radiation are on top of it. In this picture, the total stress-energy tensor reads:
\begin{align}\label{RVMstress}
T_{\mu\nu} = -g_{\mu\nu} \, \kappa^2 \, \Lambda(t) + T_{\mu\nu}^m=-g_{\mu\nu} \, \rvm(t) + T_{\mu\nu}^m\,,
\end{align}
where the superscript ``$m$'' refers generically here to matter (dust) and radiation contributions, with $\kappa^2 = 8\pi {\rm G} = M_{\rm Pl}^{-2}$ the (four-space-time-dimensional) gravitational constant, with ${\rm G}=M_P^{-2}$ the Newton constant, $M_P = 1.22 \times 10^{19}$~GeV being the four-dimensional Planck mass scale, and $\MPl=M_P/\sqrt{8\pi} = 2.43 \times 10^{18}$~GeV the reduced Planck mass (we work in units of $\hbar=c=1$ throughout).

The total energy density $\rho^{\rm total}$ is therefore
\begin{align}\label{rhorvm}
\rho^{\rm total} = \rho^\Lambda_{\rm RVM} + \rho^{\rm dust} + \rho^{\rm radiation}\,,
\end{align}
where we use the notation $\rho^\Lambda_{\rm RVM}$ to represent the RVM contribution.

The following renormalization group equation (RGE) was proposed
for  $\rvm$ in the context of the RVM as a function of the Hubble rate~\cite{rvm1,rvm2,rvm3,JSPrevs}:
\begin{equation}\label{runningrho}
\frac{d\, \rho^\Lambda_{\rm RVM} (t)}{d\, {\rm ln}H^2 } = \frac{1}{(4\pi)^2}
\sum_i \Big[a_i M_i^2 H^2 + b_i H^4 + c_i \frac{H^6}{M_i^2} + \dots
\Big].
\end{equation}
\newtext{Here $H$ plays the role of running scale $\mu$ of the RGE. The coefficients $a_i, b_i, c_i \dots$ are dimensionless
and they receive contributions
from loop corrections of fermion ($i=F$)
and boson ($i=B$) matter fields with different
masses $M_i$.  The missing term proportional to $M_i^4$ on the \textit{r.h.s.} of the above equation is forbidden since there is no fully active particle for the RGE, as all masses
are larger than the typical value of $H$ at any epoch of the cosmological evolution below the Planck mass. Therefore, the running goes slowly thanks to the decoupling terms. However, because of the dimensionality of $\rL$,
the first allowed term is a ``soft decoupling term''  $\sim M_i^2 H^2$\,\cite{rvm1}, which increases with the value of the masses and hence the effect need not be negligible\,\footnote{\newtext{\newnewtext{We note that, even if we consider the
radiation dominated epoch of the Universe, at temperature $T$, Friedman's equation (with $\rho_m\sim T^4$) implies that the requirement of satisfying} the condition $H>M_i$
roughly means $T^2/\MPl>M_i$, or equivalently
${M_i^4}/{T^4}<M_i^2/{\MPl^2}\ll1$
for any particle of mass $M_i$. Hence, at the time when the $\sim M_i^4$
contributions to the \newnewtext{running  of the vacuum energy density start becoming active,} they are \newnewtext{still} negligible compared to
the radiation contribution $\sim T^4$. We \newnewtext{therefore} conclude that, within  the RG
formulation in which the RVM is contextualized, the terms $\sim M_i^4$ remain RG-decoupled throughout the entire
cosmic history\,\cite{JSPrevs}}.}. For this reason the running is actually dominated
by the heaviest fields in the particular Grand Unified Theory (GUT) context where the considerations are made\,\cite{astroJCAP}.  This is in contrast to \newnewtext{what happens in the} usual gauge theories, like QED or QCD, where the decoupling terms are all suppressed. The next-to-leading terms  $\sim H^4$ are not suppressed by heavy masses, and \newnewtext{although}  irrelevant for the current universe, \newnewtext{nonetheless they} can play a central role in the early universe and can explain inflation\,\newnewtext{\cite{bls1,bls2,bls3,SolGo2015,bamasol}}.  The conventional  terms suppressed \`a  la \newnewtext{Appelquist} \& Carazzone\,\cite{AC1975}  appear only at the next-to-next-to-leading order, i.e. the ${\cal O}(H^6/M_i^2)$ terms of the cosmological RGE\,\cite{JSPrevs,FritzschSola2012}, which are  a factor $H^4/M_i^4\ll1$ smaller than the soft decoupling (leading) ones.  See \cite{rvm1} for the original proposal and \cite{Babic} for additional discussions. }

\newtext{\newnewtext{It is important to note that}, because of the general covariance of the
effective action, among the possible terms emerging from the
quantum effects one expects only those carrying an even number of
time derivatives of the scale factor $a$. If expressed in terms of the
Hubble rate, $H=\dot{a}/a$, \newnewtext{this} amounts to terms of the form $H^2$,
$\dot{H}$, $H^4$, $\dot{H}^2$, $H^2 \dot{H}$ etc.  Thus, the
linear terms in $H$ (and in general any term with an odd number of
derivatives of the scale factor, such as $H^3$, $\dot{H}\,H$,
$\ddot{H}$ etc) are forbidden in the RVM since they would be incompatible
with the general covariance of the effective
action\,\cite{JSPrevs}.  In particular,
at low energies only the $H^2$ and $\dot{H}$ terms are relevant for the phenomenological confrontation with the data. The
higher order ones can however be important for the early
Universe\, \newnewtext{\cite{bls1,bls2,bls3,bls4,GRF2015,bls5,bamasol}}}.

\newtext{As  indicated, in Eq.\,(\ref{runningrho})  we have identified the RG scale $\mu$ as  $\mu\sim H$,  and hence  the Hubble rate plays the role of  the typical RG-scale
in cosmology. However, a more general option would be to associate $\mu^2$
to a linear combination of $H^2$ and $\dot{H}$ (both terms being
dimensionally homogeneous). Adopting this setting and integrating
(\ref{runningrho}) up to the terms of ${\cal O}(H^4)$, or similar dimension,  it is
easy to see that we can express the result as follows:}
 \be
\rvm(H,\dot{H})
=a_0+a_1\,\dot{H}+a_2\,H^2+a_3\,\dot{H}^2+a_4\,H^4+a_5\,\dot{H}\,H^2+...
\label{GRVE} \ee
\newtext{where the coefficients $a_i$ have different dimensionalities in
natural units, and  ... denote the  possible decoupling terms (suppressed by mass powers)  which are irrelevant for our discussion.  Specifically, $a_0$ has dimension $4$ since this is
the dimension of $\rL$; $a_1$ and $a_2$ have dimension $2$; and,
finally, $a_3$, $a_4$ and $a_5$ are dimensionless. The  RVM is the extension of the $\CC$CDM model
based on a dynamical vacuum energy density of the form (\ref{GRVE}), stemming from the basic RG equation (\ref{runningrho}).
Despite \newnewtext{the fact that higher order terms} are still possible in \eqref{GRVE}, \newnewtext{nevertheless} the expression \newnewtext{as written} contains
the basic terms up to four derivatives of the scale factor, and
hence it encodes the basic ingredients of the model both for the low (i.e. the late) and the high energy (early and very early) universe.
In particular it encodes a possible description for inflation.}

\newtext{For simplicity, let us hereafter  stick to the simplest association $\mu=H$. \newnewtext{Taking  into account that $\dot{H}=-(q+1) H^2$, where $q$ is the deceleration parameter,  which assumes the values $q=1,1/2,-1$, as we move from the radiation- into the matter- and DE-dominated epochs, respectively}, we can see that the modification introduced by $\dot{H}$ is not very important and we can pick up the main effect already with the canonical association $\mu=H$ -- this is indeed substantiated in the practical analyses, see e.g. \cite{JJA,AdriaJoansigma8,JAJ,GoSolBas2015}.
In this situation, we have $a_1=a_3=a_5=0$ in (\ref{GRVE}). The
remaining coefficients can be related immediately to those in
(\ref{runningrho}), and the final result can be
cast as\,\cite{JSPrevs}}
\begin{equation}\label{rLRVM}
\rho^{\Lambda}_{\rm RVM}(H) = \frac{\Lambda(H)}{\kappa^2}=
\frac{3}{\kappa^2}\left(c_0 + \nu H^{2} + \alpha
\frac{H^{4}}{H_{I}^{2}}\right) \;,
\end{equation}
where $H_I$ is the Hubble parameter close to GUT scale,
$c_0$ is an integration constant (with mass dimension $+2$ in
natural units, i.e. energy squared), while the
the coefficients $(\nu,\alpha)$ are written as\,\cite{bls2,JSPrevs}
\begin{equation}\label{eq:nuloopcoeff}
\nu=\frac{1}{48\pi^2}\, \sum_{i=F,B} a_i\frac{M_i^2}{\MPl}
\end{equation}
and
\begin{equation}\label{eq:alphaloopcoeff}
\alpha=\frac{1}{96\pi^2}\, \frac{H_I^2}{\MPl^2}\sum_{i=F,B}
b_i\,.
\end{equation}
In fact $\nu$ and $\alpha$ can be viewed as the
reduced (dimensionless) beta-functions
of $\rho^{\Lambda}_{\rm RVM}$ at low and high energies respectively\,\cite{rvm2,rvm3,JSPrevs}.
Of course, due to the fact that
all known particles have $M_i^2\ll M_{\rm Pl}^2$,
the above coefficients are expected to be quite small in a typical GUT,
namely ${\cal O} (10^{-6}-10^{-3})$, see \cite{rvm2}.

On considering a spatially flat
Friedman-Lema\^{\i}tre-Robertson-Walker (FLRW) spacetime, favoured by observations~\cite{planck}, which we restrict our attention to  in this work, \newtext{one can show that the main
cosmological equations in the presence of  the RVM vacuum energy density \eqref{rLRVM}  acquire the form~\cite{bls1,bls2}}
\be
\label{HE}
{\dot H}+\frac{3}{2}(1+\omega)H^2
\left(1-\nu-\frac{c_0}{H^2}-
\alpha\frac{H^2}{H_{I}^2}\right)=0,
\ee
where $\omega=\rho_{m}/p_{m}$, with $\rho_m$ ($p_m$) the matter/radiation energy density (pressure), and the overdot denotes a derivative with respect to the cosmic time $t$ of the FLRW universe. In the early universe
we have relativistic matter,  $\rho_{m}=\rho^{\rm radiation}$
with $\omega=1/3$, while in the late universe matter is dominated
by dust,  $\rho_{m}=\rho^{\rm dust}$ with
$\omega=0$. Unlike the standard  $\Lambda$CDM model of cosmology ($\Lambda=$const.)~\cite{planck},
here there is an exchange between matter and
vacuum, which implies
\begin{equation}
\dot{\rho}_{m}+3(1+\omega )H\rho_{m}=-{\dot{\rho}}^{\Lambda}_{\rm RVM}\,.
\label{frie33}
\end{equation}

The global dynamics of the RVM throughout the cosmic history has been studied in detail in \cite{bls1,bls2}. According to it,
the universe \newtext{starts from a  nonsingular state characterized by an unstable initial
de Sitter vacuum phase\cite{bls4}}. It subsequently passes smoothly from an
early inflationary epoch to a radiation period (``graceful exit")
and, at the end, it goes
into the dark-matter- and dark-energy dominated epochs. The RVM evolution  also provides an explanation
of the large entropy problem\,\cite{SolGo2015,bls4,GRF2015,bls5}.
Below we present briefly the main points, for concreteness.
Focusing on the early universe era, for which $c_0/H^2\ll 1$,
the integrated form of Eq.(\ref{HE}) admits the following solution in terms of the scale factor (upon using  $d/dt=H d/da$ in it ):
\begin{equation}
\label{HS1}
 H(a)=\left(\frac{1-\nu}{\alpha}\right)^{1/2}\,\frac{H_{I}}{\sqrt{D\,a^{3(1-\nu)(1+\omega_m)}+1}}\,,
\end{equation}
where $D>0$ is a constant. It is easy to check that
that for
$Da^{4(1-\nu)} \ll 1$ the universe starts from an unstable
de Sitter era $H^2=(1-\nu) H_I^2/\alpha$ which is
powered \newtext{by the huge value of $H_I \sim \sqrt{\alpha}M_X^2/\MPl\lesssim \Big(10^{-5}-10^{-6}\Big)\,\MPl$\,\cite{SolGo2015}, where $M_X\sim 10^{16}$ GeV is the typical value of the GUT scale.  Note that the previous relation is \newnewtext{essential, since it is  equivalent to the condition} that the fluctuations from the tensor modes do not induce CMB temperature
anisotropies larger than the observed ones ($H/\MPl\lesssim10^{-5}$ in the early universe); and it is indeed satisfied for $\alpha\sim 10^{-3}-10^{-4}$, which is in the expected range for this small parameter.
After the early inflationary epoch, specifically in the case of
$Da^{4(1-\nu)} \gg 1$, we find $H^2\sim a^{3(1-\nu)(1+\omega_m)}\sim a^{-4}$ (for $|\nu|\ll1,  \omega=1/3$)  and the universe definitely enters
the standard radiation phase, \newnewtext{as expected}.}
On the other hand, in the late universe, when the term $c_{0}/H^{2}$ in
Eq.(\ref{HE}) begins to dominate over $\alpha H^{2}/H^{2}_{I}$,
the corresponding integration leads to the solution
\be
\label{Hz}
H^{2}(a) = H_0^2
\left[{\tilde \Omega}_{m0}\,a^{-3(1-\nu)}+{\tilde \Omega}_{\Lambda 0}\right],
\ee
where  ${\tilde \Omega_{m0}}=\frac{\Omega_{m0}}{1-\nu}$  and
${\tilde \Omega_{\Lambda 0}}=1-{\tilde \Omega_{m0}}=\frac{\Omega_{\Lambda 0}-\nu}{1-\nu}$, with  $\Omega_{m0}+\Omega_{\Lambda 0}=1$  the standard sum rule  (the suffix ``0''  denoting present-era quantities). The presence of the parameter $\nu$ in the scaling of the matter contribution in \eqref{Hz} is an important and characteristic prediction of the RVM that allows comparison with the data.

In fact, the RVM  agrees excellently with the current cosmological data at large scales~\cite{planck}, but also makes important predictions \cite{JJA,AdriaJoansigma8,JAJ,GoSolBas2015} that could alleviate current tensions in the data, concerning, for instance, the so-called $\sigma_8$ tension and an associated improvement in describing large-scale structure formation, compared to the $\Lambda$CDM paradigm. The model also provides better insight
into the discrepancy with the (local) value of $H_0$ between measurements by the Hubble Space Telescope, based on Cepheid observations~\cite{hst}, and those by the Planck Collaboration, based on Cosmic Microwave Background (CMB) studies~\cite{planck}.  In \cite{JJA,AdriaJoansigma8,JAJ} it is argued that the presence of the index $\nu$   in the RVM evolution of the Hubble parameter \eqref{Hz}, which affects the scaling of the vacuum energy density \eqref{rLRVM} and, thus, differentiates it from the standard $\Lambda$CDM case, leads to combined fits to SnIA+BAO+H(z)+CMB data that favour a lower value of $\sigma_8$.

\newtext{Depending on whether one considers an interaction of the dynamical DE with matter or assumes self-conservation of the DE, one can favor the lower value of $H_0$ measured by the Planck Collaboration~\cite{planck} or push this value higher. This feature has been demonstrated \newnewtext{recently in \cite{Mehdi2019}}, where it is shown that, \newnewtext{upon the assumptions that
the DE adopts the RVM form, and  does not interact with matter},  it is possible to simultaneously decrease the value of $\sigma_8$ and increase the prediction on $H_0$, \newnewtext{such that} the fitted value of  $H_0$ definitely becomes much closer to the local value determined by Riess et al. \cite{hst}.}

Remarkably, some microscopic models supporting the RVM-type evolution  (\ref{rLRVM})-(\ref{Hz}) of the energy density of the Universe have been presented in \cite{bamasol}, based on inflationary scenarios involving dynamical breaking of minimal supergavity, or in \cite{rvm3,JSPrevs} on the basis of the \newnewtext{conformal} anomaly-induced effective action.
One of the main points of the current work  is to demonstrate that RVM contributions of $H^2$ type in the vacuum energy density  arise in more generic cosmological scenarios,
inspired from string theory, in which axion fields, coupled to Gravitational anomalies in de Sitter-eras of the Universe, also result in RVM $H^2$ contributions.

However,  our work will make an important further step by presenting a consistent (albeit minimal, rather toy) scenario, of a string Universe, in which
primordial gravitational waves induce gravitational anomalies during the inflationary phase, where only the inflaton and gravitational degrees of freedom, including the Kalb-Ramond (KR) axion associated with the antisymmetric tensor field of the massless gravitational string multiplet, are present in the string low-energy effective action~\cite{gsw,string,kaloper}. The coupling of the KR axion to the gravitational anomaly leads to undiluted KR background fields at the end of inflation, which violate {\it spontaneously} Lorentz and CPT symmetry. This, in turn, plays an important r\^ole in generating lepton asymmetry in models involving (heavy) right-handed neutrinos~\cite{ms,decesare,bms,bms2}, through the decays of the latter into standard model particles in the presence of the KR background.
The lepton asymmetry can then be communicated to the baryon sector via standard Baryon (B) and Lepton (L) number violating, but B-L conserving, sphaleron processes in the
Standard-Model (SM) sector of the model~\cite{krs}.

The basic results of this approach have already appeared in a letter form in \cite{essay}. Here we discuss the details
but also present  further developments, in particular concerning the potential r\^ole of KR axions as dark matter in late eras of the Universe.

Gravitational anomalies, when present, are known to affect {\it diffeomorphism invariance} of the quantum theory, in the sense that the matter stress-energy tensor is not conserved~\cite{jackiw}. In the absence of matter/radiation degrees of freedom, as is the case of our string effective model during inflation, where we assume only degrees of freedom from the gravitational string multiplet to be present, this may not be a catastrophe.
The anomaly-induced non-conservation of the stress tensor simply accounts for the exchange of energy among the (quantum) gravitational degrees of freedom.

During the radiation and matter eras, however, gravitational anomalies should be cancelled for the consistency of the matter quantum theory, which should be diffeomorphism invariant. In our model this is provided by the generation of chiral femion matter with anomalous axial currents, such as chiral leptons in the SM sector or other chiral fermions that might exist in beyond the standard model (BSM) physics models, which {\it cancel} the gravitational anomalies during this epoch. The
coupling of the (undiluted) KR axion to right-handed {\it massive} neutrino matter during the early radiation era succeeding inflation
is essential for leptogenesis via the mechanism of~\cite{ms,decesare,bms,bms2}.

In general, chiral anomalies survive in the radiation- and matter -dominated eras, and this is crucial for providing a link of the KR axion with the DM content of the Universe at late epochs.
The reader should recall that \newnewtext{chiral anomalies} are harmless from a diffeomorphism-invariance point of view, as they do not contribute to the stress tensor of matter.
As we shall discuss in this article, the KR axion provides a source of (``stiff''~\cite{stiff}) axionic DM, and is responsible for generating, through  its coupling with the chiral anomaly, a large-scale cosmic magnetic field at late epochs, whose magnetic energy density contributes to the late-era  energy budget of the Universe, with terms of RVM type, scaling as $H_0^2$. There are models~\cite{pilaftsis}, however, in which the KR axion couples to chiral matter (such as Majorana right handed neutrinos) via shift-symmetry breaking interactions, possibly generated by non-perturbative effects (string instantons), and via shift-symmetry-preserving kinetic mixing to other axions that are abundant in string theory~\cite{arv}. In fact, such a mixing allows for the generation of a Majorana mass for the right handed neutrinos, which is a crucial feature for the aforementioned leptogenesis scenario~\cite{decesare,bms,bms2}. These string theory axions can then play the r\^ole of additional components of DM (\newnewtext{in some of these scenarios, there is also a non-perturbative generated potential for the KR axion itself, at late eras, which thus implies its potential role as a massive DM candidate).}

In the current era, where matter becomes subdominant, and the Universe enters a de Sitter phase again, dominated by dark energy, gravitational anomalies due to gravitational wave perturbations resurface, but they are much more suppressed compared to their primordial counterparts, since the current Hubble parameter $H_0$ is much more suppressed compared to the one during inflation, $H_I \gg H_0$.  \newtext{ As a matter of fact, this is also what makes possible \newnewtext{for the DE in our epoch to inherit} a ``relic'' dynamical $H^2$-component as  part of the observed DE contribution to the current energy budget of the Cosmos. Therefore, \newnewtext{in the context of the scenario described in the present article}, we naturally predict dynamical DE, which, as argued above, seems to be favoured by current observations\,\newnewtext{\cite{JJA,AdriaJoansigma8,JAJ,GoSolBas2015,GBZhao2017}}.}

In the above scenario, therefore, the matter dominance over antimatter is entirely attributed to the existence of anomalies and the associated coupling of a gravitational axion degree of freedom (the KR axion) to them. In this work we shall discuss all such issues in detail, with the aim of demonstrating the potential importance of gravitational anomalies for the dominance of matter over antimatter in the cosmos and thus for our `very existence'.  The $H^2$-RVM-type vacuum energy, associated with the anomaly contributions, plus the existence of (`stiff') axion DM,  might then constitute {\it smoking-gun evidence} for such claims.

The structure of the article is as follows: In the next section \ref{sec:2A}, we discuss the (four space-time dimensional) primordial effective action of the model, based only on gravitational degrees of freedom of the massless bosonic string multiplet. By imposing the constraint on the modification of the Bianchi identity due to the gravitational Chern-Simons (gCS) terms by means of a pseudoscalar Lagrange multiplier field in the path integral, we demonstrate how the latter acquires dynamics and becomes
equivalent to a fully fledged KR axion field. Its CP-violating coupling to the anomaly term is crucial in ensuring background solutions, which {\it break spontaneously} Lorentz and CPT symmetry, and remain undiluted at the end of the inflationary era. This is demonstrated in section \ref{sec:rvminfl}, where it is also shown that primordial gravitational waves is the primary source of gravitational anomalies during that phase in the Universe's evolution. Moreover, the anomaly contributes to the energy density of the cosmic fluid terms which have the form of ``running-vacuum-model (RVM)'' contributions, proportional to the square of the Hubble parameter, $H^2(t)$. \noveltext{In section \ref{sec:fullRVM} we discuss the potential r\^ole of the
gravitational anomaly term, averaged over the inflationary space-time, as a provider of an effective $H^4$ term in the RVM energy density, which can then be held responsible for inflation, without the need for invoking an external inflaton field, the r\^ole of which is thus played by the scalar-field (the `vacuumon') effective description of the RVM .} In section \ref{sec:cancel}, we discuss the cancellation of the gravitational anomalies during radiation/matter-dominated eras, as a result of the generation of anomalous chiral leptonic matter at the end of inflation. There remain, however, uncompensated \newnewtext{chiral anomalies} during those eras, which also furnish the cosmic-fluid energy density with RVM-like $H^2(t)$ contributions.  The presence of the (undiluted by inflation) KR axion field, plays an important r\^ole in generating leptogenesis during the radiation era (and subsequently baryogenesis) in models with heavy right-handed sterile neutrinos, which is
discussed in section \ref{sec:asym}. In section \ref{sec:modern}, we demonstrate how the KR axion background, which couples to the uncompensated chiral anomaly,  plays a r\^ole analogous to the chiral chemical potential in electrodynamics of standard axions, which has important  implications for the generation of a cosmological magnetic field at late eras of the Universe, whose energy density contributes to the axion-DM energy budget.
%Moreover, during the current de-Sitter era, gravitational anomalies re-emerge, but now with much more suppressed $H^2_0$ running-vacuum-type contributions to the Universe energy density, which can be potentially detected.
In section \ref{sec:mixing} we speculate on extensions of the model, involving mixing of the KR axion with other axions, which exist abundantly in string theory~\cite{arv} and can play the r\^ole of additional axionic DM components. We discuss the compatibility of the generation of a shift-symmetry-breaking quintessence-like potential for the KR field at late eras of the Universe with the (approximately) constant background configurations that we studied in section \ref{sec:modern}.
Finally, section \ref{sec:concl} contains our conclusions. \nickcorr{Although in this work we consider the concrete case in which the string mass scale is of the order of the (reduced) Planck mass, our results are valid in the more general case where these scales are different. A brief discussion on this is given in the Appendix.}

\section{Anomalous String Effective Actions, Inflation and Running Vacuum \label{sec:2}}

\subsection{The Primordial Effective Action with (Gravitational) Anomalies \label{sec:2A}}

The massless bosonic gravitational multiplet of a generic string theory consists of three fields~\cite{gsw}: a traceless, symmetric, dimensionless, spin-2 tensor field $g_{\mu\nu}$, that is uniquely identified with the graviton, a dimensionless spin 0 scalar  field, the dilaton $\Phi$,\footnote{The dilaton is sometimes referred to as the trace part of the graviton. This has the following meaning: If we apply the equivalence principle, so that locally the target space time, in which a string propagates, is taken - through an appropriate coordinate choice -  to be the flat Minkowski, then the graviton fluctuations are defined through the {\it linearisation}  of the metric tensor: $g_{\mu\nu} = \eta_{\mu\nu} + \kappa \, h_{\mu\nu}$, with $h_{\mu\nu}$ a mass-dimension-one tensor with respect to the Lorentz symmetry, and $\kappa^2=8\pi {\rm G}$ is the four-dimensional gravitational constant. The associated group SO(D-1,1) of transformations in D
target-space dimensions of the string contains then a traceless spin-2 tensor representation, corresponding to the graviton, the spin-1 antisymmetric part, and a trace part, which refers to as the dilaton $\kappa^{-1} \Phi$, with $\Phi$ dimensionless. In General Relativity, one imposes a `gauge fixing', in which the graviton fluctuation tensor in the linearised formalism is {\it transverse} and {\it traceless}, thus corresponding to the aforementioned spin-2 traceless part of the SO(D-1,1) representations.} with $g_s = e^\Phi$ the string coupling, and the dimensionless spin-1 antisymmetric tensor (Kalb-Ramond) field $B_{\mu\nu} = - B_{\nu\mu}$. In the closed string sector, where we restrict ourselves for concreteness for the purposes of this work, there is a $U(1)$ gauge symmetry
$B_{\mu\nu} \rightarrow B_{\mu\nu} + \partial_\mu \theta_\nu - \partial_\nu \theta_\mu$ which characterises the target-space low-energy string effective action. This implies that
the latter depends only on the field strength of the field $B_{\mu\nu}$, which is a three-form with components
\begin{equation}\label{hfield}
{H}_{\mu\nu\rho} = \partial_{[\mu}\, B_{\nu\rho]},
\end{equation}
where the symbol $[\dots ]$ denotes complete antisymmetrisation of the respective indices.
The 3-form ${H}_{\mu\nu\rho}$ satisfies the Bianchi identity
\begin{equation}\label{bianchi}
\partial_{[\mu}\, {H}_{\nu\rho\sigma]} = 0,
\end{equation}
by construction.

The bosonic part of the (four-space-time-dimensional) effective action, $S_B$, that reproduces the string scattering amplitudes to lowest non trivial order in an expansion in powers of the string Regge-slope $\alpha^\prime$ (i.e. quadratic order in derivatives),
where we restrict our attention to  from now on, reads in the Einstein frame~\cite{string,kaloper}~\footnote{Our conventions and definitions used throughout this work are: signature of metric $(+, -,-,- )$, Riemann Curvature tensor
$R^\lambda_{\,\,\,\,\mu \nu \sigma} = \partial_\nu \, \Gamma^\lambda_{\,\,\mu\sigma} + \Gamma^\rho_{\,\, \mu\sigma} \, \Gamma^\lambda_{\,\, \rho\nu} - (\nu \leftrightarrow \sigma)$, Ricci tensor $R_{\mu\nu} = R^\lambda_{\,\,\,\,\mu \lambda \nu}$, and Ricci scalar $R = R_{\mu\nu}g^{\mu\nu}$.}
\be\label{sea}
S_B  =\; \int d^{4}x\sqrt{-g}\Big( \dfrac{1}{2\kappa^{2}} [-R + 2\, \partial_{\mu}\Phi\, \partial^{\mu}\Phi] - \frac{1}{6}\, e^{-4\Phi}\, {\mathcal H}_{\lambda\mu\nu}{\mathcal H}^{\lambda\mu\nu} - \dfrac{2}{3\alpha^\prime\, \kappa^2 }e^{2\Phi}\delta c + \dots \Big),
\ee
where  ${\mathcal H}_{\mu\nu\rho} \equiv \kappa^{-1} H_{\mu\nu\rho}$ has dimension [mass]$^2$,
and the $\dots$ represent higher derivative terms, which are of higher order in $\alpha^\prime$, with
$\alpha^\prime = M_s^{-2}$ the Regge slope of the string and $M_s$ the string mass scale. The latter is not necessarily the same as the four dimensional gravitational constant $\kappa^2 = 8\pi \, {\rm G} = M_{\rm Pl}^{-2}$.

The last term on the right-hand side of (\ref{sea}) represents a (four-space-time-dimensional) vacuum energy term. In non-critical string models~\cite{aben}, such a term arises from
a {\it positive} $\delta c > 0 $ central charge surplus of {\it supercritical} strings, which owes its existence to $\sigma$-model conformal anomaly contributions from `` internal dimensjons'' of the string, the  ``external dimensions'' $D=4$ defining the four-dimensional target space-time of our Universe. In brane universe scenarios, such vacuum energy contributions could come from bulk-space terms, and they include anti-de-Sitter-type (negative) contributions~\cite{rizos}. For our purposes in this work we shall assume $\delta c =0$.  We shall also assume that the dilaton varies slowly or that it has stabilised (through some appropriate non perturbative string mechanism) to a constant value $\Phi_0$, so that we may approximate $\partial_{\mu}\Phi \, \partial^{\mu}\Phi \simeq 0$ in (\ref{sea}) throughout the current work.
This implies an (approximately) constant string coupling $g_s = g_s^{(0)} e^{\Phi_0}$. Without loss of generality, then, we may set  $\Phi_0 = 0$. The string coupling $g_s^{(0)}$ can be fixed by phenomenological considerations of the four dimensional effective field theory~\cite{gsw}.

We can then write the  action $S_B$ as:
\begin{align}\label{sea2}
S_B =-&\; \int d^{4}x\sqrt{-g}\Big( \dfrac{1}{2\kappa^{2}}\, R + \frac{1}{6}\, {\mathcal H}_{\lambda\mu\nu}\, {\mathcal H}^{\lambda\mu\nu} + \dots \Big).
\end{align}
It is known~\cite{gsw,string} that the KR field strength terms ${\mathcal H}^2$ in (\ref{sea2}) can be absorbed (up to an irrelevant total divergence) into a contorted generalised curvature
$\overline R (\overline \Gamma)$, with a ``torsional connection''~\cite{hehl} $\overline \Gamma$, corresponding to a contorsion tensor proportional to ${\mathcal H}_{\mu\nu}^\rho$ field strength,
\begin{align}\label{torcon}
{\overline \Gamma}_{\mu\nu}^{\rho} = \Gamma_{\mu\nu}^\rho + \frac{\kappa}{\sqrt{3}}\, {\mathcal H}_{\mu\nu}^\rho  \ne {\overline \Gamma}_{\nu\mu}^{\rho}~,
\end{align}
where $\Gamma_{\mu\nu}^\rho = \Gamma_{\nu\mu}^\rho$ is the torsion-free Christoffel symbol.  Exploiting local field redefinition ambiguities~\cite{string,kaloper}, which do not affect the perturbative scattering amplitudes, one may extend the above conclusion to the quaritc order in derivatives, that is,
to the ${\mathcal O}(\alpha^\prime)$ effective low-energy action, which includes Gauss-Bonnet quadratic curvature invariants.

In string theory, in the presence of gauge and gravitational fields, cancellation of anomalies, requires the modification of the right-hand-side of (\ref{hfield})
by appropriate gauge (Yang-Mills (Y)) and Lorentz (L) Chern--Simons three-forms~\cite{gsw}
\begin{align}\label{csterms}
\mathbf{{\mathcal H}} &= \mathbf{d} \mathbf{B} + \frac{\alpha^\prime}{8\, \kappa} \, \Big(\Omega_{\rm 3L} - \Omega_{\rm 3Y}\Big),  \nonumber \\
\Omega_{\rm 3L} &= \omega^a_{\,\,c} \wedge \mathbf{d} \omega^c_{\,\,a}
+ \frac{2}{3}  \omega^a_{\,\,c} \wedge  \omega^c_{\,\,d} \wedge \omega^d_{\,\,a},
\quad \Omega_{\rm 3Y} = \mathbf{A} \wedge  \mathbf{d} \mathbf{A} + \mathbf{A} \wedge \mathbf{A} \wedge \mathbf{A},
\end{align}
where we used differential form language for brevity, with $\wedge$ denoting the usual exterior (``wedge'') product among differential forms, such that  ${\mathbf f}^{(k)} \wedge {\mathbf g}^{(\ell)} = (-1)^{k\, \ell}\, {\mathbf g}^{(\ell)} \wedge {\mathbf f}^{(k)}$, where ${\mathbf f}^{(k)}$, and ${\mathbf g}^{(\ell)}$ are $k-$ and $\ell-$ forms, respectively. Above, $\mathbf{A}$ is the Yang-Mills potential (gauge field) one form, and $\omega^a_{\,\,b}$ the spin connection one form (the Latin indices $a,b,c,d$ are tangent space (i.e. Lorentz group SO(1,3)) indices). The addition (\ref{csterms})
leads to a modification of the Bianchi identity (\ref{bianchi})~\cite{gsw}
\begin{equation}\label{modbianchi}
\mathbf{d} \mathbf{{\mathcal H}} = \frac{\alpha^\prime}{8 \, \kappa} {\rm Tr} \Big(\mathbf{R} \wedge \mathbf{R} - \mathbf{F} \wedge \mathbf{F}\Big)
\end{equation}
with $\mathbf{F} = \mathbf{d} \mathbf{A} + \mathbf{A} \wedge  \mathbf{A}$ the Yang-Mills field strength two form  and $\mathbf{R}^a_{\,\,b} = \mathbf{d} \omega^a_{\,\,b} + \omega^a_{\,\,c} \wedge \omega^c_{\,\,b}$, the curvature two form and the trace (Tr) is over gauge and Lorentz group indices.
The non zero quantity on the right hand side  of \eqref{modbianchi} is the ``mixed (gauge and gravitational) quantum anomaly''.\footnote{Notice that the modifications (\ref{csterms}) and the right-hand-side of the Bianchi (\ref{modbianchi}) contain the {\it torsion-free} spin connection. In fact, it can be shown~\cite{hull,mavindex} that any potential contributions from the torsion $\mathbf{H}$ three form in the anomaly equation can be removed by adding to the string effective action appropriate counterterms order by order in perturbation theory.}

\newtext{The Bianchi identity constraint (\ref{modbianchi}) in differential form language  can be expressed in the usual tensor notation as follows:}
\begin{equation}\label{modbianchi2}
 \varepsilon_{abc}^{\;\;\;\;\;\mu}\, {\mathcal H}^{abc}_{\;\;\;\;\;\; ;\mu}
 %=   \frac{ \alpha^\prime}{64\, \kappa}\, \varepsilon_{\mu\nu\rho\sigma}  \Big(R_{ab}^{\,\,\,\,\,\,\,\mu\nu}\, R_{\rho\sigma ab} - F^{\mu\nu}\, F^{\rho\sigma} \Big)
 =  \frac{\alpha^\prime}{32\, \kappa} \, \sqrt{-g}\, \Big(R_{\mu\nu\rho\sigma}\, \widetilde R^{\mu\nu\rho\sigma} -
F_{\mu\nu}\, \widetilde F^{\mu\nu}\Big) \equiv \sqrt{-g}\, {\mathcal G}(\omega, \mathbf{A}),
\end{equation}
where the semicolon denotes covariant derivative with respect to the standard
Christoffel connection, and
\begin{equation}\label{leviC}
\varepsilon_{\mu\nu\rho\sigma} = \sqrt{-g}\,  \epsilon_{\mu\nu\rho\sigma}, \quad \varepsilon^{\mu\nu\rho\sigma} =\frac{{\rm sgn}(g)}{\sqrt{-g}}\,  \epsilon^{\mu\nu\rho\sigma},
\end{equation}
with $\epsilon^{0123} = +1$, {\emph etc.}, are the gravitationally covariant Levi-Civita tensor densities, totally antisymmetric in their indices.
The symbol
$\widetilde{(\dots)}$
over the curvature or gauge field strength tensors denotes the corresponding dual, defined as
\begin{align}\label{duals}
\widetilde R_{\mu\nu\rho\sigma} = \frac{1}{2} \varepsilon_{\mu\nu\lambda\pi} R_{\,\,\,\,\,\,\,\rho\sigma}^{\lambda\pi}, \quad \widetilde F_{\mu\nu} = \frac{1}{2} \varepsilon_{\mu\nu\rho\sigma}\, F^{\rho\sigma}.
\end{align}

Since the anomaly ${\mathcal G}(\omega, \mathbf{A})$ is an exact one loop result, one may implement the Bianchi identity (\ref{modbianchi2}) as a $\delta$-functional constraint in the quantum path integral of the action (\ref{sea2}) over the fields ${\mathcal H}$, $\mathbf{A}$, and $g_{\mu\nu}$, and express the latter in terms of a Lagrange multiplier (pseudoscalar) field~\cite{kaloper} $b(x)/\sqrt{3}$ (where the normalisation factor $\sqrt{3}$ is inserted so that the field $b(x)$ will acquire a canonical kinetic term, as we shall see below) :
\begin{align}\label{delta}
&\Pi_{x}\, \delta\Big(\varepsilon^{\mu\nu\rho\sigma} \, {{\mathcal H}_{\nu\rho\sigma}(x)}_{; \mu} - {\mathcal G}(\omega, \mathbf{A}) \Big)
\Rightarrow  \nonumber \\ &\int {\mathcal D}b \, \exp\Big[i \, \,\int d^4x \sqrt{-g}\, \frac{1}{\sqrt{3}}\, b(x) \Big(\varepsilon^{\mu\nu\rho\sigma }\, {{\mathcal H}_{\nu\rho\sigma}(x)}_{; \mu} - {\mathcal G}(\omega, \mathbf{A}) \Big) \Big] \nonumber \\
&= \int {\mathcal D}b \, \exp\Big[-i \,\int d^4x \sqrt{-g}\, \Big( \partial ^\mu b(x) \, \frac{1}{\sqrt{3}} \, \epsilon_{\mu\nu\rho\sigma} \,{\mathcal H}^{\nu\rho\sigma}  + \frac{b(x)}{\sqrt{3}}\, {\mathcal G}(\omega, \mathbf{A}) \Big)\Big]
\end{align}
where  the second equality has been obtained by partial integration, upon assuming that the KR field strength dies out at spatial infinity. Inserting (\ref{delta})
into the path integral with respect to the action (\ref{sea2}), and integrating over the ${\mathcal H}$ field, one obtains an effective action in terms of the anomaly and a
canonically normalised dynamical, {\it massless}, KR axion field $b(x)$~\cite{kaloper}
\begin{align}\label{sea3}
S^{\rm eff}_B =&\; \int d^{4}x\sqrt{-g}\Big[ -\dfrac{1}{2\kappa^{2}}\, R + \frac{1}{2}\, \partial_\mu b \, \partial^\mu b +  \sqrt{\frac{2}{3}} \, \frac{\alpha^\prime}{96\, \kappa} \, b(x) \, \Big(R_{\mu\nu\rho\sigma}\, \widetilde R^{\mu\nu\rho\sigma} - F_{\mu\nu}\, \widetilde F^{\mu\nu}\Big) + \dots \Big],
\end{align}
where the dots $\dots$ denote gauge, as well as higher derivative, terms appearing in the string effective action, that we ignore for our discussion here.\footnote{It should be noticed that, in our conventions for the Levi-Civita tensor \eqref{leviC}, the kinetic term of the $b$-field in \eqref{sea3} has the {\it opposite sign} to that of the (covariant) square of the ${\mathcal H}_{\mu\nu\rho}$ tensor in \eqref{sea2}.} We thus observe that, in view of the anomaly, the KR axion field couples to the gravitational and gauge fields. This interaction is P and T violating, and hence in view of the overall CPT invariance of the quantum theory, also CP violating. It will be quite important for our purposes in this work.
In fact, the term
$\sqrt{-g}\, \Big( R_{\mu\nu\rho\sigma}\, \widetilde R^{\mu\nu\rho\sigma} - F_{\mu\nu}\, \widetilde F^{\mu\nu} \Big)$ in (\ref{sea3}) is the well known \emph{Hirzebruch-Pontryagin topological density} and is a total derivative
\begin{align}\label{pontryaginA}
&\sqrt{-g} \, \Big(R_{\mu\nu\rho\sigma}\, \widetilde R^{\mu\nu\rho\sigma} - F_{\mu\nu}\, \widetilde F^{\mu\nu} \Big) = \sqrt{-g} \, {\mathcal K}_{\rm mixed}^\mu (\omega)_{;\mu} = \partial_\mu \Big(\sqrt{-g} \, {\mathcal K}_{\rm mixed}^\mu (\omega) \Big) \nonumber \\
&= 2 \, \partial_\mu \Big[\epsilon^{\mu\nu\alpha\beta}\, \omega_\nu^{ab}\, \Big(\partial_\alpha \, \omega_{\beta ab} + \frac{2}{3}\, \omega_{\alpha a}^{\,\,\,\,\,\,\,c}\, \omega_{\beta cb}\Big) - 2 \epsilon^{\mu\nu\alpha\beta}\, \Big(A^i_\nu\, \partial_\alpha A_\beta^i + \frac{2}{3} \, f^{ijk} \, A_\nu^i\, A_\alpha^j \, A_\beta^k \Big)\Big],
\end{align}
with Latin letters $i,j,k$ being gauge group indices, and $\sqrt{-g}\, {\mathcal K}_{\rm mixed}^\mu$ denoting the mixed (gauge and gravitational) anomaly current density.

In the early Universe, before and during inflation, we assume that only fields from the gravitational multiplet of the string exist, which implies that our effective action pertinent to the dynamics of the inflationary period, is given by (\ref{sea3}) upon setting the gauge fields to zero, $\mathbf{A}=0$. Thus, to describe the dynamics of the beginning and the inflationary period of the Universe, we use the effective action
\begin{align}\label{sea4}
S^{\rm eff}_B =&\; \int d^{4}x\sqrt{-g}\Big[ -\dfrac{1}{2\kappa^{2}}\, R + \frac{1}{2}\, \partial_\mu b \, \partial^\mu b +   \sqrt{\frac{2}{3}}\,
\frac{\alpha^\prime}{96 \, \kappa} \, b(x) \, R_{\mu\nu\rho\sigma}\, \widetilde R^{\mu\nu\rho\sigma} + \dots \Big],
\end{align}
involving only the KR axion and the gravitational field. The presence of the axion $b(x)$ represents the effects of `torsion',  in view of our previous discussion on the r\^ole of the KR field strength as a (quantum) torsion \eqref{torcon} in string theory~\cite{gsw,string,kaloper}. On ignoring the gauge sector, the topological density \eqref{pontryaginA}
becomes
\begin{align}\label{pontryagin}
\sqrt{-g} \, R_{\mu\nu\rho\sigma}\, \widetilde R^{\mu\nu\rho\sigma}
&= \sqrt{-g} \, {\mathcal K}^\mu (\omega)_{;\mu} = \partial_\mu \Big(\sqrt{-g} \, {\mathcal K}^\mu (\omega) \Big) \nonumber \\ &= 2 \, \partial_\mu \Big[\epsilon^{\mu\nu\alpha\beta}\, \omega_\nu^{ab}\, \Big(\partial_\alpha \, \omega_{\beta ab} + \frac{2}{3}\, \omega_{\alpha a}^{\,\,\,\,\,\,\,c}\, \omega_{\beta cb}\Big)\Big]~,
\end{align}
\noveltext{which is also called ``gravitational Chern-Simons (gCS)'' term, a terminology that we shall use in this work.} 
The (purely gravitational) quantity $\sqrt{-g}\, {\mathcal K}^\mu$ may be viewed as the `\emph{axial current density}' of our bosonic theory (\emph{i.e}. in the absence of fermions), as its four (covariant) divergence is related to the gravitational anomaly.  For completeness and future convenience, we also express below $\sqrt{-g} \, {\mathcal K}^\mu$  in terms of the  (standard) torsion-free Christofel  connection $\Gamma^\alpha_{\,\,\beta\gamma}$,
\begin{align}\label{kcs}
\sqrt{-g}\, {\mathcal K}^\mu &= \epsilon^{\mu\beta\gamma\delta} \, \Big(\Gamma^\nu_{\,\,\beta\sigma} \, \partial_\gamma \Gamma^\sigma_{\,\,\delta \nu} + \frac{2}{3} \Gamma^\nu_{\,\,\beta\sigma}\, \Gamma^\sigma_{\,\,\gamma\lambda}\, \Gamma^\lambda_{\,\,\delta\nu}\Big).
\end{align}

We now notice that, by partially integrating the CP violating anomaly term in (\ref{sea4}), ignoring surface terms (on account of the assumption that the gravitational field and its derivatives vanish at infinity), and using (\ref{pontryagin}), one arrives at the effective action
\begin{align}\label{sea5}
S^{\rm eff}_B &=\; \int d^{4}x\, \sqrt{-g}\Big[ -\dfrac{1}{2\kappa^{2}}\, R + \frac{1}{2}\, \partial_\mu b \, \partial^\mu b  -
 \sqrt{\frac{2}{3}}\,
\frac{\alpha^\prime}{96 \, \kappa} \, \partial_\mu b(x) \, {\mathcal K}^\mu + \dots \Big] \nonumber \\ & \equiv \quad {\mathcal S}^{\rm grav} + {\mathcal S}^b + {\mathcal S}^{\rm b-grav},
\end{align}
where ${\mathcal S}^{\rm grav}$ denotes the pure-gravity Einstein-Hilbert Ricci scalar action,
$S^{b}(b,\, g_{\alpha\beta})$ denotes the `matter' action of the $b(x)$ field, that does {\it not} contain derivatives of the graviton,
\begin{align}\label{matter}
S^b \equiv \int d^4x \, \sqrt{-g} \, \frac{1}{2}\, \partial_\mu b \, \partial^\mu b ~,
\end{align}
and
\begin{align}\label{bgrav}
{\mathcal S}^{\rm b-grav} \equiv   - \sqrt{\frac{2}{3}}\,
\frac{\alpha^\prime}{96 \, \kappa} \, \int d^4x\, \sqrt{-g} \, \Big( \partial_\mu b(x) \, {\mathcal K}^\mu \Big) =  \sqrt{\frac{2}{3}}\,
\frac{\alpha^\prime}{96 \, \kappa} \, \int d^4x\, \sqrt{-g} \,  b \, R_{\mu\nu\rho\sigma}\, \widetilde R^{\mu\nu\rho\sigma}~,
\end{align}
is the the KR-axion-gravitational anomaly term  (\ref{pontryaginA}).

The `matter' KR-axion stress-energy  tensor is calculated from \eqref{sea5} by using the standard definition of  $T^b_{\mu\nu}  $  in  General Relativity,
\begin{align}\label{stressb}
T_{\mu\nu}^b = \frac{2}{\sqrt{-g}} \, \frac{\delta S^{b}(b,\, g_{\alpha\beta})}{\delta g^{\mu\nu}}
  =   \partial_\mu b \, \partial_\nu b - \frac{1}{2}g_{\mu\nu} (\partial_\alpha b\, \partial^\alpha b).
\end{align}
To compute the metric variation of  \eqref{bgrav},
we take into account
that the variation  of the Christoffel symbol with respect to the metric tensor $g_{\mu\nu}$ is:
\begin{align}
\delta \Gamma^\beta_{\,\, \alpha\gamma} = \frac{1}{2} g^{\beta\delta} \Big( (\delta g_{\delta\gamma})_{;\alpha} + (\delta g_{\alpha\delta})_{;\gamma} - (\delta g_{\alpha\gamma})_{;\delta} \Big)~.
\end{align}
One can then easily express the infinitesimal metric variation of the Pontryagin-term $b R \widetilde R$ in terms of the so-called four-dimensional {\it Cotton-tensor} ${\mathcal C}_{\mu\nu}$~\cite{jackiw}:
\begin{align}\label{csgrav}
\delta \Big[ \int d^4x \sqrt{-g} \, b \, R_{\mu\nu\rho\sigma}\, \widetilde R^{\mu\nu\rho\sigma} \Big] = 4 \int d^4x \sqrt{-g} \, {\mathcal C}^{\mu\nu}\, \delta g_{\mu\nu} =\nickcorr{-}
4 \int d^4x \sqrt{-g} \, {\mathcal C}_{\mu\nu}\, \delta g^{\mu\nu}, 
\end{align}
where~\cite{jackiw,kaloper}
\begin{align}\label{cotton}
&{\mathcal C}^{\mu\nu} \equiv  \nickcorr{-\frac{1}{2}} \Big[v_\sigma \, \Big( \varepsilon^{\sigma\mu\alpha\beta} R^\nu_{\, \, \beta;\alpha} +
\varepsilon^{\sigma\nu\alpha\beta} R^\mu_{\, \, \beta;\alpha}\Big) + v_{\sigma\tau} \, \Big(\widetilde R^{\tau\mu\sigma\nu} +
\widetilde R^{\tau\nu\sigma\mu} \Big)\Big]\, , \nonumber \\ 
&= - \frac{1}{2} \Big[\Big(v_\sigma \, \widetilde R^{\lambda\mu\sigma\nu}\Big)_{;\lambda}  + \, (\mu \leftrightarrow \nu)\Big]\, ,
\nonumber \\ 
&v_{\sigma} \equiv \partial_\sigma b = b_{;\sigma}, \,\,v_{\sigma\tau} \equiv  v_{\tau; \sigma} = b_{;\tau;\sigma}.
\end{align}
\nickcorr{As follows from its definition \eqref{cotton}, and properties of the Riemann tensor, the Cotton tensor is traceless~\cite{jackiw}
\begin{align}\label{tracecot}
g_{\mu\nu}\, \mathcal C^{\mu\nu}= 0~.
\end{align}}

At this stage, we would like to make some generic remarks concerning conservation properties of the Cotton tensor, and thus potential problems associated with theories with gravitational anomalies~\cite{jackiw}. From \eqref{csgrav}, we may write the corresponding (generic) Einstein equation in the form
\begin{align}\label{einsteincs}
R^{\mu\nu} - \frac{1}{2}\, g^{\mu\nu} \, R  = \nickcorr{\Lambda g^{\mu\nu} + \sqrt{\frac{2}{3}}\,
\frac{\alpha^\prime\, \kappa}{12} \,  {\mathcal C}^{\mu\nu} + \kappa^2 \, T^{\mu\nu}_{\rm matter}},
\end{align}
where $T^{\mu\nu}_{\rm matter}$ is a generic matter stress tensor, including axion-like fields (like our KR above, {\it cf.} \eqref{matter}), which does {\it not} contain couplings to curvature and, in general, derivatives of the metric tensor. The latter couplings contribute only to ${\mathcal C}^{\mu\nu}$. In standard situations, \nickcorr{without gravitational anomalies}, general coordinate diffeomorphism invariance, implies the conservation of the matter stress tensor, $T^{\mu\nu}_{\rm matter \,\,\,\,; \nu}=0$, \nickcorr{given the covariant constancy of the metric, which ensures that the cosmological constant $\Lambda$ contribution to the total energy momentum tensor is conserved}. Because of the curvature tensor Bianchi identity, the Einstein tensor $R^{\mu\nu} - \frac{1}{2}\, g^{\mu\nu} \, R $,
also obeys such a covariant  conservation law, but this is {\it not} the case for the Cotton-tensor, as one can readily check~\cite{jackiw}:
\begin{equation}\label{csder}
{\mathcal C}^{\mu\nu}_{\,\,\,\,\,\,\,;\mu} = -\frac{1}{8} v^\nu \, R^{\alpha\beta\gamma\delta} \, \widetilde R_{\alpha\beta\gamma\delta}.
\end{equation}
Thus, in the presence of {\it gravitational} anomalies, the diffeomorphism invariance \nickcorr{would appear} to be in trouble, unless one deals with specific gravitational backgrounds~\cite{jackiw,blum}, as the ones pertaining to the FLRW Universe of interest to us here, for which the Pontryagin density {\it vanishes} $R_{\mu\nu\rho\sigma} \widetilde R^{\mu\nu\rho\sigma} = 0$.
\nickcorr{Indeed},  in our case, during the inflationary era, for which $\mathbf A =0$, the term $b \, R \widetilde R$ in (\ref{sea4}), yields, on account of  \eqref{csgrav}, \eqref{cotton}, a  Cotton tensor of the form~\cite{kaloper}
\begin{align}\label{setaxion}
{\mathcal C}_{\mu\nu} \,  \propto \,  \Big(\partial^\rho b \, \widetilde R_{\rho\mu\lambda\nu}\Big)^{;\lambda}  + \, (\mu \leftrightarrow \nu), \end{align}
where the dual Riemann tensor $\widetilde R_{\mu\nu\rho\sigma}$ has been defined in \eqref{duals}, and the proportionality numerical coefficients are of no interest to us, and hence we do not write them explicitly here. For a homogeneous and isotropic FLRW space-time, and axion field $b(t)$, for which only the temporal derivative is non zero, we obtain from (\ref{setaxion}) that $T^{bR\widetilde R}_{00}=0$, on account of the antisymmetry of the Riemann tensor $R_{\mu\nu\alpha\beta}=-R_{\nu\mu\alpha\beta}$ \newnewtext{and properties of its dual}. The pressure density contributions of such terms also vanish, as follows from the Bianchi identity of the Riemann curvature tensor, $R_{\mu[\nu\rho\sigma]}=0$, with $[\dots ]$ denoting antisymmetrisation of the respective indices. Thus for a FLRW universe, the Cotton tensor {\it vanishes}, consistently with diffeomorphism invariance.

\nickcorr{The ``apparent'' non conservation of the matter stress tensor in the presence of the Cotton tensor in the Einstein's equation
\eqref{einsteincs} appears to be in contradiction with the perfectly covariant form of the axion-gravitational-Chern-Simons (gCS) coupling in \eqref{sea4} under general coordinate transformations. As is standard, when evaluating anomalies in higher-order quantum corrected effective actions, 
one employs specific regularisations, such that there is some sort of conserved ``improved'' second rank tensor which plays the r\^ole of the energy momentum tensor, compatible with general covariance.} 

\nickcorr{ This also happens here, for generic space-time backgrounds.
Indeed, as can be seen from \eqref{einsteincs}, from the Bianchi identities of the Einstein tensor, 
there is a conserved {\it modified stress-energy tensor} 
\begin{align}\label{cons}
\kappa^2 \, {\widetilde T}_{b + \Lambda + {\rm gCS}}^{\mu\nu} \equiv \sqrt{\frac{2}{3}}\,\frac{\alpha^\prime\, \kappa}{12} \mathcal C^{\mu\nu} + \kappa^2 T_b^{\mu\nu}  + \Lambda g^{\mu\nu}  \quad \Rightarrow \quad  {\widetilde T}_{b + \Lambda + {\rm gCS} \,; \mu}^{\mu\nu} =0~,\end{align}
the extra terms, proportional to the Cotton tensor $C^{\mu\nu}$, describing energy exchange between the axion and gravitational field.\footnote{\nickcorr{The existence of a conserved modified stress-energy tensor proportional to higher curvature terms also characterises the 
case of dilaton-Gauss-Bonnet $\mathcal O(\alpha^\prime)$ terms 
in string effective actions~\cite{kanti}:
\begin{align}\label{seaPhi}
S_B  =\; \int d^{4}x\sqrt{-g}\Big( \dfrac{1}{2\kappa^{2}} [-R + 2\, \partial_{\mu}\Phi\, \partial^{\mu}\Phi] - c_1 \frac{\alpha^\prime}{8g_s^{(0)2}\, \kappa^2} e^{-2\Phi} 
 \Big(R_{\mu\nu\rho\sigma} \, R^{\mu\nu\rho\sigma} - 4 R_{\mu\nu}\, R^{\mu\nu} + R^2\Big) 
 + c_2 (\partial \Phi)^4 + \dots \Big)
\end{align}
where $c_i$, $i=1,2$ are numerical coefficients, and the $\dots$ denote contributions from central charge deficit, antisymmetric tensor fields, and higher derivatives.
The existence of the Gauss-Bonnet terms, lead to modified stress tensor conservation 
${\mathcal T}_{\mu\nu}^{\phi +\mathcal G\mathcal B\,;\mu} =0$, with 
\begin{align}\label{gbstress}
{\mathcal T}_{\mu\nu}^{\phi +\mathcal G\mathcal B} &=  \frac{2}{\kappa^2} \, \partial_\mu \Phi \partial_\nu \Phi -  \frac{1}{\kappa^2} \, g_{\mu\nu} (\partial_\sigma \Phi)^2
+\frac{\alpha^\prime}{\kappa^2}\,  {\mathcal P}_{\mu\nu} + \dots, \nonumber \\
{\mathcal P}_{\mu\nu}  &= \frac{c_1}{8g_s^{(0)2}} \, \Big(g_{\mu\rho}\, g_{\nu\lambda} + g_{\nu\rho}\, g_{\mu\lambda} \Big) \, \varepsilon^{\sigma\lambda\alpha\beta} \Big(\widetilde R^{\rho\gamma}_{\,\,\,\,\alpha\beta} \, \partial_\sigma [ e^{-2\Phi} ]\Big)_{;\gamma},
\end{align}
with the extra interactions describing energy exchange between dilatons $\Phi$ and gravity, in a similar spirit to our case with axions interacting with the gravitational Chern Simons term, which is also a higher curvature term. In fact, the analogy with our case goes even further, in the sense that the modifications ${\mathcal P}_{\mu\nu}$ in \eqref{gbstress} lead to non-positive contributions to the modified stress energy tensor, which in ref. \cite{kanti} were deemed important in evading the no-hair theorem, allowing for black hole solutions with (secondary) dilaton hair. In our case, the negative contribution of the gravitational Chern-Simons term proves crucial
for the consistency of our framework, see section \ref{sec:fullRVM}}.} 
The covariant conservation law \eqref{cons}, then, leads as usual to the energy conservation of the KR axion-gravity system. Any regularisation scheme employed in the computation of the anomaly should respect then this conservation law, and thus the CP-violating axion-gravity interaction terms in the effective action \eqref{sea5} are fully consistent, both formally and conceptually, as was to be expected given that such terms arise in the context of string theory~\cite{kaloper}, which is a consistent theory of quantum gravity~\cite{gsw}. } 

\nickcorr{What we say here is that, rather than restricting~\cite{jackiw,blum} the consistent space-time backgrounds by demanding the covariant conservation of the pure axion matter stress tensor, $T^{\mu\nu}_{\rm axion-matter\,\,;\,\nu}=0$, when there are gravitational anomalies, to which these axions couple, this on-shell conservation law breaks down, as a result of exchange of energy with the gravitational field. There is instead a modified stress tensor \eqref{cons} which remains conserved. This peculiarity refers only to axion fields and should be reflected in the solutions to the equations of motion for these fields in the presence of anomalous background space times. It is the purpose of this work to demonstrate the existence of such consistent solutions, albeit as we shall see they violate ``spontaneously'' Lorentz symmetry. The latter should not come as a surprise, due to the ``spontaneous'' breaking of diffeomorphism invariance by the anomalous space-time gravitational backgrounds (``vacuum''). The underlying UV complete, {\it full} quantum gravity theory should be {\it diffeomorphism invariant}, as is the case of string theory in our example.}

\nickcorr{Indeed, as we shall discuss later on ({\it cf.} subsections \ref{sec:rvminfl} and \ref{sec:fullRVM}), primordial gravitational waves during the inflationary phase of  FLRW Universes do induce non-trivial CP-violating anomalous gravitational-Chern-Simons-KR-axion  couplings, and {\it condensates} of the gravitational anomaly. Upon taking into account such condensates, the classical equation of motion for the KR axion  field are modified from the standard one in the absence of gravitational anomalies. In the anomaly-free case, the KR axion, satisfies classically $\Box b(x)=0$, where $\Box$ is the covariant D' Alembertian. This implies the classical conservation law $T^{\mu\nu}_{b\,\,\,\,\,;\mu} =0$, where $T_b^{\mu\nu}$ is given in \eqref{stressb}.  As we shall see below, in the presence of gravitational-anomaly {\it condensates}, this equation is modified to \eqref{krbeom}, which admits the non-trivial Lorentz-violating solutions \eqref{lv}, mentioned above. For such solutions, it is the modified stress tensor \eqref{cons}, taking into account the KR-gravitational-Chern-Simons interaction, that is classically conserved on account of the classical Einstein's equations. As we shall discuss in section \ref{sec:fullRVM}, the anomaly condensates induce a background FLRW Universe with a positive (de-Sitter type) cosmological constant, which drives inflation.}

\nickcorr{However, in this case things are even more subtle, in the sense that  
the anomalous gravitational contributions are obtained by averaging over quantum graviton fluctuations~\cite{stephon}, and in this sense 
Einstein's equations, which are classical equations, do not describe the graviton quantum fluctuations that induce the Chern-Simons term. Hence, there is {\it no inconsistency} as far as the underlying quantum gravity (string) theory during the inflationary era of the Universe is concerned, which at low energies is described by the effective action \eqref{sea4}, and contains only fields from the string gravitational multiplet. The latter is fully consistent and diffeomorphism invariant.}

\nickcorr{Nonetheless, as we shall see in section \ref{sec:cancel},  in the radiation or matter
eras, after the exit from inflation, the generation of chiral matter would lead to a {\it cancellation} of gravitational anomalies, as would be 
``conventionally'' required for the ``consistency'' of the matter and radiation quantum field theory, without the need to employ a generalised 
stress energy tensor \eqref{cons}. In such a case, the axion fields would only coupled at most to chiral or in general triangle anomalies, which do not contribute to the stress tensor, due to their topological form (see discussion in section \ref{sec:cancel}, after Eq.~\eqref{anom2}), and, thus, the conventional local covariant conservation of the matter/radiation stress tensor is guaranteed for any metric background. }

\nickcorr{After these important remarks} we next proceed to discuss the equation of state of the KR fluid. From \eqref{stressb}, and taking into account
the generic relation for the stress-energy tensor for an observer moving with a four-velocity $u_\mu$ with respect to an inertial frame
\begin{equation}\label{stressvel}
T_{\mu\nu} = \Big(\rho + p\Big) u_\mu\, u_\nu - g_{\mu\nu}\, p\, ,
\end{equation}
 we obtain for the energy density $\rho^b = T_{00}^{b\, {\rm rest}}$ and pressure $p^b$ defined via $T^{b\, {\rm rest}}_{ii} = -p^b g_{ii}$ (no sum over $i$) for an observer at rest with respect to the cosmic frame of a FLRW Universe, with a homogeneous and isotropic KR axion field $b(t)$ fluid:
\begin{align}\label{endenpress}
\rho^b = \frac{1}{2} (\dot {\overline b})^2, \qquad p^b = \frac{1}{2} (\dot {\overline b})^2 = \rho^b.
\end{align}
This has a {\it stiff matter}~\cite{stiff} equation of state, $w=1$
 and hence cannot by itself lead to a``running vacuum'' type of fluid. The scaling (with the Universe scale factor) of the energy density of stiff matter is
 \begin{equation}\label{stiff}
\rho^b = p^b \sim  a^{-3(1+w)} = a^{-6}, \quad w=1.
 \end{equation}
Below we shall explicitly demonstrate this by evaluating the induced energy density, as a self-consistency check of the approach.
To this end, we first observe from (\ref{sea5}), that the classical equations of motion of the KR axion field $b(x)$, imply the existence of
backgrounds $\overline b$ that satisfy
\begin{align}\label{krbeom}
\partial_{\alpha}\Big[\sqrt{-g}\Big(\partial^{\alpha}\bar{b}  -  \sqrt{\frac{2}{3}}\,
\frac{\alpha^\prime}{96 \, \kappa} \, {\mathcal K}^{\alpha}  \Big)\Big] = 0,
\end{align}
where, as we shall see, ${\mathcal K}^\mu $ will be associated with an average of the Hirzebruch-Pontryagin density (\ref{pontryagin}) over the inflationary space time, which in the presence of the CP violating anomalous interactions of (\ref{sea5}) can be non-vanishing~\cite{stephon}.
\nickcorr{By multiplying \eqref{krbeom} with $\partial^\nu b$, and taking into account \eqref{csgrav}, \eqref{cotton} and \eqref{csder}, 
the reader can easily verify that this equation implies the conservation of the improved stress tensor \eqref{cons}, as explained previously.}

In order not to disturb the homegeneity and isotropy of the inflationary space time, we may assume only (cosmic) time $t$ dependence of the KR background $\overline b (t)$, which, in view of (\ref{krbeom}), would imply that only the temporal component ($\mu=0$) of the `axial current density'  could be non-trivial, ${\mathcal K}^0 (t) \ne 0$. The general solution of (\ref{krbeom}), which we assume from now on,
is:
\begin{align}\label{krbeom2a}
\dot{\overline{b}}  =  \frac{{\mathcal C}_0}{\sqrt{-g}} + \sqrt{\frac{2}{3}}\, \frac{\alpha^\prime}{96 \, \kappa} \, {\mathcal K}^{0},
\end{align}
where $\dot{\overline{b}} =  \frac{d}{dt}\bar{b}(t)$ and ${\mathcal C}_0$ a constant.  Eq.~\eqref{krbeom2a} is a mathematically consistent relation, since both $\partial_\mu b$ and ${\mathcal K}_\mu$ are (covariant) {\it axial} four-vectors.

The relation \eqref{krbeom2a}  induces a background for the KR axion field that breaks
{\it spontaneously} Lorentz, CP and CPT symmetry. In fact the {\it masslessness} of the KR axion $b$ can be understood by viewing this pseudoscalar field as the {\it Goldstone-Boson} of the spontaneously broken Lorentz symmetry~\cite{aben}.

The term proportional to ${\mathcal C}_0$  in \eqref{krbeom2a} is expected to be suppressed in an inflationary space-time, so without loss of generality we may set from now on ${\mathcal C}_0=0$ and consider the solution
\begin{align}\label{krbeom2}
\dot{\overline{b}}  =  \sqrt{\frac{2}{3}}\, \frac{\alpha^\prime}{96 \, \kappa} \, {\mathcal K}^{0}.
\end{align}
From the anomaly equation (\ref{pontryagin}), assuming homogeneity and isotropy for the anomaly density $\sqrt{-g(t)}\, {\mathcal K}^\mu (t)$, with $t$ the cosmic time,  one has
\begin{equation}\label{pontryagin2}
 \frac{d}{dt}  \Big(\sqrt{-g}\, {\mathcal K}^0 (t) \Big) = \langle \sqrt{-g} \, R_{\mu\nu\rho\sigma}\, \widetilde R^{\mu\nu\rho\sigma} \rangle ~,
 \end{equation}
 where $\langle \dots \rangle$ denote appropriate averages over graviton fluctuations in the inflationary space time to be defined below~\cite{stephon}.

 In an unperturbed FLRW space-time, with scale factor $a(t)$, the right-hand side of (\ref{pontryagin2}) vanishes, as already mentioned, which would imply
 \begin{equation}\label{k0alone}
 {\mathcal K}^0 (t) \propto \Big(\sqrt{-g(t)}\Big)^{-1} \sim a^{-3}(t)~,
 \end{equation}
 consistent with the expected `stiff matter' scaling (\ref{stiff}) in this case, where only a massless KR axion field without potential is the only constituent of ``matter'' in the Universe.

 \subsection{Gravitational Waves during Inflation, Anomalies and a ``Running Vacuum '' \label{sec:rvminfl}}

 \joantext{In this context, another scalar field or mechanism, can be introduced to induce inflation. At the moment we assume that the new field is some conventional inflaton field, $\varphi$, imported from an external framework which the current one might be embedded into. Later on, in the next \noveltext{subsection}, we will see that such \noveltext{a} scalar field need not be a new fundamental field but just the one that enables mapping de RVM to its scalar field representation.  However, everything that we will say below does not depend on the nature of $\varphi$ and hence we postpone the discussion of the scalar picture of RVM \noveltext{to} Sec. \ref{sec:fullRVM} .}
 So, let us assume for concreteness the existence of an inflaton scalar field, $\varphi$, which is different from the KR axion $b(x)$.\footnote{In some supergravity models, the r\^ole of the inflaton might be played by the  {\it real part} of a complex scalar field, which represents the dilaton $\Phi$ (see \eqref{sea}), whose {\it imaginary part} is the axion; a slow roll dilaton, upon assuming appropriate potentials, leads then to inflation, and the cosmic time derivative of the axion field might be taken to be of the same order as that of the dilaton (slow roll for both components of the complex scalar field). This is the case assumed in \cite{stephon}.} Augmenting our effective action \eqref{sea5} by the inclusion of a scalar-$\varphi$ sector, with canonical kinetic term and a potential ${\mathcal U}(\varphi)$, we write for the complete effective action~\footnote{\newnewtext{For brevity and concreteness, we assume here that the scalar field couples minimally to gravity. Non minimal couplings to gravity are certainly interesting scenarios, which however we do not consider here, as they will not be directly relevant to (in the sense of not affecting qualitatively) the main conclusions of our work, which are the RVM-type contributions to the vacuum energy density of the Universe due to the gravitational (and chiral) anomalies and the novel matter-antimatter asymmetry induced by the undiluted KR field at the end of inflation.}}:
\begin{align}\label{effdil}
S^{\rm eff}_{b+\varphi+{\rm gravity}} &=\; \int d^{4}x\sqrt{-g}\Big[ -\dfrac{1}{2\kappa^{2}}\, R +  \frac{1}{2}\, \partial_\mu \varphi \, \partial^\mu \varphi
- {\mathcal U}(\varphi) + \frac{1}{2}\, \partial_\mu b \, \partial^\mu b  -
 \sqrt{\frac{2}{3}}\,
\frac{\alpha^\prime}{96 \, \kappa} \, \partial_\mu b(x) \, {\mathcal K}^\mu + \dots \Big],
\end{align}
where the $\dots$ denote higher derivative terms, including higher curvature terms irrelevant for our purposes here.\footnote{If quadratic terms in the (scalar) curvature, $\beta R^2$, $\beta > 0$, exist in the effective action \eqref{effdil}, then one may associate the inflaton field $\varphi$ with the scalar mode contained in those $R^2$-terms~\cite{R2scal}, in which case the potential ${\mathcal U}(\varphi)$ is that of the Starobinsky model for inflation~\cite{staro}.}
From (\ref{effdil}), we observe that the equations of motion for the KR-axion field are the same  as those obtained from the action \eqref{sea5}, {\it i.e.} they still assume the form \eqref{krbeom}, but, now, the total ``matter'' stress tensor, for the fields $\varphi(x)$ and $b(x)$, reads:
 \begin{align}\label{stressbphi}
 T_{\mu\nu}^{\varphi+b} = \partial_\mu \varphi \partial_\nu \varphi +   \partial_\mu b \partial_\nu b - g_{\mu\nu} \Big(\frac{1}{2}\, \partial_\alpha \varphi \, \partial^\alpha \varphi
 + \frac{1}{2}\, \partial_\alpha b \, \partial^\alpha b  - {\mathcal U}(\varphi)\Big),
 \end{align}
 where the reader is reminded of the fact that the anomaly terms do not contribute in a FLRW space-time, assumed on average (see, however below, when we consider gravitational-wave perturbations~\cite{stephon}). This implies that the energy density $\rho^{\varphi+b}$ and pressure $p^{\varphi+b}$ are:
 \begin{align}\label{endenpress2}
\rho^{\varphi +b} = \frac{1}{2} (\dot {\overline b})^2 +  \frac{1}{2} (\dot {\varphi})^2  + {\mathcal U}(\varphi), \qquad p^{\varphi +b} =
\frac{1}{2} (\dot {\overline b})^2 +  \frac{1}{2} (\dot {\varphi})^2  - {\mathcal U}(\varphi)~.
\end{align}
 For slow-running of both $\varphi (t)$ and $b(t)$ fields, that is  $({\dot \varphi})^2, ({\dot b})^2  \ll |{\mathcal U}(\varphi)|$ which we assume for our purposes here (and we shall check the self-consistency of this assumption explicitly in what follows), we observe then that the conditions for inflation are satisfied to leading order in small quantities,
 \begin{equation}\label{inflcond}
 p^{\varphi + b} \simeq - \rho^{\varphi+b} \simeq - {\mathcal U}(\varphi),
 \end{equation}
 provided ${\mathcal U}(\varphi) > 0$ (in our conventions).

 Naively speaking, as follows from (\ref{k0alone}), one would expect that in the case of inflation the (temporal component of the) anomaly current ${\mathcal K}^\mu$ would be completely {\it washed out} at the end of inflation, as a result of the exponential expansion of the scale factor during the inflationary phase:
 \begin{equation}\label{ainfl}
 a(t) \sim \exp(H \, t)~,
 \end{equation}
where $H \simeq$ constant denotes the (approximately) constant Hubble parameter during inflation (in units in which today's scale factor $a_0=1$, which are used throughout).

  However, as we shall demonstrate now, this is not always the case. \newtext{Indeed,  it is possible to consider scenarios displaying  `cosmological birefringence’ during inflation. This means that one can distinguish the effects from chiral gravitational components having different dispersion relations,  \newnewtext{which} explains the name.
  In what follows, we shall explore situations under which, \newnewtext{due to the above phenomenon},  the right-hand side of (\ref{pontryagin2}) might be non vanishing,  and,  as we shall discuss, under certain circumstances to be specified below,
the washing out of the anomaly triggered by inflation could be avoided.}

To this end, let one consider a spatially-flat FLRW space-time, with scale factor $a(t)$, {\it perturbed weakly} by scalar ($\widetilde \phi$, $\psi$) vector ($w_i$) and tensor ($h_{ij}$) perturbations
 \begin{align}\label{metric}
 ds^2 = (1 + 2 \widetilde \phi) dt^2 - w_i dt\, dx^i - a^2(t) \Big[\Big( (1 + 2\psi)\, \delta_{ij} + h_{ij}\Big)\, dx^i \, dx^j \Big].
 \end{align}
 Only the tensor perturbations contribute to $R \widetilde R$ terms, and hence we keep them in our subsequent discussion.

 \newtext{ Notice that the tensor perturbations constitute the nondiagonal part of the metric. In the study of the usual cosmic  perturbations of  the matter and dark energy fields the vector
part of the perturbation is set to zero and one exclusively focus on the Bardeen gravitational potentials $\widetilde \phi$ and $\psi$ since  the  non-diagonal
spatial part decouples from the rest in the form of gravitational
waves propagating in the FLRW background.  Here, however, we rather focus on the tensor part and ignore the rest since it has no impact on our considerations. In fact, it is  only during the inflationary stage that the primordial gravitational waves can provide a significant contribution.  After inflation they are washed out only to reappear in the very late universe but in a much weaker \newnewtext{form, as we shall see in section \ref{sec:modern}}. }

 In \cite{stephon}, the right-hand side of the averaged Hirzebruch-Pontryagin density (\ref{pontryagin2}) has been evaluated for metrics representing gravitational wave space-times during inflation, which is a solution of Einstein's equations in the action (\ref{sea5}) with the anomalous term, and we use here as a prototype for yielding non-zero anomalies of relevance to us. Assuming, for concreteness, gravitational waves propagating along the $z$ spatial direction, we consider the metric:
  \begin{align}\label{metric2}
 ds^2 = dt^2 - a^2(t) \Big[(1 - h_+(t,z))\, dx^2 + (1 + h_+(t,z))\, dy^2 + 2h_\times (t,z)\, dx\, dy + dz^2 \Big],
 \end{align}
 in the usual notation for the polarisation of the gravitational waves.
 For an inflationary space-time the scale factor has the exponential form (\ref{ainfl}). The CP violation, induced by axion-like couplings to the Hirzenbruch density  (\ref{pontryagin2}) in (\ref{sea4}), can be seen if one uses the chiral graviton basis:
 \begin{equation}
 h_{L,R} = \frac{1}{\sqrt{2}}\,\Big(h_+ \pm i h_\times \Big),
 \end{equation}
 where the  $-$ ($+$) sign pertains to L(R), and $h_{L,R}$ are scalar complex conjugate fields. The CP-violating topological interactions of the axion field in (\ref{sea4}) imply inequivalent behaviour of $h_{L,R}$ in the inflationary space-time.

 Taking into account that~\cite{stephon}
 $R_{\mu\nu\rho\sigma}\, \widetilde R^{\mu\nu\rho\sigma} \simeq 4\, i \, a^{-3} \Big[ \partial_z^2 h_R \, \partial_t\, \partial_z h_L + a^2 \, \partial_t^2 h_R \,
 \partial_t \partial_z h_L + \frac{1}{2} \partial_t (a^2) \, \partial_t h_R \, \partial_t\, \partial_z h_L - (L \leftrightarrow R)\Big] $, that is quadratic in the graviton perturbations, we may make the following approximation, to leading (up to second) order in small perturbations, that we shall be
working in this article,
  \begin{align}\label{approx}
 \langle \sqrt{-g} \, R_{\mu\nu\rho\sigma}\, \widetilde R^{\mu\nu\rho\sigma} \rangle \simeq \sqrt{-g} \, \langle R_{\mu\nu\rho\sigma}\, \widetilde R^{\mu\nu\rho\sigma} \rangle,
 \end{align}
which implies that one should use the unperturbed inflationary metric (with scale factor \eqref{ainfl}) inside the metric determinant $\sqrt{-g}$ on both sides of \eqref{pontryagin2}.

 The average of (\ref{pontryagin2}) over such a space-time then, up to second order in fluctuations $h_{L,R}$, has been performed in ref.~\cite{stephon}, with the result:
 \begin{align}\label{rrt}
  \langle R_{\mu\nu\rho\sigma}\, \widetilde R^{\mu\nu\rho\sigma} \rangle  = \frac{16}{a^4} \, \kappa^2\int \frac{d^3k}{(2\pi)^3} \, \frac{H^2}{2\, k^3} \, k^4 \, \Theta + {\rm O}(\Theta^3) ,
 \end{align}
 to leading order in $k \, \eta \gg 1$, where $k$ is the standard Fourier scale variable,
 and $\eta$ is the conformal time defined as~\cite{stephon}
\begin{align}\label{conf}
d\eta =  \frac{dt}{a(t)} \,  \Rightarrow \, \eta = \frac{1}{H}\, \exp (-Ht)
\end{align}
and in the last relation we took into account \eqref{ainfl}, which is valid during inflation. \newtext{We should note that $d\eta$ and $dt$ have actually opposite signs for the inflationary solution, and hence when the cosmic time increases the conformal time decreases. This is to be taken into account in the integration limits of each variable. Thus, the infinite future in conformal time is attained in the limit $\eta\to 0$.}
In \eqref{rrt}, we used the notation of \cite{stephon} to write for the (\newtext{dimensionless}) quantity $\Theta$ associated with the anomalous interactions in (\ref{sea4}):
 \begin{align}\label{theta}
 \Theta = \sqrt{\frac{2}{3}}\, \frac{\alpha^\prime \, \kappa}{12} \, H \,  {\dot {\overline b}} \, .
  \end{align}
 At this point, we make the important remark that the non trivial result \eqref{rrt} induced by the (primordial) gravitational wave pertubations will imply a non-zero result on the right hand side of \eqref{csder}, \newtext{which produces a gravitational anomaly, in the sense that the matter stress-energy tensor is no longer conserved and, for constant G,  it  implies the violation of the Bianchi identiy. Ultimately the reason for this situation is that, since quantum graviton fluctuations are invoked in the computation, there is {\it no guarantee} that the classical Einstein's equation \eqref{einsteincs} will continue to hold, and this is implied here by the non-conservation of the classical KR-axion stress tensor. Finally, we note that the nonvanishing of  \eqref{rrt}  is \newnewtext{due to the fact that} inflation produces
a violation of the  CP-symmetry out of equilibrium, and this fulfils Sakharov's necessary conditions for baryogenesis,  \newnewtext{which} will have implications for our subsequent discussion \newnewtext{on the generation of matter-antimatter asymmetry in our model, in section \ref{sec:asym}}.}

Above, we assumed slow roll for ${\overline b}$,
\begin{equation}\label{slowroll}
{\dot {\overline b}} \ll  H/\kappa,
\end{equation}
so that $|\Theta | \ll 1$, which justifies neglecting ${\mathcal O}(\Theta^3)$ terms in (\ref{rrt})~\cite{stephon} (the reader should recall that, during inflation, the Hubble parameter $H$ is assumed approximately constant). This necessitates an $\alpha^\prime =1/M_s^2$, with $M_s$ the string mass scale, such that $\alpha^\prime H^2 \ll 1$ during inflation, for which the scale factor $a(t)$ appearing in \eqref{rrt} assumes the de-Sitter form \eqref{ainfl}. A natural choice, which we adopt in this work, is to assume large string mass scales $M_s$ near the reduced four dimensional Planck mass scale, i.e.\nickcorr{\footnote{\nickcorr{In general~\cite{gsw}, the string scale $\alpha^\prime$ is an independent parameter from the four-dimensional
Planck scale $\kappa^2$. We shall discuss the phenomenology of this more general case briefly later on in the article and in Appendix \ref{sec:appA}.}}}
\begin{equation}\label{akappa}
\alpha^\prime \sim \kappa^2 = M_{\rm Pl}^{-2},
\end{equation}
given that the inflationary Hubble scale is expected from phenomenology~\cite{planck} to be $H \kappa  < 10^{-4}$ (we use here
bounds for single field inflation models). Here we take  for concreteness $H$ in the range
\begin{equation}\label{Hinfl}
\frac{H}{\MPl}\in \Big[ 10^{-5} , 10^{-4} \Big)~.
\end{equation}
 From (\ref{krbeom2}), then, the slow-roll conditions on $\overline b(t)$, (\ref{slowroll}), should also characterise $\langle {\mathcal K}_0 \rangle$, as a consistency check.

While staying in the FLRW frame, it is convenient to pass into conformal time $\eta$ \eqref{conf} to study the solutions of \eqref{pontryagin2}.
We also use an ultraviolet cutoff  $\mu$ for the modes, such that their physical momentum $k/a$ is cut off by~\cite{stephon}
\begin{equation}\label{cutoff}
k\, \eta  < \mu /H.
\end{equation}

\noveltext{Indeed, let us note that the leading contributions to the momentum $k$ integral on the right-hand side of \eqref{rrt} come from modes $1 \ll k\, \eta < \mu/H$~\cite{stephon}. On using \eqref{krbeom2} and \eqref{theta}, and taking into account that $\eta$
runs in the opposite direction of the cosmic time $t$,
we obtain from \eqref{rrt}, to leading order in the CP violating quantity $\Theta$ \eqref{theta}:
\begin{align}\label{rrt2}
  \langle R_{\mu\nu\rho\sigma}\, \widetilde R^{\mu\nu\rho\sigma} \rangle  &= \frac{1}{\pi^2} \Big(\frac{H}{M_{\rm Pl}}\Big)^2 \, \mu^4\, \Theta   \nonumber \\
 &= \frac{2}{3\pi^2} \frac{1}{96 \times 12} \,  \Big(\frac{H}{M_{\rm Pl}}\Big)^3 \, \Big(\frac{\mu}{M_{\rm Pl}}\Big)^4 \,  M_{\rm Pl}\, \times \, \, {\mathcal K}^0 (t).
\end{align}
Using this result, then from \eqref{pontryagin2}, \eqref{rrt} and \eqref{conf} we get~\cite{essay}:
\begin{align}\label{k01}
&\frac{d}{dt} \Big(\sqrt{-g}\,  {\mathcal K}^0 (t(\eta)) \Big ) = - (\eta H) \, \frac{d}{d\eta} \Big(\sqrt{-g}\,  {\mathcal K}^0 (t(\eta)) \Big) \nonumber \\
&= \Big[5.86 \times 10^{-5} \,  \Big(\frac{H}{M_{\rm Pl}}\Big)^3 \, \Big(\frac{\mu}{M_{\rm Pl}}\Big)^4 \,  M_{\rm Pl}\Big]\, \times \, \Big(\sqrt{-g} \, {\mathcal K}^0 (t(\eta))\Big).
\end{align}}
The slow-roll nature (\ref{slowroll}) of ${\mathcal K}^0(t)$, follows immediately from
(\ref{k01}), already from the beginning of inflation $t=0$ (or equivalently ({\it cf.} (\ref{conf})), $\eta=H^{-1}$),
as a consequence of the fact that during inflation $H \ll M_{\rm Pl}$ ({\it cf.} \eqref{Hinfl}).
This is a self-consistency check of our approach in adopting the solution (\ref{krbeom2}). The end of inflation occurs for $t \gg M_{\rm Pl}^{-1}$, and for all practical purposes we set it here formally at $t \to \infty$ (i.e. for conformal time (\ref{conf})  $\eta \to 0$). Thus, in conformal time units the duration of the inflationary period is $\Delta \eta \sim H^{-1}$.

 On assuming that $H$ remains approximately constant during the inflation period, (\ref{k01}) can be integrated over $\int^\eta_0 d\eta^\prime $.
 %In view of the cut-off (\ref{cutoff}), the integral over $\eta$ is dominated by large values of $\eta$, corresponding to the early eras of the inflationary period, for which $\eta H \simeq 1$.
 With the above in mind, we can estimate from \eqref{k01}
\noveltext{\begin{align}\label{k02}
{\mathcal K}^0 (t(\eta)) &=  \frac{1}{\sqrt{-g(t(\eta))}}\, {\mathcal K}^0_{\rm begin} (t(\eta=H^{-1})) \, \exp\Big[ -5.86 \times 10^{-5} \,  \Big(\frac{H}{M_{\rm Pl}}\Big)^2 \, \Big(\frac{\mu}{M_{\rm Pl}}\Big)^4 \, {\rm ln}(H\, \eta)\Big] \nonumber \\
& \sim {\mathcal K}^0_{\rm begin} (t(\eta=H^{-1})) \, \exp\Big[  - 3H\, t(\eta) \, \Big( 1  -  1.95 \,  \times 10^{-5} \,  \Big(\frac{H}{M_{\rm Pl}}\Big)^2 \, \Big(\frac{\mu}{M_{\rm Pl}}\Big)^4 \Big)\Big] \nonumber \\ &
\nickcorr{\equiv {\mathcal K}^0_{\rm begin} (t(\eta=H^{-1})) \, \exp\Big[  - 3H\, t(\eta) \, \mathcal A \Big]}~,
\end{align}}\newtext{where we used \eqref{conf} to write $\ln{(H\, \eta)}=-Ht$ and (\ref{ainfl}) to express   $ 1/\sqrt{-g(t)} \sim a^{-3}(t)=\exp{ [ - 3H\, t(\eta)]}$  so as to integrate this expression as part of the exponential. Finally,  as already mentioned, we have set the beginning of inflation at $t=0$ ($\eta = H^{-1}$), which is assumed immediately after the Big Bang, and its end at $t \to +\infty$ ($\eta \to 0$).}

The value ${\mathcal K}^0_{\rm begin} (t(\eta=H^{-1})$, which on account of (\ref{krbeom2}) corresponds to an initial condition for the cosmic time derivative of the KR axion, $\dot{\overline{b}}(0)$, is a boundary condition to be determined phenomenologically, as we shall discuss later on. In our normalisations (\ref{conf}), the initial scale factor $a(t(H^{-1}) )=1$, and thus $\sqrt{-g(t(H^{-1}))}=1$.

 The reader should compare \eqref{k02} with \eqref{k0alone}. The presence of gravitational waves during the inflationary phase may lead to a decrease in general, or even complete elimination, of the exponential washing out effects of inflation as $t \to +\infty$.
Indeed, the factor $\mathcal A$ in the exponent on the right-hand-side of \eqref{k02}) reads:
\noveltext{\begin{equation}
{\mathcal A} \equiv  1  -  1.95 \,  \times 10^{-5} \,  \Big(\frac{H}{M_{\rm Pl}}\Big)^2 \, \Big(\frac{\mu}{M_{\rm Pl}}\Big)^4 =
1 - \Big(\frac{H}{M_{\rm Pl}}\Big)^2 \, \Big( 0.664 \, \frac{\mu}{10\, M_{\rm Pl}}\Big)^4~.
\label{factor}
\end{equation}
Due to the slow running of $H$ during inflation, ${\mathcal A} $ is {\it approximately constant}. In inflationary scenarios where
$H \ll M_{\rm Pl}$ (\eqref{Hinfl}), and taking into account that a {\it natural range} of the cutoff $\mu$ is $\mu \lesssim M_{\rm Pl}$, one would expect, in general,
${\mathcal A} \simeq 1$, in which case the anomaly would be washed out at the end of inflation $t \to +\infty$.
However, one observes that
\begin{equation}\label{A=0}
{\mathcal A}=0 \quad \stackrel{({\it cf.} \eqref{factor})}{\Rightarrow}  \quad
\frac{H}{M_{\rm Pl}} = \Big(15.06 \, \frac{M_{\rm Pl}}{\mu}\Big)^2.
 \end{equation}
 If one insists on phenomenologically acceptable ranges of $H \ll M_{\rm Pl}$, e.g. \eqref{Hinfl}, then we observe from \eqref{factor},  \eqref{A=0} that {\it transplankian modes} should be necessarily  involved to ensure that the factor ${\mathcal A}=0$, since the cut-off in that case should exceed the Planck scale
 \begin{equation}\label{transpl}
 \mu \sim 10^{3} \, M_{\rm Pl}.
 \end{equation}}
 This provides, through \eqref{krbeom2}, a self-consistent and necessary condition for ${\dot b}$ to be approximately constant during inflation, \noveltext{which implies
 a {\it spontaneous violation} of the Lorentz symmetry by the KR background. The scale $\mu$ of this violation \eqref{transpl}, being transplanckian, does not affect the effective potential of the low-energy effective field theory at inflation, the latter defined for modes below the Planck scale.}

 \newnewtext{Having said that, we feel like remarking that the appearance of transplanckian modes, might indicate to many a potential breakdown of an effective field theory, or the weak gravity conjecture, i.e. that
 the effective quantum field theory we are dealing with cannot be consistently coupled to the full quantum gravity if \eqref{transpl} is valid.
 We, however, adopt a different interpretation, in that \eqref{transpl} offers a sign that these gravitational waves are indeed of quantum gravity origin and are generated deeply in the transplankian region but appear to us as classical gravitational waves below the Planck scale, which is the only region we can deal with at the semiclassical level.\nickcorr{\footnote{\nickcorr{Nonetheless, transplanckian values for the cutoff $\mu$ can be avoided 
in the more general case, where $\alpha^\prime (= M_s^{-2}) \ne \kappa^2 (= \MPl^{-2})$, upon appropriately restricting the range of values of $\alpha^\prime$, as explained in Appendix \ref{sec:appA}. }}}
In this respect, we also mention that transplanckian values of the inflaton field are also considered in inflationary scenarios, but still a classical general relativity treatment applies in such cases~\cite{transpl}.}\footnote{\noveltext{Nonetheless, we should remark at this point that, independently of our considerations here, it was pointed out in~\cite{lyth} that the predictions of \cite{stephon} for leptogenesis due to primordial chiral fermions depend heavily on the ultraviolet completion of the theory, in our case the full string theory, given that mainly modes in the deep quantum-gravity/string-theory regime contribute to the lepton asymmetry; moreover, as argued in \cite{lyth,paban}, by performing proper ultraviolet regularization, including
higher-than-quadratic-order derivative terms, one may effectively obtain much smaller lepton number than the one claimed in \cite{stephon}, since the cutoff $\mu$ is effectively replaced by the Hubble constant during the de Sitter phase. Par contrast, in our approach, there are no primordial fermions, and leptogenesis during the radiation era occurs in a completely different way~\cite{bms,bms2} to be discussed in section \ref{sec:asym}, due to the presence of a constant Lorentz Violating axial background of the KR field. The latter is induced by the gravitational anomaly \eqref{rrt}, and, as we shall show below, remains undiluted at the end of inflation, provided  transplanckian modes \eqref{transpl} are included.
Thus, although the induced CP violation, required for a non-zero (average) value of
the gravitational anomaly, and thus leptogenesis, is generated by gravitational waves, and one needs the full string/quantum gravity  theory
to determine the initial value of the KR axion at the Big bang ($t=0$), nevertheless, the low-energy effective field theory approach suffices for a description of the generation of a lepton asymmetry during the radiation epoch.  As we shall discuss in section \ref{sec:asym} below, the latter is proportional to
 the KR axion background itself, whose value at the exit from the inflationary era is treated as a phenomenological parameter in our scenario, since an exact prediction would depend on
 the details of the underlying microscopic (non-perturbative) string theory model, which, at present, are not known.  Our considerations therefore are
different from those of \cite{lyth,paban}, in that, in our model, the lepton asymmetry can be computed  in terms of the gravitational-anomaly-induced (Lorentz-violating) KR axion background (\joantext{in fact, the reader can easily verify that such backgrounds constitute also solutions of the axion equations of motion of the one-loop effective action of \cite{paban}, but no predictions on their magnitude can be made in that framework, given that the coefficients of the various terms can only be computed if the UV complete theory is known}).  Incidentally, for a connection of the transplanckian problem to Lorentz violation, but from a rather different perspective than ours, see
also \cite{collins}.}}

 The relations \eqref{transpl} or \eqref{transpl2}, provide, through \eqref{krbeom2}, the self-consistent and necessary conditions for ${\dot b}$ to be approximately constant during inflation,  which thus remains {\it undiluted} at the end of the inflationary period of the string Universe: 
 \begin{align}\label{lv}
 \dot{\overline b} = \sqrt{\frac{2}{3}}\, \frac{\alpha^\prime}{96 \, \kappa} \, {\mathcal K}^{0} \simeq {\rm constant}~.
 \end{align}

 We can now use the above result to provide a phenomenologically consistent estimate of ${\mathcal K}_{\rm begin}^0(t=0)$. In principle, without details of the model for inflation it is not possible to do this. The KR field is an independent field from the inflaton $\varphi$, and thus in principle, although both are slow running, the only constraint is that ${\dot {\overline b}}$ has to be much smaller than $|{\mathcal U}(\varphi)|$, in order not to upset the inflationary condition \eqref{inflcond}. A reasonable scenario, which allows a self consistent
 phenomenology, is to assume that these two rates are of the same order of magnitude. Such a case characterises, for instance, the scenario of \cite{stephon}, inspired from string-inspired conformal supergravity models, where the axion is just the imaginary part of a complex scalar field, whose real part is the dilaton. In our case, the KR axion originates from the same gravitational multiplet of strings as the graviton and dilaton, and thus the above assumption is also reasonable.
 Taking into account the phenomenological value for the slow-roll parameter for (single-field) inflation $\epsilon$, as inferred from cosmological CMB observations~\cite{planck}, we then write
 \begin{align}\label{slowrollepsi}
 \epsilon = \frac{1}{2} \frac{1}{(H M_{\rm Pl})^2}\, {\dot \varphi}^2 \sim \frac{1}{2} \frac{1}{(H M_{\rm Pl})^2}\, {\dot {\overline b}}^2 \sim
 10^{-2},
 \end{align}
which implies~\footnote{In the next section \ref{sec:lepto}, we shall see that such an  order of magnitude for $\epsilon$,
 or equivalenty $\dot{\overline b}$ at the end of inflation \eqref{slowrollepsi2}, also leads to phenomenologically acceptable leptogenesis in the radiation era, according to the mechanism of~\cite{bms2}.}
\begin{align}\label{slowrollepsi2}
 {\dot {\overline b}} \sim  \sqrt{2\,\epsilon} \, M_{\rm Pl} \, H \sim  0.14 \, M_{\rm Pl} \, H~,
 \end{align}
which can be integrated to give
\begin{align}\label{slowrollepsi3}
 {\overline b}(t) \, \sim \,  \, \overline b(0) + \sqrt{2\,\epsilon} \, M_{\rm Pl} \, H\, t~,
 \end{align}
where $\overline b(0)$ is an initial value of the KR axion field, at the beginning of inflation, immediately after the Big Bang. We shall come back to the phenomenology of this initial value later on, in section \ref{sec:fullRVM}.

This allows, through \eqref{krbeom2} and \eqref{akappa}, to express the (approximately constant, during inflation) anomaly ${\mathcal K}^0 \sim {\mathcal K}_{\rm begin} (t=0)$  as~\cite{essay}:
\begin{align}\label{anomeps}
 {\mathcal K}^0 \sim {\mathcal K}_{\rm begin} (t=0) \sim 16.6 \, H \, M_{\rm Pl}^2.
 \end{align}
From \eqref{endenpress2}, \eqref{slowrollepsi}, then, we can express the contributions of the anomaly to the energy density of the string-inspired Universe as:
 \begin{align}\label{enpressphib2B}
 \rho^{\varphi+b} \simeq 3M_{\rm Pl}^4 \Big[3.33  \times 10^{-3} \, \Big(\frac{H}{M_{\rm Pl}}\Big)^2  + \frac{{\mathcal U}(\varphi)}{3M_{\rm Pl}^4}\Big].
  \end{align}
Inflation occurs as long as ${\mathcal U}(\varphi) \gg 10^{-2} \, (H\, M_{\rm Pl})^2 $.
The terms depending explicitly on $H$ in \eqref{enpressphib2B} constitute {\it running vacuum}-like corrections~\cite{JSPrevs} to the classical inflationary (almost constant) potential ${\mathcal U}$.  In case, for instance, the inflationary potential is that of Starobinsky, with parameter $\beta$, which arises naturally in string-inspired models that contain higher-curvature corrections in their effective low-energy actions,  the dynamical vacuum model energy density assumes the form~\cite{bamasol}
\begin{align}\label{rvm1}
\rho_{\rm RVM} (H) = 3 M_{\rm Pl}^4 \Big(c_0 + \nu \, \Big(\frac{H}{M_{\rm Pl}}\Big)^2 + \beta \, H^4\Big), \quad \beta > 0\,.
\end{align}
\newtext{As we can see, this expression  is of the generic running vacuum form \eqref{rLRVM} that we have studied in the previous section.}
In our case,   
\begin{align}\label{nuvalue}
\nu \sim 3.33 \times 10^{-3} \ll 1~, 
\end{align}
and $c_0 \ll \Big(\frac{H}{M_{\rm Pl}}\Big)^2$ may be considered as part of ${\mathcal U}(\varphi)$ so we can ignore it safely when we talk about quantities during the inflationary era. \newtext{The neglected term resurfaces of course in the late universe and becomes the leading contribution to the DE.}

\subsection{Anomaly induced Inflation through Running Vacuum \label{sec:fullRVM}}

In this section we wish to discuss in some detail what was already announced at the beginning of the previous subsection, namely the fact that the scalar field $\varphi$ that we have introduced there need not be a fundamental external inflaton but it can be identified with the field $\phi$ (different from $\varphi$) that defines the scalar field representation of the RVM in its full fledged form \eqref{rLRVM} or \eqref{rvm1}. This form contains both $H^2$ and the higher power $H^4$, the latter being essential to trigger inflation in the RVM.    In what follows we wish, first of all, to note that our gravitational anomaly framework actually predicts the full RVM form of the vacuum energy density, in which the higher power $H^4$ is generated by the gravitational Chern-Simons (gCS) anomaly term, that is, the last term on the right-hand side of the string effective action (\ref{sea4}). 

This comes about upon averaging such an effective action over the inflationary spacetime, i.e. when we consider the vacuum expectation value  of 
$\langle b(x) R_{\mu\nu\rho\sigma}(x) \widetilde{R}^{\mu\nu\rho\sigma}(x)\rangle$  in the inflationary background.  \nickcorr{This is 
viewed as a condensate of graviton fluctuations, which is formed in  the context of a UV complete theory of quantum gravity, such as string theory in the present example. 
From a formal point of view, such condensates appear dynamically by first averaging the (quantum gravity) 
partition function corresponding to the low-energy effective action \eqref{sea4} over gravitational perturbations about a de Sitter background, and then looking for {\it local minima} of this action, characterised by semiclassical (Einstein type) equations with respect to the gravitational field. 
In general this is a complicated process, where the full string theory (or UV complete quantum-gravity)  dynamics plays a r\^ole, and at present 
a complete formal treatment is not available. Nonetheless, for our purposes here, we adopt a phenomenological approach, in which we postulate the existence of this condensate in a low-energy effective action framework, basing this assumption on our previous results on the induced anomaly by means of primordial gravitational wave perturbations of the de Sitter background during inflation \eqref{rrt}. We assume that such primordial gravitational wave perturbations set the dominant scale for the condensate.}

\nickcorr{Once such a condensate is formed  
we may expand the gCS term \eqref{bgrav} in the effective action \eqref{sea4} over quantum fluctuations about it, by writing formally
\begin{align}\label{order}
\mathcal S^{{\rm b-grav}}  = \sqrt{\frac{2}{3}}\, \frac{\alpha^\prime}{96 \, \kappa} \, \int d^4x \, \sqrt{-g}\,  \Big( \langle \, \overline b(x) \,  R_{\mu\nu\rho\sigma}\, \widetilde R^{\mu\nu\rho\sigma} \, \rangle 
+ \,\mathbf{ :}\,   b(x) \, R_{\mu\nu\rho\sigma}\, \widetilde R^{\mu\nu\rho\sigma} \, \mathbf{:} \Big)~, 
\end{align}
where $\mathbf{:} \dots \mathbf{:}$ denotes proper quantum ordering of (quantum field) operators, which, in the path-integral language, is interpreted as indicating terms with the appropriate subtraction of the UV divergencies, via regularization by means of the UV cut-off $\mu$. This quantum-ordered term can give rise (via its variation with respect to the gravitational field) to a quantum-ordered Cotton tensor \eqref{cotton}, which is traceless  ({\it cf.} \eqref{tracecot}). }

\nickcorr{The reader should take notice of the fact that, as typical with condensates in field theory, the quantity $\langle \, \overline b(x) \,  R_{\mu\nu\rho\sigma}\, \widetilde R^{\mu\nu\rho\sigma} \, \rangle $ does {\it not} depend on the metric tensor, which thus leads to the addition of  a  Dark Energy (DE) type term in the effective action (\ref{sea4}), which describes the effects of the gravitational anomaly condensate:
\begin{align}\label{lambda}
\mathcal S_\Lambda  &= \sqrt{\frac{2}{3}}\,
\frac{\alpha^\prime}{96 \, \kappa} \, \int d^4 x \sqrt{-g}\, \langle \overline b \, R_{\mu\mu\rho\sigma}\, \widetilde R^{\mu\nu\rho\sigma} \rangle \nonumber \\ & \simeq   \int d^4 x \, \sqrt{-g}\, \Big(\nprooftext{5.86 \times 10^7 \, \, \sqrt{2\, \epsilon} \, 
\Big[\frac{|\overline b(0)|}{\MPl}} + \sqrt{2\, \epsilon} \,  \mathcal N \Big] \, H^4 \Big) \, 
%\nonumber  \\ & 
\equiv \nftext{-} \int d^4x \, \sqrt{-g} \,\nprooftext{ \frac{\Lambda}{\kappa^2}}~.
\end{align}
Above, the symbol $\simeq $ indicates and order of magnitude estimate, and we used \eqref{k01},  \eqref{transpl}, \eqref{lv} and \eqref{slowrollepsi3}, and  took into account that $H\, t $ is bounded from above by $(H\, t)_{\rm max}$, a maximum order of magnitude, evaluated at the end of the inflationary period, for which $(H\, t)_{\rm max} = H\, t_{\rm end} \sim {\mathcal N} = 60-70 $, with ${\mathcal N}$ the number of e-foldings. We also set $\epsilon \sim 10^{-2}$, as required by inflationary phenomenology ({\it cf.} \eqref{slowrollepsi}). In a sense, the term \eqref{lambda} is equivalent to a quantum-gravity-induced ``trace'' of the Cotton tensor, which, as we have seen above, is {\it classically} traceless \eqref{tracecot}. Such a $\Lambda$-type-term cannot arise in a classical general-relativistic treatment, and, hence, it was not considered in the analysis of \cite{jackiw}.
In fact, such a VEV acts as a new effective (induced) contribution to the vacuum energy density \eqref{endenpress2}.
}

\nickcorr{We next notice that, if we consider \nprooftext{ $\overline b(0) < 0$ and} transplanckian values for $|\overline b(0)| \gg \MPl $ (in analogy with what happens with the inflaton field in conventional large-field inflationary scenarios), then the quantity $\Lambda > 0$ in \eqref{lambda} does not change order of magnitude during the entire inflationary period, for which $H \simeq $ constant, and thus it can be approximated by a constant. In fact, for this purpose, it suffices to assume 
\begin{align}\label{b010}
|\overline b(0)| \gtrsim \sqrt{2\, \epsilon}\, {\mathcal N} \, \MPl \sim 10 \, \MPl~.
\end{align}
Hence, the term \eqref{lambda} behaves as a positive-cosmological-constant (de Sitter) type term, which is responsible for inducing inflation. Quantum fluctuations of the condensate are then responsible for deviations from scale invariance, providing a novel mechanism for cosmological perturbations to be explored further and compared with data in a future work. }

\nickcorr{We would now like to demonstrate the r\^ole of the anomaly-condensate-induced dark energy density 
\eqref{lambda} in ensuring that the temporal ($00$) component of the conserved modified stress-energy tensor $\widetilde T^{\mu\nu}_{b+{\rm gCS} + \Lambda}$, \eqref{cons}, which would correspond to the total energy of the system, is {\it positive}, thus implying stability. To this end, we consider \eqref{cons}, and assume a non-zero vacuum expectation value \eqref{rrt} of the anomaly term, due to gravitational waves, and an isotropic and homogeneous temporal component of the Cotton tensor $\mathcal C^{00}(t)$. Anticipating the latter to be proportional to $\Theta^2 \ll 1$ ({\it cf.} \eqref{rrt})), one obtains from \eqref{csder}:\footnote{\nickcorr{For the remainder of this subsection we treat $M_s=(\alpha^\prime)^{-1/2}$ and $\MPl = \kappa^{-1}$ as independent, see Appendix \eqref{sec:appA}, in order to demonstrate the generic nature of our results regarding the r\^ole of the condensate; we revert back to the 
special case $\alpha^\prime \sim \kappa$, adopted in our work so far, from the next section onwards.}}
\begin{align}\label{solution}
\mathcal C^{\mu 0}_{\,\,\,\,\,\,;\mu} &= \frac{d}{dt} \mathcal C^{00} + 4 H\, \mathcal C^{00}  \simeq - \frac{1}{8} \, \dot{\overline b} \, \langle R^{\alpha\beta\gamma\delta} \, \widetilde R_{\alpha\beta\gamma\delta}\rangle 
\, \nonumber \\ & \simeq  \, - \frac{1}{8} \sqrt{\frac{2}{3}}\, \frac{\alpha^\prime \, \kappa}{12} \, H \, 
\frac{1}{\pi^2} \Big(\frac{H}{M_{\rm Pl}}\Big)^2 \, \mu^4 \, {\dot {\overline b}}^2,
 \end{align}
in a mean field approximation, to lowest order in a perturbative $\Theta$ expansion, whereby in the left-hand side of the equation we considered a (spatially-flat) FLRW background space-time. In arriving at \eqref{solution} we used \eqref{tracecot}. We also remind the reader that the notation $\overline b$ denotes the KR background, satisfying \eqref{lv}.  
Assuming a (approximately)  constant in time $\mathcal C^{00}$ and homogeneity and isotropy ({\it i.e}. setting $\mathcal C^{0i} =0$) we find  from \eqref{solution} the consistent solution 
\begin{align}
 \mathcal C^{00} \simeq -\epsilon \, \sqrt{\frac{2}{3}}\, \frac{\alpha^\prime \, \kappa}{192} \,  
\frac{1}{\pi^2} \, \mu^4 \, \, H^4 \, < \, 0,
\end{align}
where we used \eqref{slowrollepsi2} keeping, though, the slow-roll parameter $\epsilon$ generic for the moment.
From \eqref{einsteincs}, this contributes to the energy density of the vacuum a {\it negative} term,\footnote{For the benefit of the reader, we note that the negativity of $\mathcal C^{00}$ is robust against a change of signature of the coefficient of the gCS term in \eqref{sea4}, given that the latter will be compensated by a corresponding change of signature of the 
 averaged anomaly \eqref{rrt}, which is proportional to that coefficient.} in a similar spirit to the Gauss-Bonnet-dilaton coupling~\cite{kanti}, also appearing in string-effective actions, which, like the gravity-anomaly term \eqref{bgrav}, also involves terms quadratic in the Riemann curvature  tensor:
\begin{align}\label{envac}
\rho^{{\rm gCS}} &= \sqrt{\frac{2}{3}}\, \frac{\alpha^\prime}{12\, \kappa}\, \mathcal C^{00} \simeq  -
\frac{2}{3}  \, \frac{1}{\pi^2 \times 192 \times 12} \,  \epsilon \, \Big(\frac{\mu}{M_s}\Big)^4\, H^4 \nonumber \\ & \simeq -
2.932 \times 10^{-5} \,  \epsilon \, \Big(\frac{\mu}{M_s}\Big)^4\, H^4 \, <\, 0.
\end{align}
Using \eqref{A=0}, we then obtain in order of magnitude\footnote{An important remark we would like to make is that the condition \eqref{A=0} 
is assumed to be valid as an order of magnitude estimate, and does {\it not} imply that the cutoff $\mu$ varies with $H$ as $H^{-1/2}~$. The quantity $\mu$ is independent of $H$ and a constant in time. This implies that the gCS term varies as $H^4$, in contrast to the $\rho^b$ term that varies as $H^2$. However, for our solution under which \eqref{A=0} is valid, both terms are of the same order of magnitude.}
\begin{align}\label{envac2}
\rho^{{\rm gCS}}  \simeq  - 1.484 \,  \epsilon \, \MPl^2 \, H^2,
\end{align}
%
%Using \eqref{transpl}, we then obtain in order of magnitude 
%\begin{align}\label{envac2}
%\rho^{\rm g\mathcal C\mathcal S}  \simeq  -2.93 \times 10^{7} \, \epsilon \, H^4 = - 3M_{\rm Pl}^4 \Big[ 0.977 \times 10^7 \Big(\frac{H}{\MPl} \Big)^4\Big]\, \epsilon,
%\end{align}
%where we keep the $\epsilon$ generic here, for reasons that will become clear later. 
}

\nickcorr{From \eqref{cons}, and the first equality of \eqref{solution}, we also obtain
\begin{align}\label{cons3}
&\frac{d}{dt}(\rho^b + \rho^{{\rm gCS}}) + 3 H \Big( (1+w_b)\, \rho^b + \frac{4}{3}\,\rho^{{\rm gCS}} \Big)  \simeq 0 
\nonumber \\ & \Rightarrow \quad
\rho^b \simeq -\frac{2}{3}\, \rho^{{\rm gCS}}~, 
\end{align}
where the last result holds if $\frac{d}{dt}(\rho^b + \rho^{{\rm gCS}} ) \simeq 0$ and we took into account that the equation of state of the pure $b$-fluid is $w_b=1$, as follows from \eqref{stressb}. Thus, we see from \eqref{cons3} that the negative value of the 
$\rho^{{\rm gCS}}$ is essential for the consistency of the approach, since 
it is only then that the energy conservation of the total stress energy tensor \eqref{cons} leads to consistent results, given the positivity of $\rho_b$. From \eqref{envac2} and \eqref{cons3} we then obtain 
\begin{align}\label{ben}
\rho^b \simeq 0.9895 \, \epsilon \, \MPl^2 \, H^2.
\end{align}
The KR axion stress tensor $T^{\mu\nu}_b$ \eqref{stressb} in \eqref{einsteincs}, on the other hand, will contribute $H^2$ terms to the vacuum energy density but of the {\it same order of magnitude} as the $\sim H^4$ terms of the gravitational anomaly, due to \eqref{cons3}):
\begin{align}\label{enpressphib2}
 \rho^{b} = \frac{1}{2} (\dot {\overline b})^2 \simeq  \epsilon \, M_{\rm Pl}^2 \, H^2~,
 \end{align}
where we used the first equality in \eqref{slowrollepsi2}. Comparing with \eqref{ben} we can then see the consistency of our approach,
for every value of the slow roll parameter $\epsilon < 1$ and every value of $H$. We can then adopt the range of values for these parameters dictated by the data~\cite{planck}, \eqref{slowrollepsi} and \eqref{Hinfl}, respectively. The 1$\%$ discrepancy between \eqref{enpressphib2} and \eqref{ben} is to be expected, according to our discussion in Appendix \ref{sec:appA} ({\it cf.} \eqref{A=0N}), which implies that the result \eqref{envac2} for $\rho^{g\mathcal C\mathcal S}$  should be mulitplied by an uncertainty factor  $(1 - \frac{\xi}{3\mathcal N})$ 
in the range $0.9889 \lesssim (1 - \frac{\xi}{3\mathcal N}) \lesssim 0.9905$. This is perfectly justified when taking into account also theoretical uncertainties in our estimate \eqref{rrt} of the gravitational-anomaly condensate. }

\nickcorr{We now remartk that, as follows from \eqref{envac}, \eqref{cons3}, the sum of the respective energy densities turns out to be {\it negative} 
\begin{align}\label{negative}
\rho^b + \rho^{{\rm gCS}} 
  = \frac{1}{3}\, 
\rho^{{\rm gCS}} 
  \, \simeq - 0.496  \, \epsilon\, \MPl^2 H^2 < \, 0,
\end{align}
indicating that, if there were no other contributions to the energy density of the KR axion-gravity system, the gravitational anomaly would induce an instability in the de Sitter vacuum. }

\nickcorr{However, as already mentioned, the term \eqref{lambda} in the energy density, induced by the anomaly condensate, leads to an additional $\Lambda$-de-Sitter-type contribution to the modified stress-energy tensor \eqref{cons}, with an equation of state $\rho_\Lambda = -p_\Lambda $, 
which 
%\begin{align}\label{cons4}
%\kappa^2 \, {\widetilde T}_{b + g\mathcal C\mathcal S + \Lambda}^{\mu\nu} \equiv \sqrt{\frac{2}{3}}\,\frac{\alpha^\prime\, \kappa}{12} \mathcal C^{\mu\nu} + \kappa^2 T^{\mu\nu} +  \Lambda \, g^{\mu\nu} ~, \qquad \Lambda \equiv \langle b (x) \, R_{\mu\mu\rho\sigma}\, \widetilde R^{\mu\nu\rho\sigma} \rangle \simeq 
%5.86 \times 10^7 \, \epsilon\, \mathcal N \, H^4 > 0~.
%\end{align}
does not modify its conservation \eqref{cons}, but corresponds to a (positive) contribution to the total energy density
\nprooftext{$\rho^\Lambda \simeq 5.86. \times 10^7 \, \sqrt{2\, \epsilon} \, \frac{|\overline b(0)|}{\MPl} \, H^4$.}
For $\epsilon \sim 10^{-2}$, $\mathcal N ={\mathcal O}(60-70)$ and $\overline b(0) \gtrsim 10 \, \MPl$, it dominates the total energy density, 
\begin{align}\label{toten}
\rho_{\rm total} = \rho^b + \rho^{{\rm gCS}} + \rho^\Lambda \simeq 
3\MPl^4 \, \Big[ -1.7 \times 10^{-3} \Big(\frac{H}{\MPl}\Big)^2 
+ \frac{\sqrt{2}}{3} \, \nprooftext{\times \, 5.86 \, \times \, \frac{|\overline b(0)|}{\MPl}\, \times 10^6} \, \Big(\frac{H}{\MPl}\Big)^4 \Big] > 0~,
\end{align}}
 
\nickcorr{Before closing the current subsection, we would like to compare the expression  \eqref{toten} with the form of the RVM energy density  \eqref{rLRVM}.  For the conventional RVM, the expecation is that $\nu, \alpha$ are positive~\cite{rvm1,rvm2,rvm3,JSPrevs,SolGo2015}. 
On comparing \eqref{toten} with \eqref{rLRVM}, by identifying $\rho_{\rm total}$ and $\rho^{\Lambda}_{\rm RVM}(H)$, we make the following observations for our model:
\begin{itemize} 
\item{(i)} In our string-inspired model for the early Universe we have $c_0=0$.  Such a term may appear in the late eras of the Universe, e.g. through the generation of a potential for the $b(x)$ field, as we shall discuss in section \eqref{sec:mixing}.
\item{(ii)} 
As a result of the negative contributions of the Cotton tensor 
to the energy density $\rho_{\rm total}$, the coefficient of the $H^2$ terms in \eqref{toten} would imply, on account of \eqref{rLRVM}, a $\nu < 0$ in the early Universe, where gravitational anomaly contributions dominate. However there is no contradiction with the spirit of the RVM. Indeed, in  our case, the Cotton tensor is {\it not} a vacuum contribution, as it is associated with {\it gravitational-wave excitations} of the FLRW metric background space-time. For the background space-time, the Cotton tensor vanishes, as we have already mentioned~\cite{jackiw}. 
On the other hand, the KR axion is associated with the spin-one antisymmetric tensor field of the massless gravitational multiplet of strings~\cite{string}, which in the case of the (phenomenologically relevant) superstring constitutes the ground state, due to the absence of tachyon modes from the spectrum. 
In this sense, the RVM should be associated with the contributions of the $b$-axion field stress tensor $T^{\mu\nu}_b$ \eqref{stressb} alone, ignoring the Chern-Simons terms, which, on account of \eqref{slowrollepsi}, \eqref{slowrollepsi2} leads ({\it cf.} \eqref{rLRVM}) to a {\it positive} $\nu$ coefficient \eqref{nuvalue}, as mentioned previously. 
In the radiation and matter dominated eras, where the gravitational anomalies cancel~\cite{essay}, as we shall discuss in section \eqref{sec:cancel}, this is also the case. 
\item{(iii)} On the other hand,  
we find  that the coefficient $\alpha $ is {\it positive} already during the inflationary era, and of order:
\begin{equation}\label{eq:alphavalue}
\alpha= \frac{\sqrt{2}}{3} \,  \nprooftext{\times \, 5.86 \, \times \,  \frac{|\overline b(0)|}{\MPl} \, \times 10^6} \left (\frac{H_I}{\MPl}\right)^2\sim \nprooftext{ 2.8\,  \times 10^{-2} \, \frac{|\overline b(0)|}{\MPl}}~,
\end{equation}
assuming a (typical) Hubble parameter $H_I$ during inflation of order \eqref{Hinfl}. 
Notice that the value of $\alpha$ does not depend on the specific magnitude of the string scale, but only on the ratio $\mu/M_s$ (see discussion in Appendix \ref{sec:appA}, and Eq.~\eqref{transpl2}). From \eqref{toten}, and  \eqref{Hinfl}, then, one easily sees that we may identify the total energy density with a GUT-like potential $V \sim M_X^4$ corresponding to an energy scale $M_X$:
\begin{align}\label{mxscale}
\rho_{\rm total} &\simeq \rho^\Lambda \sim M_X^4 \simeq  \nprooftext{8.3\, \times \, \frac{|\overline b(0)|}{\MPl}\, \times10^{-10}}\, \MPl^4 \nonumber \\&  \Rightarrow \quad M_X \simeq \nprooftext{1.3 \, \times 10^{16} \,  \Big(\frac{|\overline b(0)|}{\MPl} \Big)^{1/4}}~ {\rm GeV},
\end{align}
which, for \nprooftext{$|\overline b(0)| \gtrsim 10 \, \MPl$} ({\it cf}. \eqref{b010}) is in agreement with generic RVM predictions based on GUT models~\cite{SolGo2015}. 
\end{itemize}}

\nickcorr{The next point is also of crucial interest for us. The quantum fluctuations of the gravitational fields that produce the anomaly condensate $\Lambda$-term \eqref{lambda} could be describe by an effective action of a {\it composite} scalar mode, $``\phi''$, consisting of a coherent superposition of quantum $b$-axion and graviton modes. The (gravitational) interactions among those fields, will result in self interactions of the condensate field, and thus an effective potential, that can in principle be computed. In practice, as the full string theory is in operation here, such a task is currently not feasible. In simpler situations, for instance dynamically broken supergravity scenarios, a low-energy effective potential of condensates of gravitino fields has been computed in \cite{houston}, and the situation resembled the Starobinsky model of inflation~\cite{staro}, under the conditions discussed in detail in that work.}

Interestingly enough, the  $\sim H^4$ behavior can be equivalently mapped to a scalar field behavior.  Such \noveltext{a} scalar field picture will be called the ``vacuumon picture'' of the RVM since the field $\phi$ is called the \noveltext{{\it vacuumon}~\cite{vacuumon}}. To implement the  mapping of the RVM to the vacuumon picture one has the following correspondence with the total density and pressure\,\cite{bls1,bls2,vacuumon}:
\begin{eqnarray}\label{map}
\rho_{\rm tot}\equiv \rho_{\phi}={\dot \phi}^{2}/2+V(\phi)\,\ \ \ \ \ \ \ \ \ \
p_{\rm tot}\equiv p_{\phi}={\dot \phi}^{2}/2-V(\phi)\,,
\end{eqnarray}
with
\begin{equation}
\dot{\phi}^{2} =-\frac{2}{\kappa^{2}}\dot{H} \;, \label{ff3}
\end{equation}
%%%%%%%%%%%%%%%%%%%%%%%%%%%%%%%%%%%%%%%%%%%%%%%%%%%%%%%%%%%%%%%%%%%%%%%%%%%%%%%%%%%%%%%%%
and
\begin{eqnarray}
\label{Vz} V&=&\frac{3H^{2}}{\kappa^{2}}\left(
1+\frac{\dot{H}}{3H^{2}}\right)= \frac{3H^{2}}{\kappa^{2}}\left(
1+\frac{a}{6H^2}\,\frac{d H^2}{da}\right) \;,
\end{eqnarray}
%%%%%%%%%%%%%%%%%%%%%%%%%%%%%%%%%%%%%%%%%%%%%%%%%%%%%%%%%%%%%%%%%%%%%%%%%%%%%%%%%%%%%%%%%%
 is the effective potential of the vacuumon scalar field $\phi$. After we  have realized that the higher order term $\sim H^4$ of the RVM density  \eqref{rLRVM} can indeed be generated thanks to the gravitational anomaly term, one can just use  the vacuumon picture. In particular, using Eq. (\ref{Vz}),  one can compute the effective potential associated to the  RVM density, whose explicit form was given in the aforementioned references, with the result
\begin{equation}
\label{Pott}
U(\phi)=\frac{H^{2}_{I}}{\alpha \kappa^{2}}\; \frac{2+{\rm
cosh}^{2}(\kappa \phi)} {{\rm cosh}^{4}(\kappa \phi)}  \;.
\end{equation}
\noveltext{In this scenario, if the potential \eqref{Pott} would be the true potential to describe the dynamics of the quantum fluctuations  of the scalar anomaly-condensate in our case, this would be the potential  assumed in Eq.(\ref{enpressphib2B}).}

\nickcorr{However, there are some subtle issues in the approach of \cite{bls1,bls2,vacuumon} that prevent one from extending it straightforwardly to the case examined in this article. The scalar field $\phi$ in \eqref{map} and \eqref{Pott} is a classical field, which is used to describe the temporal evolution of the classical RVM vacuum.  It is by no means equivalent to the true quantum scalar mode encoded in the quantum fluctuations of the condensate \eqref{order}, which, as we mentioned above, needs to be computed within the proper string theory framework.  That scalar condensate mode would be the true ``vacuumon'' field, which should be used in the inflationary phenomenology of cosmological perturbations in our scenario. Hence, the true effective potential of this composite ``vacuumon'', including properly all the quantum corrections, might be very different from the ``classical'' potential \eqref{Pott} used in \cite{vacuumon} to describe the classical RVM evolution. Nonetheless, our arguments above indicate that,  in the present string-inspired RVM scenario, where gravitational anomaly condensates coupled to KR axions
from the massless bosonic gravitational string multiplet, induce dynamically de Sitter space times, there could be such a fully-fledged vacuumon quantum field, that also represents the fluctuations of the RVM and thus could be used for the inflationary phenomenology of the model.}

\nickcorr{If we use the (correct) vacuumon representation, then, its aforementioned  effective potential would contain the same information as if one would use the RVM density \eqref{rLRVM}.
Borrowing the correspondence formula (\ref{ff3}) between the two pictures we find that the slow roll parameter for the vacuumon is
\begin{align}\label{slowrollepsi3b}
 \epsilon = -\frac{\dot{H}}{H}=\frac{1}{2} \frac{1}{(H \MPl)^2}\, {\dot \phi}^2 \simeq
 10^{-2},
 \end{align}
and as we can see it takes exactly the same form as for the inflaton case in Eq.\, (\ref{slowrollepsi}).  The upshot is that the averaged  gCS anomaly term over the de Sitter spacetime leads to a $\sim H^4$ contribution to the effective vacuum energy  density of the RVM and there is no need to introduce any \textit{ad hoc} inflaton to trigger inflation by hand, \noveltext{given that} inflation can be entirely driven by this term  ~\cite{bls1,bls2,bls3,bls4,GRF2015,bls5}.}

Thus, we can stay exactly with the same fundamental fields \noveltext{as the ones we started with in} the effective action of bosonic string theory in Sec. \ref{sec:2A}.  The RVM density  \eqref{rLRVM} appears to be an effective description of the same \noveltext{physical} context when it is averaged over the inflationary spacetime. Such \noveltext{a} description can alternatively be formulated within the vacuumon picture and in this case it \noveltext{is} a scalar field (the vacuumon) which mimics the $\sim H^4$ behavior (and \noveltext{thus} the inflaton behavior) through  an appropriate effective potential. The vacuumon, therefore, is not an external scalar field but just an internal degree of freedom associated \noveltext{with} the gCS anomaly,  \noveltext{leading to} the scalar field representation of the higher order  $\sim H^4$ term in the original averaged effective action over the de Sitter background. This fact allows us to entirely reproduce the same considerations as in the previous section but without invoking any new scalar field, which would be extraneous to our original  massless bosonic gravitational multiplet of string theory \noveltext{(as this would require an appropriate dilaton potential, in case the dilaton is identified with the inflaton, which however cannot be generated at tree level in string loop perturbation theory, but requires higher string loops, which we do not have control of)}.  The RVM formulation is therefore fully self-consistent for the description of the cosmic evolution.

\section{Post Inflationary Era and Anomalous Matter over Antimatter Dominance \label{sec:lepto}}

\subsection{Chiral Fermionic Matter and Cancellation of Gravitational Anomalies \label{sec:cancel}}

At the end of inflation, the proper decay of the running vacuum to matter and radiation components will reheat the universe and lead to
the appearance of fermions among other matter.  If such fermions have anomalous axial currents, then matter-antimatter asymmetry in the observable universe could be due to such an anomaly in the post-inflationary era through the mechanism advocated in~\cite{decesare,bms,bms2}, as we now proceed to explain.\footnote{We remind the reader that in our approach we do not discuss the role of (primordial) fermionic excitations during inflation, since we assume that only bosonic gravitational degrees of freedom describe the string-inspired Universe. Thus the considerations of \cite{stephon} for generating sufficient leptogenesis only through the gravitational anomaly induced by gravitational waves \newtext{do not apply here}, given that the relevant fermionic chiral matter in our model is generated only at the end of inflation, \newtext{not during inflation}. For completeness, we mention though that there are works in the literature~\cite{shapiro,hehl} which discuss the possibility that primordial fermionic torsion contributions in torsional versions of General Relativity (in which the spin connection and vielbein are treated as independent fields), result, through appropriate fermion condensates, in inflation. We shall not discuss such scenarios here.}

To this end, we first assume that  the space-time after inflation has the ordinary FLRW form (in the radiation era), since any primordial gravitational wave perturbations would have been washed out during inflation. This would imply that the gravitational anomaly (\ref{pontryagin2}) would vanish at large scales for such space-time backgrounds. However, locally gravitational wave perturbations are present, and could jeopardise the local diffeomorphism invariance of the
radiation (and matter) quantum theory, according to our previous discussion.
We now postulate that the generation of {\it chiral matter} at the end of inflation leads to a {\it cancellation} of the gravitational anomalies, even {\it locally}. \newtext{Otherwise diffeomorphism invariance would be violated locally in the presence of matter.} However, and this will turn out to be crucial for linking KR axions to DM in our scenario, as we shall discuss later, we assume that \newnewtext{U(1) {\it chiral anomalies}}~\cite{adler} remain {\it uncompensated}. These do not contribute to stress tensor of matter, unlike the gravitational ones, hence there is no fundamental reason for the matter theory to be chiral-anomaly free,  \newtext{only the gauge symmetry must be anomaly free so as to preserve the Ward identities}. Thus, we postulate the following relation during the radiation (and matter) eras~\cite{essay}:
\begin{eqnarray}
   \label{anom2}
&& \partial_\mu \Big[\sqrt{-g}\, \Big(  \sqrt{\frac{3}{8}} \kappa\, J^{5\mu}  -  \sqrt{\frac{2}{3}}\,
\frac{\kappa}{96} \, {\mathcal K}^\mu  \Big) \Big] \!=\!   \sqrt{\frac{3}{8}} \kappa\,  \frac{e^2}{8\pi^2}  \, \sqrt{-g}\,  {F}^{\mu\nu}\,  \widetilde{F}_{\mu\nu}
\nonumber  \\ && = -\sqrt{\frac{3}{8}} \kappa\,  \frac{e^2}{4\pi^2}  \epsilon^{0ijk}\, F_{0i} \, F_{jk} =
  - \sqrt{\frac{3}{8}} \kappa\,  \frac{e^2}{2\pi^2}  \, \sqrt{-g} \, E^i \, B^j \, g_{ij}  \; ,
\end{eqnarray}
where we used \eqref{leviC}, \eqref{duals}; $E^i$ (${B^i}$) denote the electric (magnetic) cosmic fields in curved space, respectively (from the third equality in \eqref{anom2}, the reader can readily see the topological nature (i.e. independence of the metric) of the chiral anomaly, 
\nickcorr{which thus, unlike the gravitational anomaly, does not have any contributions to the stress-energy tensor of the KR axion field}\nickcorr{\footnote{\nickcorr{In this work, for simplicity, we consider only chiral U(1) anomalies. In general, one may face situations in which there are also QCD triangle anomalies, which would amount to adding to the right-hand sides of the first and subsequent equalities of \eqref{anom2} a term of the form 
$ +  \sqrt{\frac{3}{8}} \,\frac{\alpha^\prime}{\kappa}\, \Big(\frac{\alpha_s}{8\pi}\, \sqrt{-g} \, G_{\mu\nu}^a \, \widetilde G^{a\mu\nu} \Big)$, 
where $G_{\mu\nu}^a$ denotes the gluon field strength, with $a=1, \dots 8$ an adjoint SU(3) colour index, and $\alpha_s$ is the strong interactions fine structure constant. This term, like the chiral U(1) anomaly one, is also topological and does not yield any contributions to the stress-energy tensor of the KR field.}}}); $J^{5\mu} = \sum_{j} \overline \psi_j \, \gamma^\mu \, \gamma^5 \, \psi_j $ is the axial current, with the summation being over appropriate fermion species $\psi_j$ of the matter sector, e.g. charged chiral quarks or leptons in the SM sector.

The reader is reminded that the appearance of the square of the QED coupling $e$ (electron charge) on the right-hand-side of (\ref{anom2}), is a result of the fact that the chiral anomaly (like the gravitational anomalies) is a one-loop exact effect~\cite{adler}, with the chiral fermions circulating in the loop.
For concreteness and brevity, in \eqref{anom2} we assumed the circulation of a single chiral fermion of charge equal to the electron charge $e$. In realistic applications, one should replace
$e^2$ on the right-hand side of \eqref{anom2} by an `effective' squared charge:
\begin{align}\label{gench}
e^2 \quad  \Rightarrow \quad e_{\rm eff}^2 = e^2 \, {\mathcal N},
\end{align}
where ${\mathcal N}$ is a model dependent numerical constant, which depends on the number and kind of fermions circulating in the loop, and is proportional to the square of their electric charges normalised to the electron charge $e$. For instance, for QCD chiral anomalies, of $N_f$ species light quarks, with electric charges $q_I$, $I=1, \dots N_f$, each of which comes in $N_c$ colours (for ordinary QCD, $N_c=3$), one has~\cite{zhit} ${\mathcal N}= \frac{N_c}{N_f} \sum_{I=1}^{N_f} \Big(\frac{q_I}{e}\Big)^2$. The generalisation \eqref{gench} will be understood in what follows.

We stress once more that, in our approach, the U(1) photon and fermion fields are produced by the decay of the running vacuum at the end of the inflationary era~\cite{bls1}. During the exit phase from inflation, there is also the KR axion, which is undiluted,  (\ref{krbeom2}), \eqref{slowrollepsi2}. As we shall discuss below, this field plays an important r\^ole in {\it both} the cancellation of the gravitational anomaly and inducing leptogenesis during the radiation era~\cite{decesare,bms,bms2}.

Let us see these effects in a detailed manner by discussing the low-energy (string-inspired) effective action during the radiation era. First we remark that, upon  inclusion of fermionic matter at the end of inflation,  the contorsion interpretation of the antisymmetric tensor field strength~\cite{string,kaloper,ms,decesare}, ${\mathcal H}_{\mu\nu}^\rho$, implies a {\it minimal} coupling of this field to the axial fermion current, given that  the corresponding Dirac Lagangian for fermions in torsional gravitational backgrounds~\cite{hehl,shapiro} contains the generalised spin-connection $\overline \omega_{ab\mu}= \omega_{ab\mu} + K_{ab\mu}$,  $K_{abc} =\frac{1}{2} \, ({\mathcal H}_{cab}  - {\mathcal H}_{abc} - {\mathcal H}_{bca}) = - \frac{1}{2} {\mathcal H}_{abc}$:
\begin{align}\label{fermions}
S_{Dirac} &= \,  \int d^4x \sqrt{-g} \, \Big[ \frac{\imath}{2} \,\Big(\overline \psi_j \gamma^\mu {\overline {\mathcal D}}(\overline \omega)_\mu \, \psi_j - ( {\overline {\mathcal D}}(\overline \omega)_\mu \, \overline \psi_j  )\, \gamma^\mu \, \psi_j \Big) - m^{(j)}\, \overline \psi_j \, \psi_j \Big], \nonumber \\
& =\,  \int d^{4}x\sqrt{-g}\bar{\psi}_j\Big(\frac{\imath}{2}\Gamma^{a} \stackrel{\leftrightarrow}{\partial_{a}} - m\Big)\psi_j - \int d^{4}x\sqrt{-g} \, ({\mathcal F}_a + B_a)\, \bar{\psi}_j\gamma^{5}\Gamma^{a}\psi_j  \nonumber \\
& \equiv \; S_{Dirac}^{Free} + \int d^{4}x\sqrt{-g}\, (B_{a}  + {\mathcal F}_a)\,J^{5\, a}~,
\end{align}
with Latin indices $a, b, c, \dots$ denoting tangent-space indices, raised and lowered by the Minkowski metric $\eta^{ab}$ of the tangent space
(at a point with coordinates $x^\mu$) of a space-time with metric $g_{\mu\nu} (x) = e_\mu^a (x) \, \eta_{ab} \, e_\nu^b (x)$, with  $e^{a}_\mu (x)$  the vielbeins and  $e^\mu_a (x)$ their inverse. $\Gamma^a$ is a tangent-space Dirac matrix, such that $\gamma^\mu (x) = e^\mu_{a} (x) \, \Gamma^a$,
and we used the standard notation for $\overline \chi \,  \stackrel{\leftrightarrow}{\partial_{a}}\psi = \overline \chi \partial_a \psi - \overline{\partial_a \chi}\,  \psi $. The covariant derivative is defined as ${\overline {\mathcal D}}_a  = \partial_a  - \frac{\imath}{4} \, \overline \omega_{bca}\, \sigma^{bc}$, $\sigma^{ab} = \frac{\imath}{2}[\Gamma^a, \Gamma^b]$,  ${\mathcal F}^d  =   \varepsilon^{abcd} \, e_{b\lambda} \,  \partial_a \, e^\lambda_c$,
$B^d= -\dfrac{1}{4}\,\varepsilon_{abc}^{\;\;\;\;\;\;d}\,{\mathcal H}^{abc}$,  and $J^{5 \, \mu} = \bar{\psi}_j\,  \gamma^{\mu} \,\gamma^{5}\psi_j$, and
correspondingly $J^{5 \, a} = \bar{\psi}_j \, \Gamma^{a} \,\gamma^{5}\psi_j$. In arriving at \eqref{fermions} we used standard
properties of the flat-(tangent) space $\Gamma^a$-matrices.

Adding (\ref{fermions}) to (\ref{sea2}), implementing the constraint \eqref{delta} via a Lagrange multiplier pseudoscalar field $b(x)$,\footnote{It is crucial for the reader to notice that we keep only the gravitational part of the anomaly, setting the non-Abelian gauge fields ${\mathbf A}$ to zero; we stress that we do {\it not} include Abelian U(1) Chern-Simons terms in the modified Bianchi identity \eqref{modbianchi2}, as we anticipate the existence of chiral U(1) anomalies {\it only} in the fermion sector of the model.} canonically normalised as before, and integrating over the field ${\mathcal H}$ in the path integral,
we easily arrive at an effective action (using \eqref{akappa}):
\begin{align}\label{sea6}
S^{\rm eff} =&\; \int d^{4}x\sqrt{-g}\Big[ -\dfrac{1}{2\kappa^{2}}\, R + \frac{1}{2}\, \partial_\mu b \, \partial^\mu b -  \sqrt{\frac{2}{3}}\,
\frac{\kappa}{96} \, \partial_\mu b(x) \, {\mathcal K}^\mu
\Big] \nonumber \\
&+ S_{Dirac}^{Free} + \int d^{4}x\sqrt{-g}\, \Big( {\mathcal F}_\mu + \frac{\kappa}{2} \, \sqrt{\frac{3}{2}} \, \partial_{\mu}b \Big)\, J^{5\mu}    - \dfrac{3\kappa^{2}}{16}\, \int d^{4}x\sqrt{-g}\,J^{5}_{\mu}J^{5\mu}  + \dots \Big] + \dots,
\end{align}
where the $\dots$ in (\ref{sea6}) indicate gauge field kinetic terms, as well as terms of higher order in derivatives, of no direct relevance to us here. The reader should notice the four fermion axial-current-current term in \eqref{sea6}, which is characteristic of Einstein-Cartan theories with torsion~\cite{hehl,shapiro}, the latter being provided here~\cite{string,kaloper} by the (totally antisymmetric) quantity $\epsilon_{\mu\nu\rho\sigma} \partial^\sigma b$ which is dual to the Kalb-Ramond antisymmetric tensor field strength ${\mathcal H}_{\mu\nu\rho}$, as discussed in section \ref{sec:2} ({\it cf}. \eqref{torcon}).

We also remark that gravitational-wave local perturbations during the radiation and matter (dust) eras lead in general to a non-trivial background ${\mathcal F}_\mu$ in \eqref{sea6}; however,
such perturbations are much more suppressed during the radiation (and matter) eras as compared with their primordial counterparts; in the subsequent discussion in this session, we consider a pure FLRW background as a sufficient approximation of the Universe at large scales in late eras. For such a pure FLRW metric $g_{\mu\nu}$ background (and in general spherically-symmetric space-times with diagonal metrics~\cite{mukho}) one has that ${\mathcal F}_a =0$.

The KR axion $b(x)$ background field equation of motion then, obtained from \eqref{sea6}, reads:
\begin{align}\label{bEB}
& \partial_{\alpha}\Big[\sqrt{-g}\Big(\partial^{\alpha}\bar{b}  -  \sqrt{\frac{2}{3}}\,
\frac{\kappa}{96} \, {\mathcal K}^{\alpha}  + \sqrt{\frac{3}{8}} \kappa\,  J^{5\, \alpha}  \Big)\Big] = 0 \quad \Rightarrow  \nonumber \\
& \partial_{\alpha}\Big[\sqrt{-g}\, \partial^{\alpha}\bar{b} \Big] =
\sqrt{\frac{3}{8}} \kappa\,  \frac{e^2}{2\pi^2} \, a^5(t) \,  E^i B^j \delta_{ij}~,
\end{align}
where, in the second line, we used \eqref{anom2} and the FLRW metric, $g_{ij}=a^2(t)\, \delta_{ij}$, $i,j=1,2,3$. The alert reader should have noticed that one would had arrived at the same equation, had one used the absence of gravitational anomalies in a background FLRW space-time, but of course our result emerging from anomaly cancellation is more general as it is independent of any metric perturbations (such as gravitational waves) that would jeopardise the diffeomorphism invariance of the  radiation/matter quantum field theory.

Nonetheless, for the purposes of our discussion in this section, we do assume on average a FLRW space-time during the radiation era at large scales, for which gravitational wave perturbations are suppressed. In this case, the chiral anomaly term on the right-hand side of \eqref{bEB} is associated with the covariant derivative of the axial fermion current~\cite{adler}
\begin{eqnarray}
   \label{anom}
J^{5\mu}_{\,\,\,\,\,\,\,\,;\mu}  \!&= \frac{1}{\sqrt{-g}} \, \partial_\mu \Big(\sqrt{-g} \, J^{5\, \mu}\Big) = &\! \frac{e^2}{8\pi^2} {F}_{\mu\nu}\,  \widetilde{F}^{\mu\nu}  =   - \frac{e^2}{2\pi^2} \,  a^2(t)\, E^i B^j \delta_{ij}.
\end{eqnarray}

Assuming homogeneous and isotropic situations at large (cosmological) scales, we only consider cosmic time dependent backgrounds ${\overline b}(t), \langle J^{5\, 0}(t)\rangle$. We denoted the background for the fermion axial current by $\langle \dots \rangle$, as we may also assume thermal averages (in our treatment we assume the existence of chiral currents, as, e.g., is the case of the SM chiral (left-handed) leptonic current, $J^5_L=\sum_f \Big(\overline \ell^{(f)}_{L} \, \gamma^\mu \, \ell^{(f)}_L + \overline \nu^{(f)} \gamma^\mu \nu^{(f)} \Big)$, with $\ell^{(f)}_L \, (\nu^{(f)})$  the charged leptons (active neutrinos), and $f$ a generation number. In models beyond the SM, other chiral fermions might play a r\^ole, as well).\footnote{Expansion of quantum fermionic axial currents  around such backgrounds is performed by writing
$J^{5\,0} = \langle J^{5\,0} \rangle + {\rm quantum~fluctuations}$ in \eqref{sea6}. We ignore the quantum fluctuations for our (classical) treatment in this session. This implies that, when we consider quadratic expressions of the axial current appearing in \eqref{sea6} (and in the stress tensor computed from it, see below \eqref{fst}) we should use
\begin{equation}\label{class}
J_0^{5} \, J^{5}_0  \simeq \langle J_{0}^{5} \rangle^2 \quad > \quad 0,
\end{equation}
\emph{etc}., which will be understood in what follows. However, it should be mentioned for completeness that, when one considers fully quantum corrections, including fermion path integration, as essential when dealing with fermions, then spatial components of the axial current $J_i^5$ should in general be considered in fermionic terms, and in general one may face a situation where {\it quantum fermion condensates}, $\ll J_\mu^5 J^{5\mu} \gg \quad  \ne  \quad \ll J^5_\mu \gg \, \ll J^{5\,\mu} \gg$, could arise, which could take on negative values (constant in cosmic time, for some period of the (early) Universe)
\begin{equation}\label{Fcondensate}
 \quad \ll J^5_\mu J^{5\, \mu} \gg \quad < \quad 0.
\end{equation}
This can lead to inflation (in the sense of equations of state of the form $p \simeq -\rho$)  in models where primordial fermions are considered~\cite{shapiro,hehl}.
For our purposes, where primordial fermionic matter excitations are assumed not to be present in the effective action \eqref{effdil} during the inflationary era, we shall consider the case \eqref{class}, where only the temporal component $\langle J_0^5 \rangle $ of the axial current of some chiral matter is non zero during radiation and matter eras.}

Some discussion is required at this stage concerning the space-time dependence of the electromagnetic fields, $\mathbf{E}(x)$ and $ \mathbf{B}(x)$ (with bold face notation referring to three vectors) entering \eqref{anom2}, \eqref{anom}. It is clear that one cannot have just time dependent fields, since, on account of Maxwell's equations, ${\mathbf \nabla} \times {\mathbf E} = - \dot{{\mathbf B}}$, with ${\mathbf \nabla}$ the spatial gradient. To have non trivial chiral anomalies at large (cosmological) scales, one may adopt the simplified (but concrete) example considered in \cite{frohlich}, according to which one has a monochromatic configuration of magnetic and electric fields, corresponding to a single mode of momentum $k > 0$, such that~\footnote{The relative sign differences  between \eqref{solEB} and the corresponding solution of \cite{frohlich} are due to the opposite sign of the term coupling the KR-axion with the chiral anomaly in \eqref{sea6} from that of the corresponding term in the action of \cite{frohlich}.}
\begin{align}\label{solEB}
{\mathbf B}(t,z) &= B(t) \, \Big(-{\rm sin}(kz), \, {\rm cos}(kz), \, 0\Big), \nonumber \\ {\mathbf E}(t,z) &= -\frac{1}{k} \, \dot{\mathbf B}(t,z) = -\frac{1}{k} {\dot B}(t) \, \Big(-{\rm sin}(kz), \, {\rm cos}(kz), \, 0\Big).
\end{align}
Such configurations have been argued in \cite{zhit} to play a r\^ole in providing a source for the dark energy in the Universe. We shall take a different point of view in the current work, where we shall argue that such configurations can lead to a source of (stiff~\cite{stiff}) dark matter, through the solution \eqref{bEB} of the KR background.

The important thing to observe~\cite{frohlich} is that the chiral anomaly corresponding to \eqref{solEB} has {\it only} time dependence for a FLRW metric with a scale factor $a(t)$:
\begin{align}\label{cati}
\sqrt{-g(t)} \, E^i (t,z) \, B^j (t,z) g_{ij}(t) = - a^5(t) \, \frac{1}{2\, k}\, \frac{d}{dt} (B^2(t)).
\end{align}
In such a case, the general solution of \eqref{bEB} is :
\begin{align}\label{bsolrad}
\dot{\overline{b}} &= \frac{{\mathcal C}_0}{a^3(t)} -  \sqrt{\frac{3}{8}} \kappa\,  \frac{e^2}{4\pi^2} \, \frac{1}{a^{3}(t)} \,  \int^t  dt^\prime \, a^5(t^\prime) \, \frac{1}{2\, k}\, \frac{d}{dt} (B^2(t)) \nonumber \\
&= \frac{{\mathcal C}_0}{a^3(t)} +  \frac{1}{k} \, \sqrt{\frac{3}{2}} \kappa\,  \frac{e^2}{4\pi^2} \, \frac{1}{a^{3}(t)} \, B^2(t_0) \,  \, \int^t dt^\prime \, {\dot a}(t^\prime)
\nonumber \\
& = \frac{{\mathcal C}_0}{a^3(t)} +  \frac{1}{k \, M_{\rm Pl}} \, \sqrt{\frac{3}{2}} \, \frac{e^2}{4\pi^2} \, \frac{1}{a^{2}(t)} \, B^2(t_0) ~ .
\end{align}
where ${\mathcal C}_0$ is a constant, which we shall determine later on by using
continuity requirements for the $b$-field at the interface between the inflation and radiation eras.
To arrive at the middle equality in \eqref{bsolrad}, we took into account that the amplitude $B(t)$ of the magnetic field intensity scales with the scale factor as~\cite{primordial}
\begin{align}\label{Bscalet}
B(t) = \frac{{B}(t_0)}{ a^{2}(t)} \, ~,
\end{align}
where $t_0$ is
the age of the Universe, and, thus, $B(t_0)$ denotes today's value.

During the radiation era, as follows from Einstein's equations, the scale factor behaves as $a(t) \sim \Big(2\, \sqrt{\Omega_0^{\rm rad}}\, H_0\, t \, \Big)^{1/2}$, whilst the Hubble parameter is given by $H(t)=1/(2\,t)$, with the subscript ``0'' indicating present-day quantities.
Hence, \eqref{bsolrad} yields
\begin{align}\label{bsolc}
\dot{\overline{b}} = \frac{{\mathcal C}_0}{a^3(t)}  + \frac{1}{\sqrt{\Omega_0^{\rm rad}}}\, \sqrt{\frac{3}{2}} \,  \frac{e^2}{4\pi^2}  \, H(t)\,\frac{B^2(t_0)}{k\, M_{\rm Pl} \, H_0}~.
\end{align}
Notice that the chiral anomaly contributions to the KR background field are proportional to the Hubble parameter $H(t)$ during the radiation era. If one considered the solution with ${\mathcal C}_0=0$, then such corrections would contribute purely $H^2$ -running vacuum type corrections \eqref{rvm1} to the energy density~\cite{bls1}. However, in view of the smallness of cosmic magnetic fields in the Universe, including possible primordial ones~\cite{primordial,frohlich,zhit}, we expect such terms to be suppressed compared to the $a^{-3}(t)$ term in the early universe, when ${\mathcal C}_0 \ne 0$, a case relevant for leptogenesis~\cite{bms2}, as we shall discuss below.

At present, we note that, on using \eqref{anom}, for homogeneous and isotropic backgrounds, we can equivalently write the solution \eqref{bsolrad} as
\begin{align}\label{bsold1}
\dot{\overline{b}} = \frac{{\mathcal C}_1}{a^3(t)} -  \sqrt{\frac{3}{8}} \kappa\,  \langle J^{5\, 0}\rangle,
\end{align}
where ${\mathcal C}_1 \ne {\mathcal C}_0$ (in general) is another integration constant.  For our purposes, and in the spirit of our treatment in section \ref{sec:2}, we take ${\mathcal C}_1=0$, hence
\begin{align}\label{bsold}
\dot{\overline{b}} =  -  \sqrt{\frac{3}{8}} \kappa\,  \langle J^{5\, 0}\rangle.
\end{align}

We shall determine next the (classical) energy momentum tensor, and check on the self consistent condition to obtain a total equation of state compatible with radiation dominance, that we used in order to arrive at the above results. To this end, we first notice that,
the fermion equations of motion (species $j$), derived from \eqref{sea6}, are:
\begin{eqnarray}\label{hde}
\imath \, e^\mu_{\, a}\,  \Gamma^a \, \nabla_\mu \psi_{j} - m^{(j)} \psi_j   +  \kappa \, \sqrt{\frac{3}{8}} \, \partial_a \overline b\, \Gamma^a \, \Gamma^5  \, \psi  - \frac{3\, \kappa^2}{8} \, \Big(\overline \psi_\ell \,   \Gamma_a \, \Gamma^5 \, \psi_\ell \Big)  \Gamma^a  \, \Gamma^5\, \psi_j = 0~,
\end{eqnarray}
where  $\nabla_\mu$ denotes the gravitational covariant derivative on spinors of the species $j$ with respect to the  torsion-free connection. One can then write
the (classical) stress tensor for the fermions as~\cite{shapiro,hehl}:
\begin{align}\label{fst}
T_{\mu\nu}^F = \frac{\imath}{2} \sum_{j} \Big( \overline \psi_j \,\gamma_{(\mu} \nabla_{\nu)} \, \psi_j- (\nabla_{(\mu}\overline \psi_j) \gamma_{\nu)} \psi_j \Big) -
\dfrac{3\kappa^{2}}{16}\, g_{\mu\nu}\,J^{5}_\alpha \, J^{5\, \alpha},
\end{align}
Above, we took into account that in the radiation/matter phase of the Universe, in which gravitational anomalies are assumed cancelled, the topological, chirally anomalous $b$-axion-fermionic matter coupling
terms in (\ref{sea6}), do not contribute to the covariant stress tensor, whose conservation is thus not affected, par contrast to the inflationary phase, where
the gravitational anomaly is present.

Solutions to the equations \eqref{hde} have been discussed in \cite{shapiro}.
It is important to notice that during the radiation era, the fermions, like all other matter species in the model, are relativistic and hence, cannot be simply assumed to have only temporal derivatives, i.e. spatial derivatives $\partial_i  \psi $ should also be considered. This complicates the detailed expressions for the stress tensor. However, for our purposes here, we may simply follow the approach of \cite{shapiro}, and estimate that such extra contributions will simply be absorbed in the energy density (and pressure) of free radiation $\rho_0^{\rm rad}$ ($p^{\rm rad}$), which dominate both the KR-axion-$b$ contributions and those from the  self-interactions of the fermions induced by the axial current-current $\langle J_0^5 \rangle^2$ interactions due to the $H$-torsion.

On account of \eqref{bsold}, then, the energy density for the fermions  acquires the form (we ignore mass terms during the radiation era, as the species are assumed relativistic)
\begin{align}\label{edf}
T_{00}^{F} & \simeq   ({T^{\rm Free}_{\rm Dirac}})_{00}  -
\dfrac{3\kappa^{2}}{16}\,\langle J^{5}_0 \rangle^2 = \dfrac{3\kappa^{2}}{16}\,\langle J^{5}_0 \rangle^2  - \kappa \, \sqrt{\frac{3}{8}}\, \dot{\overline{b}}\, \langle J_0^5 \rangle + \dots  \simeq  \dfrac{9\kappa^{2}}{16}\,\langle J^{5}_0 \rangle^2 + \dots,
\end{align}
where the $\dots$ denote pure radiation contributions from the kinetic terms which scale with the scale factor as $a^{-4}(t)$.
On the other hand, the energy density of the KR axion reads
\begin{align}\label{edb}
T_{00}^{b}  &=  \frac{1}{2} (\dot{\overline{b}})^2  = \dfrac{3\kappa^{2}}{16}\,\langle J^{5}_0 \rangle^2~.
\end{align}

The spatial  and time-space  components of $T_{ij}^{F,b}$ \eqref{fst}, computed from \eqref{sea6}, are~\cite{shapiro}
\begin{align}\label{spatialT}
T_{ij}^{F}  = g_{ij} p^F = g_{ij} \Big( \dfrac{3\kappa^{2}}{16}\,\langle J^{5}_0 \rangle^2 + \dots) ~, \quad
T_{ij}^{b}  =  g_{ij} p^b = g_{ij} \frac{1}{2} (\dot{\overline{b}})^2, \quad T^{F,b}_{0i}=0~,
\end{align}
where again the $\dots$ denote relativistic $\sim a^{-4}(t)$ contributions from the free kinetic terms of the fermions.

The total energy density $\rho^{\rm tot}$  and pressure $p^{\rm tot}$ are then given by
\begin{align}\label{totep}
T^{\rm tot}_{00} = \rho^{\rm tot}  = T_{00}^F + T_{00}^b + \rho^{\rm rad}, \quad  T_{ij}^{\rm total} = g_{ij} \, p^{\rm tot} = g_{ij} \Big(T_{ij}^F + T_{ij}^b + p^{\rm rad}\Big),
\end{align}
where the superscript ``rad'' denotes the conventional contributions from free relativistic species in the model, including photons,
with an equation of state $p_{\rm rad} = \frac{1}{3} \rho_{\rm rad}$, scaling as $a^{-4}(t)$.

By comparing \eqref{edf} with \eqref{spatialT}, the reader can readily verify that this is also the total equation of state for the axial current-current contributions in the fermion fluid,
\begin{align}\label{fermeos}
p^F = \frac{1}{3} \rho^F~.
\end{align}
However, as follows from \eqref{bsolc}, for ${\mathcal C}_0 \ne 0$, the scaling of $p^F$ and $\rho^F$ is not purely $a^{-4}$ (as would be the case with ${\mathcal C}_0=0$), but contains a superposition of terms with different scalings,   $\sim a^{-6}$, $\sim a^{-5}$ and $\sim a^{-4}$. We would like to stress that \eqref{fermeos} is the result of the solution \eqref{bsold} and the fact that, in our string inspired model, the KR axion is a fully fledged dynamical field.\footnote{The situation should be contrasted with the corresponding case of torsional
space time studied in \cite{shapiro}, where the equation of state characterising the torsion-induced fermion-self interaction contributions to the stress tensor was that of stiff matter~\cite{stiff} $p^F=\rho^F$.}

On the other hand, the KR axion component is characterised by a ``stiff matter''~\cite{stiff} equation of state
\begin{align}\label{beos}
p^b = \rho^b.
\end{align}
but again, on account of \eqref{bsolc}, the scaling of $p^b$ and $\rho^b$ is not $a^{-6}$ alone, each containing a superposition of terms $\sim a^{-6}$, $a^{-5}$ and $a^{-4}$.

On account of the conservation of the total stress tensor $T^{\rm tot}_{\mu\nu}$ \eqref{totep}, which is respected in the presence of \newnewtext{chiral anomalies}, as already explained, one may write
\begin{align}\label{totcons}
{\dot \rho}^{\rm tot} + 3 H \Big(\rho^{\rm tot} + p^{\rm tot}\Big) =0 \quad \Rightarrow \quad
 \frac{d}{dt} \Big(\rho^F + \rho^{\rm rad} \Big) + 4 H \, \Big(\rho^F  + \rho^{\rm rad} \Big) = - \frac{d}{dt}\rho^b - 6 H \, \rho^b ~,
\end{align}
where we used \eqref{fermeos}, \eqref{beos}.

If one recalls that  the cosmic electromagnetic fields are expected to be suppressed~\cite{primordial,frohlich,zhit}, one may make the reasonable assumption that it is the first term on the right hand side  of \eqref{bsolc} which dominates, at least
during the early stages of the radiation era, implying a scaling ({\it cf.} \eqref{bsold})
\begin{align}\label{bscale}
{\dot b} = -  \sqrt{\frac{3}{8}} \kappa\,  \langle J^{5\, 0}\rangle  \simeq  \frac{{\mathcal C}_0}{a^3(t)}~.
\end{align}
On making the further physically reasonable assumption that it is the radiation fields that {\it dominate} over the KR contributions in  the
stress tensor during the radiation era (and thus drive the scaling $a(t) \sim t^{1/2}$ of the Universe),  $\rho^{\rm rad} \gg \rho^F, p^{\rm rad} \gg p^F$,
one obtains a self consistent (approximate) vanishing of both sides of \eqref{totcons} {\it separately}, i.e. the following equations
\begin{align}\label{totcons2}
& \frac{d}{dt} \Big(\rho^F + \rho^{\rm rad} \Big) + 4 H \, \Big(\rho^F  + \rho^{\rm rad} \Big) \simeq   \frac{d}{dt} \Big(\rho^{\rm rad} \Big) + 4 H \, \Big(\rho^{\rm rad} \Big) = 0, \nonumber \\
& \frac{d}{dt}\rho^b + 6 H \, \rho^b  \simeq 0~,
\end{align}
which provide a self consistency check of the approach.

Continuity requires to match the background \eqref{bscale} with \eqref{krbeom2} (under \eqref{akappa}) at the temperature just at the exit of inflation, $T_i$, which, we take to be the
Gibbons-Hawking temperature~\cite{gh}
\begin{align}\label{ghtemp}
T_i = \frac{H}{2\pi}
\end{align}
with $H \simeq H_I \sim 10^{-5}\, M_{\rm Pl}$ the value of the Hubble constant during the inflationary period \eqref{Hinfl}. On taking into account, then, that, during the radiation era,  the temperature($T$)-cosmic time($t$) relation assumes the (standard Cosmology) form, $t  = 0.3 \sqrt{8\pi} \, g_\star^{1/2} \, M_{\rm Pl} T^{-2}$, where $g_\star $ (assumed approximately temperature independent) denotes the total number of relativistic degrees of freedom of the model under consideration, this implies:
\begin{align}
{\mathcal C}_0^\prime = 3.5 \times 10^{11} \, M_{\rm Pl}^2,
\end{align}
where we absorbed $T$-independent numerical constants in the definition of the constant ${\mathcal C}_0 \Rightarrow {\mathcal C}^\prime _0$ in \eqref{bscale}.
The scaling of the background \eqref{bscale} with the temperature, then, during the radiation era, is:
\begin{align}\label{bscale2}
\dot{\overline b} \simeq  3.5 \times 10^{11} \, M_{\rm Pl}^2 \, \Big(\frac{T}{M_{\rm Pl}}\Big)^3 .
\end{align}
As we shall see
in the next subsection, such backgrounds can produce phenomenologically correct leptogenesis.

\subsection{KR-axion-Induced Leptogenesis and Matter-Antimatter Asymmetry in the Universe \label{sec:asym}}

Indeed, as discussed in \cite{decesare,bms,bms2}, the presence of the background  \eqref{bscale} could lead, in principle to {\it Leptogenesis},  as it breaks {\it spontaneously} Lorentz, CP and CPT  symmetry. In \cite{bms2} we have discussed the generation of matter-antimatter asymmetry in the presence of backgrounds of the KR field precisely of the form
\eqref{bscale2}, which are considered slowly varying during the (short) freeze-out era of leptogenesis, as explained in that work.

In particular, we have considered lepton-number asymmetry originating from tree-level decays of heavy sterile (right-handed, Majorana) neutrinos (RHN)  into SM leptons. The relevant part of the Lagrangian is given by:
  \be
\label{smelag}
\mathcal{L}= {\mathcal L}_{\rm SM} + i\overline{N}\slashed{\partial}N-\frac{m_N}{2}(\overline{N^{c}}N+\overline{N}N^{c})-\overline{N}\slashed{B}\gamma^{5}N-\sum_f \, y_{f}\overline{L}_{f}\tilde{\phi}^dN+h.c.
\ee
where ${\mathcal L}_{\rm SM}$ denotes the SM Lagrangian,
$N$ is the RHN field, of (Majorana) mass $m_N$,  $\tilde \phi$ is the SU(2) adjoint of the Higgs field  $\phi$ ($\tilde{\phi}^d_i \equiv \varepsilon_{ij}\phi_j~, \, i,j=1,2,$ SU(2) indices, \newnewtext{is the SU(2) dual of the Higgs field}),
 and $L_{f}$ is a lepton (doublet) field of the SM sector, with $f$ a generation index, $f=e, \mu, \tau$, in a standard notation for the three SM generations; $y_f$ is a Yukawa coupling, which is non-zero and provides a non-trivial (``Higgs portal'') interaction between the RHN and the SM sectors. In the models of \cite{decesare,bms,bms2} a single sterile neutrino species
suffices to generate phenomenologically relevant lepton asymmetry, and hence from now on
we restrict ourselves to the first generation ($f=e$, setting $y_e = y$).  \newtext{The  quantity $\slashed{B}=\gamma^\mu B_\mu$  appearing in the axial current  term of  (\ref{smelag}) is defined in terms of the  four vector}
\begin{align}\label{background}
B_\mu = M_{\rm Pl}^{-1} \, \dot{\overline b}\, \delta_{\mu0}\,.
\end{align}
It denotes the Lorentz- (LV), CP- and CPT (CPTV) - Violating  background \eqref{bscale2}, with  $B_\mu$ having only a temporal component.
For such (slowly varying in the cosmic frame) backgrounds, as our case here, the Lagrangian (\ref{smelag}) assumes the form of a Standard Model Extension (SME) Lagrangian in a Lorentz and CPTV background~\cite{sme}.

At this stage we should make an important remark. As the reader should have noticed, in our model, the background \eqref{background}  has a derivative form, $B_\mu \propto \partial_\mu b$, which, by partial integration, implies a coupling of the KR axion to the derivative of the axial current in the effective action \eqref{smelag}. In our model, the RHN are massive in the radiation epoch, where leptogenesis occurs, and hence the classical axial current is {\it not} conserved, since its four divergence equals $i \, m_N \,(\overline{N^c} \gamma_5 N + \overline{N} \gamma_5 N^c )$, as follows from the (Majorana) equation of motion of the free RHN fields. Therefore, the non trivial coupling of the KR axion to the RHN current is guaranteed, independent of any potential anomalies, thus consistent with the cancellation of gravitational anomalies by the chiral matter in the radiation and matter dominated eras, advocated in our scenario.\footnote{Gravitational anomalies may play a r\^ole in a dynamical generation of the RHN (Majorana) mass, as in the scenario of \cite{pilaftsis}, involving mixing of the KR field with other string theory axions. Such mechanisms can be consistently embedded in our framework, specifically in the early radiation epoch, just after inflation, when chiral matter is generated. However, their discussion falls beyond the scope of the current work.}

In the context of the model \eqref{smelag}, a lepton asymmetry is generated due to the CPV and CPTV tree-level decays of the RHN $N$ into SM leptons in the presence of the background \eqref{background}~\cite{decesare,bms,bms2}:
\begin{eqnarray}\label{4channels}
{\rm Channel ~I}&:& \qquad  N \rightarrow l^{-}h^{+}~, ~ \nu \, h^{0}~,  \\ \nonumber
{\rm Channel ~II}&:& \qquad  N \rightarrow l^{+}h^{-}~,~  \overline \nu \, h^{0}~.
\end{eqnarray}
where $\ell^\pm$ are charged leptons, $\nu$ ($\overline \nu$) are light, ``active'', neutrinos (antineutrinos) in the SM sector,
$h^0$ is the neutral Higgs field, and
 $h^\pm$ are the charged Higgs fields, which, at high temperatures, above the spontaneous electroweak symmetry breaking, of interest in this scenario, do not decouple from the physical spectrum.  As a result of the non-trivial $B_0 \ne 0$ background (\ref{background}), \eqref{bscale2}, the decay rates of the Majorana RHN between the channels I and II are different, resulting in a Lepton asymmetry~\cite{bms2},
 \begin{align}\label{lepto}
 \frac{ \Delta L^{TOT}(T=T_D)}{s} \sim  q\,  \dfrac{\Phi_{0}}{m_{N}}~,  \quad q > 0,
 \end{align}
 where $s$ is the entropy density of the Universe, $T_D$ denotes the temperature at which
 this asymmetry freezes out (`freezeout point'), that is when the total decay width $\Gamma$ for the decays \eqref{4channels} equals the Hubble rate of the Universe, $H (T_D) \simeq \Gamma $, and the quantity $\Phi_0 $ is  defined as~\cite{bms2}:
 \begin{align}\label{bphi}
 B_0 (T) = {\Phi}_0 \, \Big(\frac{T}{m_N}\Big)^3
 \end{align}
 The lepton asymmetry \eqref{lepto}  can then be communicated to the baryon sector via Baryon-minus-Lepton-number ($B-L$) conserving sphaleron processes in the SM~\cite{krs}, thus producing the observed amount of baryon asymmetry (baryogenesis)  in the Universe, by requiring that
the lepton asymmetry \eqref{lepto} is of ${\mathcal O}(8 \times 10^{-11})$, as indicated by (cosmological) observations~\cite{planck}.
The number $q > 0$ expresses theoretical uncertainties in the analytical derivation of the lepton number asymmetry in \cite{bms2}, where the Pad\`e approximants method was used to solve the pertinent system of coupled Boltzmann equations associated with \eqref{4channels}. The precise value of $q$ depends on the freezeout point. Using \eqref{bscale2}, we may write
\begin{align}\label{phidef}
\Phi_0 = 3.5 \times 10^{11} \, \Big(\frac{m_N^3}{M_{\rm Pl}^2}\Big).
\end{align}
By demanding phenomenologically acceptable values of the lepton asymmetry \eqref{lepto} of order ${\mathcal O}(8 \times 10^{-11})$, one can then
infer from \eqref{phidef} that:
\begin{align}\label{mnnew}
m_N \simeq \frac{1.5}{\sqrt{q}} \times 10^{-11} \, M_{\rm Pl} \simeq  \frac{3.7}{\sqrt{q}}  \times 10^7 \, {\rm GeV}
\end{align}
The reader should bear in mind that in the semi-analytic method of \cite{bms2} only the following combination of parameters, involving $m_N$, enters the series expansions of the solutions
about a point $x = m_N/T$ used to approach (via Pad\'e approximants) the freezout point $x_D = 0.1$:
\begin{align}\label{quant}
{\mathcal I} \equiv {y}^2 \frac{M_{\rm Pl}}{ m_N}.
\end{align}
We now notice that the ratio $y^2/m_N$ appears in the expression for the SM active neutrino $\nu$ masses via the (type-I) seesaw mechanism~\cite{seesaw},\footnote{One needs more than one flavours for heavy neutrinos in that case, which can be easily accommodated in the framework of \cite{decesare,bms,bms2}.}
\begin{align}\label{active}
m_\nu \sim |y|^2 v^2/m_N.
\end{align}
In \cite{bms2},  the Yukawa coupling $y \sim 10^{-5}$ and $m_N \sim 10^5$~GeV~\cite{decesare,bms,bms2}
gave phenomenologically relevant values for  $m_\nu$. Such parameters correspond to (cf. \eqref{quant})
\begin{align}\label{Avalue}
{\mathcal I} \sim 10^{3}
\end{align}
which we keep fixed in our approach, so that the considerations of \cite{bms2} apply, and moreover one obtains the same (phenomenologcally consistent) active neutrino masses via seesaw as in \cite{bms2} (cf. \eqref{active}).

Additionally,  the assumption that $T_D \simeq m_N$ was made in \cite{bms2}, which we also maintain here.
In such a case~\cite{bms2} $q = {\mathcal O}(10)$, and from \eqref{mnnew} one obtains
\begin{align}\label{mnvalue}
m_N \simeq 1.17 \times 10^7~{\rm GeV} \, ,
\end{align}
that is, the sterile neutrino mass and, hence the freezeout temperature, in our case are two orders of magnitude higher than their counterparts considered in \cite{decesare,bms,bms2}.\footnote{It should be noted that the freezeout temperarure could be up to one order of magnitude higher than $m_N$, due to model dependence when calculating it from the equality of the total decay rate \eqref{4channels} with the Hubble parameter. In such a case, one may have $q \simeq {\mathcal O}(100)$ or larger in the lepton asymmetry equation \eqref{lepto}, implying a
$m_N ={\mathcal O}(10^6)$ or smaller, in the ball park of the sterile neutrino mass of \cite{decesare,bms,bms2}. Thus, the above numbers should be considered with a  theoretical uncertainty of a couple of orders of magnitude. The unambiguous conclusion, though, is that, in this scenario, there is phenomenologically relevant leptogenesis during the early radiation era.}

From \eqref{quant}, then, the corresponding Yukawa coupling
assumes the value $|y| \simeq 4.8 \times 10^{-5}$ (just a factor of 5 larger than that in \cite{bms2}), while from \eqref{bphi} one obtains for the background field at freezeout~\cite{bms2}: $B_0 (T=T_D \simeq m_N \sim 10^7~{\rm GeV}) = {\mathcal O}({\rm keV})$, which induces phenomenologically relevant leptogenesis at $T \sim 10^7$ GeV.

Before closing this section, we also remark that the value \eqref{mnvalue} is compatible with the upper bound on the sterile neutrino masses required in minimal scenarios for Higgs-mass stability ({\it naturalness}) in type-I seesaw models ~\cite{vissani}, that is, assuming no new physics (such as supersymmetry) at TeV scale. Indeed, the Higgs-mass-squared corrections coming from  one-loop contributions in the electroweak symmetry broken phase, due to the Higgs portal interactions in \eqref{smelag}, involving (in their generality) the three active and three sterile neutrinos propagating in the loop, read~\cite{vissani}: $\delta m_H^2 = \sum_{\alpha=e,\mu,\tau}\, \sum_{I=1}^3  \frac{1}{(4\pi)^2} y_{\alpha I} ^2 m_I^2 $. In order
 to ensure Higgs-mass stability/naturalness, then, one must have: $\delta m_H^2 \lesssim m_H^2$, where $m_H$ is the Higgs mass. In our single sterile neutrino case, considered above, we may eliminate the Yukawa coupling $y$, using the type-I see-saw formula \eqref{active}, to obtain the following criterion for mass stability:
\begin{align}\label{nbounds}
m_N \lesssim \Big(m_H^2 v^2 (4\pi)^2 \, m_\nu^{-1} \Big)^{1/3}.
\end{align}
Using the cosmological bound~\cite{planck} for the sum of the three active
neutrino masses
$\sum_{i=1}^3 m_{\nu \, i} < 0.12 $~eV,
and translating it (on account of the neutrino oscillation data on the active neutrino mass differences, assuming normal or inverted hierachies~\cite{pdg}) into an upper bound   for the single active neutrino  we consider here, $m_\nu \lesssim 0.04$~eV, we may replace the $m_\nu$ in \eqref{nbounds} by this upper bound, to obtain a {\it sufficient} condition for the satisfaction of the Higgs mass stability,
$m_N \lesssim 10^8$~GeV.  A similar estimate is obtained~\cite{vissani} in the case where there are three active and at least two sterile neutrino flavours. In that case, one may use the atmospheric oscillation experiments {\it measurement} for the observed active neutrino mass differences~\cite{pdg}, $\Delta m^2_{\rm atm} \sim 2.4 \times 10^{-3}~{\rm eV}^2$,
and the type-I seesaw generalisation of \eqref{active}, giving non-zero masses to at least two of the active neutrinos, to determine the allowed upper bound  for $m_N$ for Higgs mass stability from experimental data. Indeed,  by setting  $m_\nu \sim [\Delta m^2_{\nu\, {\rm atm}}]^{-1/2}$ in \eqref{nbounds}, one obtains $m_N \lesssim \Big(m_H^2 v^2 (4\pi)^2 \, [\Delta m^2_{\nu\, {\rm atm}}]^{-1/2} \Big)^{1/3} \sim  10^8~{\rm GeV}$.
On the other hand,  assuming two of the active neutrinos nearly degenerate, with the third one having much smaller mass, one may face a situation where $m_\nu \sim {\mathcal O}(10^{-1})$~eV, implying $m_N \lesssim 10^7$ GeV.

As already mentioned, such naturalness bounds can be
bypassed, if new physics, {\it e.g.} supersymmetry, exists at some scale below $10^7$ GeV, in which case the RHN contributions to the Higgs-mass quantum corrections might be cancelled by, say, loops of sneutrinos, if the masses of the latter are similar to those of the RHN. In our string-inspired case, such extra contributions might well exist, but here we consider minimal seesaw scenarios, which suffice for our purposes.

\section{Modern Era and Re-appearance of the Gravitational Anomalies \label{sec:modern}}

After freezout, during the radiation era, the temperature of the Universe continues to drop at a rate $a(t) \sim 1/T$, until the expansion of the Universe is such that the $a^{-2}(t)$ term, due to the chiral anomalies, in the solution for the KR axion background  \eqref{bsolc} dominates over the $a^{-3}(t)-$scaling term. Such dominance lasts until more or less the matter-radiation equality era, after which matter (mostly DM) begins to dominate, and this sets the dawn of the matter-dominance epoch, which according to data~\cite{planck} ends  at redshifts $z \simeq 0.7$, succeeded by the current de Sitter phase. As follows from Einstein's equations, during matter dominance, the scale factor behaves as $a(t) = a_{m}(t) \sim \Big(\frac{3\, \sqrt{\Omega_{m 0}}\, H_0}{2}\, t\Big)^{2/3} $. Taking into account, as standard in Cosmology, that it is only the relativistic degrees of freedom that contribute to the constant entropy density of the Universe during its entire evolution, implies that  the matter-dominated era scale factor  is inversely proportional to temperature $T$, $a_m(t) \sim T^{-1}$, as is the case during radiation dominance.

During the matter dominated era, then, as follows from \eqref{bsolrad} upon imposing the continuity assumption for the KR background and its derivatives, the $a^{-3}(t) \sim T^3$ term may be considered subdominant~\cite{bms2}, with the dominant behaviour being provided by the $a^{-2} \sim T^2$ chiral anomaly term (below, for convenience, we express the temperature in units of $M_{\rm Pl}$, and absorb any proportionality $T$-independent constants appearing in the expression of $a_{\rm m}(t)$ in the definition of $B(t_0) \rightarrow B^\prime(t_0)$):
\begin{align}\label{bsolmatt2}
\dot{\overline{b}}\,\Big|_{{\rm matter~era}} & \simeq  \frac{1}{k \, M_{\rm Pl}} \, \sqrt{\frac{3}{2}} \, \frac{e^2}{4\pi^2} \, \frac{1}{a_{\rm m}^{2}(t)} \, B^2(t_0)  \quad \Rightarrow
\nonumber \\
\dot{\overline{b}}\,\Big|_{{\rm matter~era}} &\simeq \sqrt{\frac{3}{2}} \, \frac{e^2}{4\pi^2} \, \frac{{B^\prime}^2(t_0)}{k \, M_{\rm Pl}^3} \, T^2.
\end{align}

From \eqref{bsolmatt2}, and the above discussion, we therefore conclude that at the late stages of radiation era and during matter dominance, the presence of
a chiral anomaly implies a softer ($\sim T^2$) temperature dependence of the KR axial background, as compared to the $T^3$ scaling in the case of \cite{bms2}, where chiral anomalies were ignored. In our case, any such $T^3$-scaling contribution to this background is subdominant, as follows by continuity requirements at the interface between the end of radiation- and beginning of matter- domination eras.

During the current epoch, where matter has started to fade away, and a cosmological constant-like (de Sitter) phase, seems, according to data~\cite{planck}, to start dominating the (accelerated) expansion of the Universe, the presence of late epoch gravitational waves would lead once more, following the reasoning of section \ref{sec:rvminfl}, to the resurfacing of gravitational anomalies of the type \eqref{rrt}, \eqref{theta}; these
can no longer be cancelled by the diluted chiral matter. However, now, the approximately constant Hubble parameter of the current-era de Sitter phase equals the Hubble constant today, $H \sim H_0$, which is much smaller than its counterpart during inflation. Hence any gravitational anomalies would be strongly suppressed.
 The slow roll conditions for the KR axial background $\dot{\overline b}$ are valid for scaling $\sim T^2$, which \newtext{prompts us to conjecture a behaviour today\,\cite{essay}}
 \begin{align}\label{krbtoday}
 \dot{\overline b}_{\rm today} \sim \sqrt{2\epsilon^\prime}\, H_0 \, M_{\rm Pl},
 \end{align}
 in analogy to \eqref{slowrollepsi2}.\footnote{We do not discuss here the behaviour (vs the cosmic time) of the gravitational anomaly during the entirety of the late de Sitter era. In fact, the gravitational anomaly in the current era most likely will not be constant, and thus will be washed out at the end of the new inflationary period, which however cannot be predicted, as the microscopic string theory dynamics leading to this era is not known. Given the strongly suppressed Hubble parameter today $H_0$
 as compared to its counterpart $H_I$ during the inflationary period,  $H_0 \simeq 10^{-55}\, H_I $, in order to ensure a constant gravitational anomaly \`a la \eqref{factor}, \eqref{A=0}, one would need enormously (and unnaturally)
 transplanckian cutoff values for the momentum modes, although it must be said that, since the detailed string dynamics is unknown, one cannot make definite statements on the subject. Fortunately, such issues do not affect our current study.} In general, $\epsilon^\prime \ne \epsilon$.

 An estimate of $\epsilon^\prime$ can be provided by matching the value of $\dot{\overline b}_{\rm today}$  \eqref{krbtoday} with that of
 \eqref{bsolmatt2}, upon setting $a_{\rm m}(t_0) =1$.
 We thus obtain,
 \begin{align}\label{krtoday}
 \sqrt{\epsilon^\prime} \simeq \frac{\sqrt{3}}{2} \, \frac{e^2}{4\pi^2} \, \frac{{B}^2(t_0)}{k \, M_{\rm Pl}^2\, H_0}.
 \end{align}
% where we have set $T = T_0 =T_{\rm CMB} \simeq 0.23$~meV, the current temperature of the Universe, corresponding to the CMB background~\cite{planck}, assuming to characterise the beginning of the de Sitter era, at which point the matching between \eqref{bsolmatt} with \eqref{krbtoday} is made.
We proceed now to estimate the momentum scale $k$ of the monochromatic solution \eqref{solEB}. This comes from Maxwell's equations in the presence of the chiral anomalies, which for homogeneous and istotropic KR backgrounds $\dot b(t)$ read~\cite{frohlich,zhit}:
\begin{align}\label{maxwell}
{\mathbf \nabla} \times {\mathbf B}(t) = \sigma {\mathbf E} -  \dot{\overline b} \, \sqrt{\frac{3}{8}} \, \kappa \frac{e^2}{4\pi^2}  {\mathbf B}(t),
\end{align}
where $\sigma$ is the conductivity of charged chiral matter (we used Ohm's law and identified the electric current density as ${\mathbf j} = \sigma {\mathbf E} $). From the solution \eqref{solEB}, one has
\begin{align}
{\mathbf \nabla} \times {\mathbf B}(t)  = -k {\mathbf B}(t),
\end{align}
and, thus, Eq.~\eqref{maxwell} becomes
\begin{align}\label{maxwell2}
k{B}(t) = -\frac{\sigma}{k} \dot{B}(t) +  \dot{\overline b} \, \sqrt{\frac{3}{8}} \, \kappa \frac{e^2}{4\pi^2}  {B}(t).
\end{align}
The reader should bear in mind that the classical KR background $\dot{\overline b}$ plays a r\^ole analogous to the chiral chemical potential $-\mu_5$~\cite{frohlich,zhit,andrade}.\footnote{We note, however, that there are important subtle physical differences between a bare $\mu_5$ and the temporal component of an axial vector background (axial potential), such as $\dot{\overline b}$, coupled to the axial fermion current.  The latter, {\it unlike} $\mu_5$,  does {\it not} contribute~\cite{kaplan,dvorn} to the so-called {\it chiral magnetic effect} (CME)~\cite{cme}, that is the excitation of an electric current density in the presence of an external magnetic field, with a coefficient proportional to $\mu_5$, ${\mathbf j}_{\rm CME} = \frac{e^2}{2\pi^2} \mu_5 \, {\mathbf B}$, which is an effect associated with the chiral anomaly. Indeed, if one uses energy conservation arguments~\cite{kaplan} or calculates the electric current density from first principles using, {\it e.g.}, the relativistic quantum mechanics approach~\cite{dvorn} in the presence of {\it both} a chiral chemical $\mu_5$ and the axial potential, then only the $\mu_5$ contributes to the current. In our case, the non-contribution of the axial potential to the CME is consistent~\cite{bms2} with the fact that the contributions of the KR torsion (and thus the axial KR potential $\dot{\overline b}(x)$) to the chiral anomaly can be removed by the addition of appropriate renormalisation group counterterms to the string effective action, order by order in perturbation theory~\cite{hull,mavindex}. Such issues will not be directly relevant for our purposes in this work, though, which is based on \eqref{maxwell}, \eqref{maxwell2} and the associated magnetogenesis. In this respect, the reader should notice that the last term on the right-hand side of \eqref{maxwell}, \eqref{maxwell2}, which has a form similar to the chiral magnetic effect, does contribute to the magnetic field evolution, but cannot be considered as a contribution to the electric current, for reasons explained.} However, for us, in contrast to the considerations in \cite{zhit}, the KR axion is a fully fledged quantum field.

Ignoring chiral matter  in our case, as it becomes subdominant in the modern-era de Sitter phase,  is equivalent to setting $\sigma \to 0$ in \eqref{maxwell2}.\footnote{We note at this stage, that, had we kept charged chiral matter today, and thus the conductivity $\sigma$ term in \eqref{maxwell2}, then the equation  would have admitted  a growing solution~\cite{frohlich}: $B(t)=B_0\, \exp(\delta \, t)$, with $\delta = \frac{k}{\sigma} \Big(\frac{e^2}{4\pi^2} \sqrt{\frac{3}{8}} \, \kappa\, \dot{\overline b} - k\Big) > 0$, for sufficiently low $k$, where $\dot{\overline b}$ is given by \eqref{krbtoday}. This would have led to the well-known instabilities in the presence of a chiral chemical potential~\cite{frohlich}, which in the approach of \cite{zhit} have been linked to the creation of a cosmological magnetic field, whose energy density was identified with the dark energy of the Universe in the current era. In our approach, where chiral charged matter is not dominant in the current de Sitter era, we differ from this interpretation, associating the KR axion with a source of dark matter, as we discuss below.}
Taking into account \eqref{Bscalet},
which we were using so far self-consistently, and utilising \eqref{krtoday}, we obtain from \eqref{maxwell2}:
\begin{align}\label{keps}
k \simeq \sqrt{\epsilon^\prime} \, {\frac{\sqrt{3}}{2}} \, H_0 \frac{e^2}{4\pi^2},
\end{align}
which, on account of  \eqref{maxwell2} and \eqref{krbtoday}, leads to:
\begin{align}\label{epsprime}
\epsilon^\prime \simeq \frac{B(t_0)^2}{M_{\rm Pl}^2 \, H_0^2} = \frac{2}{3} \frac{\rho_0^B}{\rho_c^{(0)}},
\end{align}
that is, $\epsilon^\prime$ is of order of the energy density of the magnetic field today $\rho_0^B=\frac{1}{2}\, B^2(t_0)$ in units of the critical density of the universe $\rho_c^{(0)} = H_0^2 \, M_{\rm Pl}^2/3$.

On account of \eqref{epsprime} and \eqref{krbtoday}, and the stiff equation of state \eqref{beos} of the (massless) KR axion, which dominates the `matter' part of the action in the de Sitter era, we then observe that the latter field can provide a source of (stiff) DM, with vacuum energy density of the order of the magnetic field energy density.
 \begin{align}\label{rhobtoday}
 \rho^b (= p^b)\Big|_{\rm today(de-Sitter-era)} = \frac{1}{2} \Big(\dot{\overline b}\Big|_{\rm today}\Big)^2 \simeq B^2(t_0).
 \end{align}
 Unfortunately, in our low-energy effective string theory, there is no way of estimating $B(t_0)$ from first principles. In the context of the underlying string model, this in principle can be done by an appropriate choice of the ground state, but in view of the landscape afflicting string theory, at present such a task does not seem feasible. Thus, we have to resort to
 phenomenological arguments.

 To this end, we first notice, that, as with the inflationary phase, it is {\it not} the massless KR  field which drives the late era de Sitter phase. There must be some other mechanism, by means of which an approximately constant potential $U$ appears dynamically during the late epochs of the Universe, which resembles quintessence, thus driving the latter de Sitter era. In such a case, one may assume that the kinetic energy of the KR axion field
 $K_b = \frac{1}{2} {\dot b}$ is roughly one order of magnitude smaller than $U$,  a typical situation of other cosmological fields, such as quintessence, which would allow the total equation of state to be approximated by that of de Sitter space-time $w\simeq -1$. In such a case, by identifying the two slow-roll parameters for the KR field, in the early and late de Sitter eras of the string Universe ({\it cf.} \eqref{krbeom2}, \eqref{krbtoday})
 \begin{align}
 \epsilon \sim \epsilon^\prime ={\mathcal O}(10^{-2})
 \end{align}
one can get the DM content in the right ballpark~\cite{planck}:
\begin{equation}\label{eq:UbTb}
  \Omega_{m0}=\frac{\rho_{m0}}{\rho^{(0)}_c}\simeq \frac{U}{\rho^{(0)}_c}\simeq 10\, \frac{K_b}{\rho^{(0)}_c}\simeq 10\, \epsilon={\cal O}(0.1)\,,
\end{equation}
where $\rho_{m0}$ is the current energy density of DM in the universe.  \newtext{\newnewtext{Above} we used the fact that, according to \eqref{slowrollepsi} and  \eqref{krbtoday},
the slow-roll parameter of $b(x)$ measures the ratio of its kinetic energy, $K_b\sim (1/2)\,\dot{b}^2$,  to the critical energy density of the Universe, $\rho_c=(M_{\rm Pl} H)^2/3$.}

 On account of \eqref{epsprime}, then, this also determines the current energy density of the cosmological magnetic field, $\rho_0^B$. Moreover, we observe that the temporal component of the KR background \eqref{background},
 $B_0 = \dot{\overline b} M_{\rm Pl}^{-1}$ in the current era \eqref{krbtoday}, is of order
 \begin{equation}\label{backtoday}
 B_0 \Big|_{\rm today} \sim 2.435 \times 10^{-34}\, {\rm eV}.
  \end{equation}
 We note that this is about fourteen orders of magnitude larger than the corresponding background found in \cite{bms2}, in the absence of chiral anomalies.

In view of the r\^ole of the almost constant $B_0 \sim \dot{\bar b}$ background as a CPT and Lorentz-symmetry violating background in the effective theory, which, as mentioned above,
falls within the framework of the Standard Model Extension~\cite{sme}, it is imperative to check the phenomenological consistency of \eqref{backtoday} with the current bounds of such backgrounds~\cite{smebounds}: $B_0 < 10^{-2} \,{\rm  eV}$ for the temporal component,  and (much more stringent) $B_i < 10^{-31}$~GeV, for the spatial components. The predicted value in our model \eqref{backtoday} satisfies comfortably those bounds, even if one takes into account the relative motion of our Laboratory frame with respect to the cosmic Robertson-Walker frame, which we take to be the CMB frame. Indeed, if the Lab frame moves with a certain velocity~\cite{conklin} $|\vec v| \ll c $ (with $c$ the speed of light {\it in vacuo}) with respect to the CMB frame, then, according to Special Relativity, we shall also observe spatial components of $B_\mu$ in the Lab frame of order
\be\label{bilab}B_i = \gamma \frac{v_i}{c}\, B_0,\quad i=1,2,3,
\ee
with $\gamma \sim 1 $ the Lorentz factor. As can be inferred from studies of the CMB anisotropies, a typical order of magnitude of the velocity of the Earth (where precision tests of the SME are made) with respect to the CMB background is~\cite{conklin,planck}: $|\vec v| ={\mathcal O}(390 \pm 60)$~Km/sec. From \eqref{backtoday}, \eqref{bilab} then,  we observe that all bounds for the Lorentz- and CPT- Violating KR background $B_\mu$, $\mu=0, \dots 3,$ are comfortably satisfied.

\section{Massive KR-Axion Dark Matter  \label{sec:mixing}}

We would like to close our study by making some further remarks on the nature of the KR axion as a source of DM.
In our approach so far, the KR axion has been treated as exactly massless, not having any potential, and thus respecting the shift symmetry. Axions in such conditions are usually viewed as Goldstone Bosons of a spontaneously broken global (shift) symmetry (such as an accidental Peccei-Quinn symmetry for  QCD axions~\cite{qcdaxion}). If the symmetry is broken explicitly, however, by non-perturbative quantum (instanton) effects, as happens, for instance, in the case of the QCD axion,
then a (small) periodic axion potential is developed. In this sense the axion acquires a
small mass, which implies its potential r\^ole as a light DM candidate.
Concretely, the QCD axion field $\theta(x) \equiv a(x)/f_a$, with $f_a$ the (mass-dimension-one) axion decay constant, estimated phenomenologically to lie in the range~\cite{qcdaxion}
$10^9~{\rm GeV} < f_a < 10^{12}~{\rm GeV}$, has anomalous couplings with the gluons of the form
$\frac{g_s^2\, \theta (x)}{32\pi^2} G_{\mu\nu}^A \, G^{A\, \mu\nu}, \, A=1,\dots 8$ (an adjoint SU(3) colour index), with $g_s$ the strong-interaction coupling, and $G^{A\, \mu\nu}$ the gluon field strength tensor. The global Peccei-Quinn $U(1)$ symmetry is associated with shifts $a(x) \rightarrow a(x) + {\mathcal \epsilon}$, ${\mathcal \epsilon} $= constant. The non-perturbative potential induced by instanton effects, which breaks this shift symmetry, has
schematically the form $V(\theta) = \Lambda_{\rm QCD}^4 \Big(1 -{\rm cos}(\theta)\Big)$, where $\Lambda_{\rm QCD}$ is the QCD scale.
Minimisation of the potential fixes the strong-CP-violating angle $\langle \theta \rangle =0$.  \newtext{The non-perturbatively generated QCD axion mass squared is $m_a^2=\left.\partial^2 V/\partial a^2\right|_{a=0}=\Lambda^4_{\rm QCD}/f^2_a$  and hence   $m_a = \Lambda^2_{\rm QCD}/f_a$}.\footnote{In more precise estimates, $\Lambda_{\rm QCD}^2$ is replaced by $m_\pi f_\pi$, where $m_\pi$ ($f_\pi$) is the pion mass (decay constant),
and the potential is appropriately modified~\cite{qcdaxion}.}

In our case, as mentioned previously, the KR massless axion, which is dual in four space-time dimensions to the antisymmetric tensor field strength,  $\partial_\mu b \sim \varepsilon_{\mu\nu\rho\sigma}\, {\mathcal H}^{\nu\rho\sigma}$,
might be viewed~\cite{aben} as the Goldstone mode of the spontaneous breaking of Lorentz symmetry induced by the constant background \eqref{slowrollepsi2} or \eqref{krbtoday}.
However, it is possible that the shift symmetry of the KR axion is broken by some non-perturbative stringy effects, which are also responsible for generating a potential for it, at least in the current cosmological epoch. In such models, there might be a slow-roll behaviour of the axion, which is thus viewed as a quintessence field~\cite{axioninfl},  driving the current de Sitter phase of the Universe.

There are also models
which involve non-trivial interactions of the KR axion with standard axions, that exist abundantly in string theory~\cite{arv}, which may thus provide additional components of axionic DM.
Below we shall discuss such a toy model, in which the field $b(x)$ acquires a potential $U$ (and a mass) in the current epoch, and the expression \eqref{krbtoday} is still a consistent solution of the equations of motion.
Ingredients of such a model have been considered in \cite{pilaftsis}, in an attempt to propose alternative, beyond see-saw~\cite{seesaw}, mechanisms for radiative generation of right-handed Majorana neutrino masses, that appear, e.g., in the Lagrangian \eqref{smelag} and are crucial for leptogenesis. The model couples the bosonic action \eqref{sea6}, involving the KR axion field in the presence of a gravitational anomaly,
to stringy or ordinary (including QCD) axion fields ${\mathcal A}_i(x)$, $i=1,\dots n$, through a kinetic mixing term~\cite{pilaftsis}
\begin{align}\label{mixing}
S^{b-a}_{\rm mixing} = \sum_{i=1}^n \gamma_i \, \int d^4 x \sqrt{-g} \, \partial_\mu {\mathcal A}_i \, \partial^\mu b(x),
\end{align}
 where the (dimensionless) mixing coefficients $0 \ne |\gamma_i| < 1$.  The axions $ {\mathcal A}_i$ are assumed to have canonically normalised kinetic terms and shift-symmetry breaking non-trivial Yukawa couplings with Right-handed Majorana neutrinos, which can be generated by non-perturbative string instanton effects.
 The details of the potential of the ${\mathcal A}_i$ fields are irrelevant for the radiative Majorana neutrino mass generation~\cite{pilaftsis}. For our purposes, it suffices to concentrate on one such axion field ${\mathcal A}(x)$. In general, we assume that the ${\mathcal A}$ axion also couples to the (gravitational) anomaly with some dimensional coupling, which we take to be
 \begin{equation}
 \int d^4 x\, \sqrt{-g}\,  \frac{ f_{\mathcal A}\, \alpha^\prime}{96 \, \kappa} \, {\mathcal A}(x) \,R_{\mu\nu\rho\sigma} \, {\widetilde R}^{\mu\nu\rho\sigma} = -
  \int d^4 x\, \sqrt{-g}\,  \frac{ f_{\mathcal A}\, \alpha^\prime}{96 \, \kappa}\, \partial_\mu {\mathcal A}(x) \, {\mathcal K}^\mu~,
 \end{equation}
 where $f_{\mathcal A}$ is a dimensionless constant, which depends on the microscopic details of the theory, in particular on stringy degrees of freedom circulating in the anomalous chiral fermion loop. In our current approach so far we have assumed for concreteness $\alpha^\prime \sim \kappa^2$, but in realistic string-theory models one may keep the Regge slope as an independent parameter, to be fixed phenomenologically, and this is what we adopt for the remainder of this section.

The equations of motion of the $b(x)$ and ${\mathcal A}(x)$ fields read (the overline above the fields, denote classical solutions, as per our previous notation):
\begin{align}\label{axions}
&& \partial_{\alpha}\Big[\sqrt{-g}\Big(\partial^{\alpha}\bar{b}  -  \sqrt{\frac{2}{3}}\,
\frac{\alpha^\prime}{96 \, \kappa} \, {\mathcal K}^{\alpha}  +  \gamma \,  \partial^\alpha \bar {\mathcal A} \Big)\Big] = \sqrt{-g}\,  \frac{\delta U(b, \bar {\mathcal A}, \dots)}{\delta b}{\Big|_{b=\bar b, \, {\mathcal A}=\bar {\mathcal A}}}~, \nonumber \\
&& \partial_{\alpha}\Big[\sqrt{-g}\Big(\partial^{\alpha}\bar{{\mathcal A}}  -  \sqrt{\frac{2}{3}}\,
\frac{f_{\mathcal A}\, \alpha^\prime}{96 \, \kappa} \, {\mathcal K}^{\alpha}  +  \gamma \,  \partial^\alpha \bar b \Big)\Big] =  \sqrt{-g}\,\frac{\delta U(b, {\mathcal A}, \dots)}{\delta {\mathcal A}}{\Big|_{b=\bar b, \, {\mathcal A}=\bar {\mathcal A}}}~,
\end{align}
where we included a potential $U(b, {\mathcal A}, \dots)$, assumed to be generated in the late cosmological eras, which breaks explicitly the shift symmetry of the axions.
Above we ignored fermion and gauge anomalies contributions, as  we assume that in the current de Sitter era, fermion matter and radiation are not dominant, while only ${\mathcal A}-$ and $b-$axion DM dominates.

We do not discuss here the details of the generation of the potential $U(b, {\mathcal A}, \dots)$, apart from noting that a cosmological-constant-type Dark energy contribution is included for phenomenological reasons. One may use quintessence-like potentials, of the form used for axion inflation~\cite{axioninfl}, which contain mass terms for the $b(x)$ field,
so that the latter can play the r\^ole of an ordinary massive axionic DM component.
The important point is that, in the presence of an axion kinetic mixing parameter $\gamma \ne 0$ \eqref{mixing}, within the context of a homogeneous and isotropic cosmological situation where the fields depend only on the cosmic time at large scales, the solution \eqref{krbeom2} is still valid despite the presence of the potential $U(b, {\mathcal A}, \dots)$.  In that case, the equations \eqref{axions} reduce to:
\begin{align}\label{axions2}
&\gamma \, \frac{d}{dt} \Big[\sqrt{-g}\Big(  \dot{\bar {\mathcal A}} \Big)\Big] = \sqrt{-g}\, \frac{\delta U(b, {\mathcal A}, \dots)}{\delta b}{\Big|_{b=\bar b, \, {\mathcal A}=\bar {\mathcal A}}}~, \nonumber \\
& \sqrt{\frac{2}{3}}\,
\frac{d}{dt} \, \Big(  \sqrt{-g}\,\frac{(f_{\mathcal A} - \gamma) \, \alpha^\prime}{96 \, \kappa} \, {\mathcal K}^{0} \Big)= \sqrt{-g}\, \Big( \frac{1}{\gamma}\, \frac{\delta U(b, {\mathcal A}, \dots)}{\delta b}
- \frac{\delta U(b, {\mathcal A}, \dots)}{\delta {\mathcal A}}\Big)\Big|_{b=\bar b, \, {\mathcal A}=\bar {\mathcal A}}~,
\end{align}
Gravitational wave perturbations contribute to the anomaly as in the inflationary period, but with a much smaller Hubble parameter $H_0 $.  We stress that, in a FLRW space-time, {\it massive} $b(x)$ fields necessitate the presence of a non-trivial
$\frac{\delta U(b, {\mathcal A}, \dots)}{\delta b} \ne 0$, and thus $\gamma \ne 0$.

In general, an approximately constant solution \eqref{krbtoday} of a massive $b$ axion is consistent with the above equations.
Let us see this in a concrete but simple case, in which $0 \ne f_{\mathcal A}=\gamma  < 1$, which implies  ({\it cf.} \eqref{axions2}):
\begin{equation}\label{equalpot}
\Big(\frac{1}{\gamma}\,\frac{\delta U(b, {\mathcal A}, \dots)}{\delta b}
- \frac{\delta U(b, {\mathcal A}, \dots)}{\delta {\mathcal A}}\Big)\Big|_{b=\bar b, \, {\mathcal A}=\bar {\mathcal A}} =0~.
\end{equation}
Using \eqref{equalpot}, we observe that the first of Eqs.~\eqref{axions2} becomes
\begin{align}\label{aeqs}
3\, H\, \dot{\bar{\mathcal A}} + \ddot{\bar{\mathcal A}} =  \frac{\delta U(b, {\mathcal A}, \dots)}{\delta {\mathcal A}}\Big)\Big|_{b=\bar b, \, {\mathcal A}=\bar {\mathcal A}}
\end{align}
We remain agnostic as to the precise underlying microscoptc string theory that produces the potential $U(b, {\mathcal A})$ through stringy instanton effects. Below therefore we resort to phenomenological plausibility arguments.
For concreteness, we assume that the axion ${\mathcal A}$ field induces the {\it late de Sitter phase} through a non-perturbatively generated (periodic) potential of a form used in inflationary scenarios~\cite{axioninfl}, which can be embedded in concrete string/brane theory models:
\begin{align}\label{UbA}
U (b, {\mathcal A}) = c_0\, M_{\rm Pl}^4 + {{\mathcal M}_1}^4 \, \Big(1 -  {\rm cos}\Big[\frac{b}{{\mathcal M}_b} -\frac{\mathcal A}{{\mathcal M}_{ \mathcal A}}\Big]\Big)
+
\dots ~, \quad c_0 > 0~,
\end{align}
where ${\mathcal M}_i > 0$, $i=1, {\mathcal A}, b$ are appropriate mass scales, to be fixed phenomenologically. The term $c_0\, M_{\rm Pl}^4 > 0$ acts as a (positive) cosmological constant term in the current era, under the slow-roll condition  for the axion fields, which are assumed weak in the current epoch (see discussion below).\footnote{Alternatively, one could also consider the potential:
\begin{align}\label{UbA2}
U (b, {\mathcal A}) =  {{\mathcal M}_1}^4 \, \Big(1 -  c_2^2\, {\rm cos}\Big[\frac{b}{{\mathcal M}_b} -\frac{\mathcal A}{{\mathcal M}_{ \mathcal A}}\Big]\Big) + \dots
~, \quad  0 \ne c_2^2 < 1~,
\end{align}
in which the dominance of the (positive) cosmological constant (${{\mathcal M}_1}^4 \, (1 -  c_2^2) > 0$), driving the current-era de Sitter phase, arises from a weak-field expansion about the origin in field space ${\mathcal A}=b=0$, corresponding to the trivial local maximum of the potential, under the assumption of slow-roll for the axion fields ${\mathcal A}(x), b(x)$.
For the purposes of our discussion in this section, both potentials \eqref{UbA} and \eqref{UbA2} are qualitatively equivalent. }
The $\dots $ in \eqref{UbA} indicate terms involving shift-symmetry-breaking couplings with other fields, e.g., the aforementioned chiral Yukawa  coupling with right handed fermions, $y\, b(x) \, \overline \psi_R^c \, \psi_R $~\cite{pilaftsis}. At late epochs, like the current one and beyond, where the Universe enters a de Sitter phase again, we assume that such fermionic matter is completely diluted, or equivalently that the corresponding Yukawa couplings (that are in general also temperature dependent) are negligible. Hence we ignore them for the purposes of our subsequent discussion.

The reader should note that the potential \eqref{UbA} is characterised by non-diagonal mass terms for the $b$ and $\mathcal A$ fields, with the corresponding mass eigenstates obtained by diagonalisation. The massive nature of the axions $b$ and ${\mathcal A}$, then, allows them to play the r\^ole of multicomponent DM in the current era.

The condition \eqref{equalpot} is satisfied for the potential \eqref{UbA},
provided
\begin{align}\label{mgma}
\frac{{{\mathcal M}_1}^4}{\gamma \, {\mathcal M}_b}  \,
 {\rm sin}(\frac{b}{{\mathcal M}_b} -\frac{\mathcal A}{{\mathcal M}_{\mathcal A}})  =  -\frac{{{\mathcal M}_1}^4}{{\mathcal M}_{\mathcal A}}  \,
 {\rm sin}(\frac{b}{{\mathcal M}_b} - \frac{\mathcal A}{{\mathcal M}_{\mathcal A}} )  \, \Rightarrow \, {\mathcal M_b} =  - \frac{{\mathcal M}_{\mathcal A}}{\gamma},
  \end{align}
where, for consistency with the condition ${\mathcal M}_i > 0, i = {\mathcal A}, b$,  we should take $\gamma < 0$ (the reader is reminded that $|\gamma| < 1$, but it can have either
sign~\cite{pilaftsis}).

We shall look for self-consistent solutions of \eqref{mgma} in which $A/{\mathcal M}_{\mathcal A} \ll 1$, to satisfy the weak-field requirement.
We shall also assume ${\mathcal M}_1 \ll {\mathcal M}_{\mathcal A}$.
Taking the kinetic mixing parameter  $ 0 \ne |\gamma| \ll 1$, for concreteness, we observe \eqref{mgma}  that ${\mathcal M}_b \gg {\mathcal M}_{\mathcal A}$, so that \eqref{aeqs} can be approximately written, to leading order in small quantities, as
\begin{align}\label{approxeq}
{\mathcal A}^{\prime\prime} + 3 \, \tfrac{H}{M_{\rm Pl}} \, {\mathcal A}^\prime \simeq - \frac{{\mathcal M}_1^4}{{\mathcal M}_{\mathcal A}^2\, M^2_{\rm Pl}} \, {\mathcal A}~,
\end{align}
where the prime denotes differentiation with respect to the dimensionless variable $x=t \, M_{\rm Pl}$. The general solution of \eqref{approxeq} is:
\begin{align}\label{gensol}
{\mathcal A}(x) = e^{-\tfrac{3H}{2M_{\rm Pl}} x}\,
\Big(\tilde C_1 \, e^{-x\, \sqrt{-\tfrac{{\mathcal M}_1^4}{{\mathcal M}_{\mathcal A}^2\, M^2_{\rm Pl}} + \tfrac{9H^2}{4M^2_{\rm Pl}}}} +
\tilde C_2\, e^{+x\,\sqrt{-\tfrac{{\mathcal M}_1^4}{{\mathcal M}_{\mathcal A}^2\, M^2_{\rm Pl}} + \tfrac{9H^2}{4M^2_{\rm Pl}}}}\Big)~,  \quad x=t \, M_{\rm Pl}~,
\end{align}
where the constants $\tilde C_i, \, i=1,2$, are determined by boundary conditions.
In the current era, $H = H_0$. Then, due to the smallness of $H_0$, we may assume
for concreteness that the arguments of the square roots in the exponents on the right-hand-side of \eqref{gensol} are negative. Upon imposing suitable boundary conditions, then, we arrive at
a dumped oscillatory solution with (increasing) cosmic time, familiar from massive axion DM cases,
\begin{align}\label{approxeqA}
{\mathcal A}(t) = {\mathcal A}_0 \, e^{-\tfrac{3H}{2}t} \, {\rm sin}\Big(t\, \sqrt{\tfrac{{\mathcal M}_1^4}{{\mathcal M}_{\mathcal A}^2}- \tfrac{9H^2}{4}}\Big), \quad {\mathcal A}_0 \ll {\mathcal M}_{\mathcal A}, \quad \tfrac{{\mathcal M}_1^4}{{\mathcal M}_{\mathcal A}^2} > \tfrac{9H^2}{4},
\end{align}
with the quantity
\be\label{effm}
m_{\mathcal A}^2 \equiv \tfrac{{\mathcal M}_1^4}{{\mathcal M}_{\mathcal A}^2}- \tfrac{9H^2}{4}
\ee
playing the r\^ole of an effective axion mass-squared in an expanding Universe (above we kept $H$ general, since the expression \eqref{approxeqA} is valid beyond the current era). The condition ${\mathcal A}_0 \ll {\mathcal M}_{\mathcal A}$ guarantees weak fields.

Slow-roll conditions for both axions ${\mathcal A}$ and $b$, which in this model behave as massive DM fields in the modern era, can thus be arranged by suitable choices of the parameters. The order of $\dot{\bar {\mathcal A}}$ and $\dot b$ today is bounded from above by current cosmological observations~\cite{planck}. Without loss of generality, and assuming that the axions constitute the dominant form of DM today, one may assume ({\it cf}. \eqref{krbtoday})
\begin{align}\label{equalorder}
|\dot{\mathcal A}|_{\rm today} \,  \sim \, |\dot b|_{\rm today} = {\mathcal O}\Big(\sqrt{2\epsilon^\prime}\, H_0 \, M_{\rm Pl}\Big)~,
\end{align}
which can be easily achieved by an appropriate choice of the parameters.

 The energy density of the $b-{\mathcal A}$ fluid at the current (approximately de Sitter) era is then given by:
\begin{equation}\label{rho0}
\rho_{\rm today}^{b-a} = \frac{1}{2} (\dot{\bar {\mathcal A}})^2 + \frac{1}{2} (\dot{\bar b})^2 + \frac{\gamma}{2} \, \dot{\bar {\mathcal A}}\, \dot{\bar b} + U(\bar b, \bar {\mathcal A}, \dots)\Big|_{\rm today}, \quad \gamma \ll 1.
\end{equation}
In view of \eqref{krbeom2} and \eqref{equalorder}, and the fact that a cosmological constant term is present in the (slowly-varying) potential $U(b, {\mathcal A}, \dots)$ ({\it cf.} \eqref{UbA} or \eqref{UbA2}), one can readily see that the energy density \eqref{rho0} in the present epoch
acquires a ``running vacuum'' form \newnewtext{ \eqref{rLRVM}}, with $H^2$ contributions associated with the resurfaced gravitational anomalies.
On account of
the current constraints on DM energy density~\cite{planck}, and the r\^ole of both axion fields as massive DM components with a quintessence-like potential,
we thus observe that Eq.~\eqref{equalorder} is consistent with the identification of the slow roll parameters of the $b$ axion between
the inflationary and current eras, \eqref{slowrollepsi} and \eqref{krbtoday} respectively,
$\epsilon^\prime \sim \epsilon = {\mathcal O}(10^{-2})$, as assumed in our model, following the argumentation leading to \eqref{eq:UbTb}.

This completes our discussion. We stress once more that,
unfortunately, at present, the above analysis provides only plausibility arguments, {\it not} a concrete mechanism for mass generation for the KR axion, due to the lack of knowledge of the underlying microscopic string/brane model that could generate the (non-perturbative) potential $U(b, {\mathcal A})$. Nonetheless, we believe that the arguments are sufficiently interesting
to foster further research in this direction. The fact that our model promotes axionic DM as the dominant species of DM in the Universe, makes it relevant for current DM studies, in particular in models in which the effective DM mass \eqref{effm} is small, so that the respective DM is ultralight. Such ultralight DM constitutes currently the subject of intense research,  proposing, for instance, the use of precision atomic  or laser interferometric devices or other quantum sensors, to falsify particle-physics models involving scalar of pseudoscalar (axion) DM particles with masses smaller than $10^{-21}$~eV~\cite{sensors}.

\section{Conclusions \label{sec:concl}}

In this article we have provided a string-inspired theoretical framework
in which, during the early phase
of the Universe, there are important
contributions to the vacuum energy density
which are related to the CP-violating {\it gravitational anomalies} of a
primordial space-time of string theory. The latter are induced by primordial gravitational waves  during the inflationary era, in the presence of Lorentz- and CPT-violating
backgrounds of the KR axion field of the massless Bosonic string multiplet.  The KR field itself, though, does {\it not} cause or drive
inflation, which is due to other independent
mechanisms.

During the primordial inflationary era, we assume that only (stringy) gravitational degrees of freedom are present. Hence, the gravitational anomalies, whose presence in general would cause diffeomorphism-invariance breaking in the quantum theory, do not constitute any inconsistency, as would be the case if matter were present, since the anomalies describe the exchange of energy solely among (quantum) gravitational degrees of freedom. \nickcorr{Moreover, there is a second rank modified stress tensor which is conserved and describes any exchange of energy between the KR axion field and gravity. The stress tensor of this KR axion alone, which would be the ``matter'' stress tensor if anomalies were absent, is not conserved in their presence.}
It is important to mention that
the inflationary epoch can be described using the formalism of an
effective ``{\it running vacuum}'' model (RVM) with $H^2$ type
contributions to the vacuum energy density, which owe their existence to the gravitational anomaly. \joantext{Furthermore, as we have shown, in our string-inspired theoretical framework inflation can be correctly initiated and terminated (graceful exit) with the help of the gravitational Chern-Simons term, whose average over de Sitter spacetime induces also an additional, higher order, power $\sim H^4$ contribution to the vacuum energy density. This higher order term triggers inflation within the context of the RVM, as has been proven in detail in the literature~\cite{bls1,bls2,bls3,bls4,GRF2015,bls5}.  It follows that the entire history of the universe can be described in an effective RVM language upon starting from the fundamental massless bosonic gravitational multiplet of a generic string theory.  We believe that this is an interesting and remarkable result of our work, which, \noveltext{to the best of our knowledge, was never put forward in the literature prior to the present work.} Thanks to this result, the effective language of the RVM can be used in a very practical way to compute the main traits of the cosmic evolution starting from inflation and going through the standard radiation- and matter-dominated epochs until the late time universe,  i.e. the incipient DE epoch around our time, and finally into the future.  }

Because of the anomalous coupling of the KR axion to gravitational anomalies, the field remains undiluted at the end of inflation.
During the radiation/matter eras, chiral fermionic matter generated at the end of inflation cancels the gravitational anomalies, thus restoring diffeomorphism invariance in the radiation/matter quantum field theory, as required for consistency. We have found that, as the Universe passes from inflation
to radiation-dominated epoch, the presence of the undiluted CP and (spontaneously) CPT-violating KR
axion background, may lead to baryogenesis
via leptogenesis, in models involving heavy right handed (sterile) neutrinos. The lepton asymmetry is generated by CP-(and CPT-) Violating decays of the sterile neutrinos intro standard model particles in the presence of the KR background. Baryon-lepton-number-conserving sphaleron processes in the Standard-Model sector of the theory can then communicate the lepton asymmetry to baryons, thus leading to baryogenesis.
Therefore, the aforementioned process
could provide an efficient
way to understand the underlying physical mechanism for the
dominance of matter over antimatter in the early Universe. Moreover, during the radiation/matter dominance, uncompensated {\it gauge chiral} anomalies of the fermionic-matter axial (chiral) currents, also lead to $H^2(t)$ ``running vacuum''-type contributions to the energy density of the Universe.

\joantext{In the late universe such running vacuum contribution  involves an additive constant term,  which was was neglected in the early universe, and hence the effective or ``running'' cosmological term within the RVM is of the form $\Lambda(H)=c_0+\nu H^2$, where the value of $c_0$ is close (but not exactly equal) to the cosmological constant term of the $\CC$CDM, and $\nu H^2$  (with $|\nu|\ll1)$ is the running part of the DE density, a characteristic feature of the model.   The RVM is, therefore, finally testable in a very concrete way. It provides \noveltext{a specific} mechanism for inflation, which is different from the conventional one based on the inflaton \noveltext{field}~\cite{bls1,bls2,bls3,bls4,GRF2015,bls5}, \noveltext{but also} furnishes a mildly varying vacuum contribution which surfaces in the late universe and can be perceived as a form of dynamical dark energy. \noveltext{Such a form of DE } leads to an overall improvement of the fit of the cosmological observations as compared to the case  of a rigid $\CC$-term\,\cite{JJA,AdriaJoansigma8,JAJ,GoSolBas2015}. In addition,  that dynamical component of the DE can help alleviate some of the tensions presently existing in the $\CC$CDM concerning $\sigma_8$\,\cite{AdriaJoansigma8} and $H_0$\,\cite{Mehdi2019}. }

During the current de-Sitter era, the dilution of any matter, and the dominance again of the stringy gravitational degrees of freedom, including the KR axion, leads, through late-epoch gravitational waves, to the resurfacing of gravitational anomalies. We have also discussed how the KR field in the present era can act as a source for Dark Matter in models involving large-scale cosmic magnetic fields, generated by the the chiral anomalies. The magnetic energy density contributes to the late-era  energy budget of the Universe, with terms of RVM type, scaling as $H_0^2$. Moreover, there are scenarios in which the KR field mixes with other axion fields, abundant, e.g., in string models, thus providing models for multi-component DM.

\newnewtext{Before closing, we feel making a last but rather important remark. Since our effective field theoretic running-vacuum model of quantum gravity, upon which we based our studies here, is inspired from string theory, it would be interesting to discuss it in the context of the recent conjectures on the incompatibility of de Sitter vacua (characterised by a rigid positive cosmological constant) with the `swampland criteria', and in general of how one can couple quantum field theories  to quantum gravity models, especially in view of our transplanckian regime of modes entering \eqref{transpl}~\cite{swampland}. We leave this interesting topic for future works, as we did not discuss here microscopic string theory models leading to inflation. We remark, though, that the dynamical nature of the running vacuum leads to deviations of our model from the standard $\Lambda$CDM, as far as the nature of the vacuum energy is concerned, which is dynamical in our case; in this respect, compatibility of some of the models falling in our framework with the `swampland criteria' is to be expected.}

To summarise our findings:  the (gravitational) anomaly played an important dual r\^ole for  {\it our existence}: first, it induced a non-diluted axion background of DM at the end of inflation into the radiation epoch, {which itself induces leptogenesis}; and,
second,  it fostered the subsequent generation of chiral  matter  from the decay of the running vacuum, thus cancelling the unbalanced gravitational anomaly and restoring general covariance in our Universe. \noveltext{As demonstrated in our work}, the gravitational or chiral anomalies lead to mildly running dark energy, as a smoking gun evidence of their presence!

So, paraphrasing the famous quote by Carl Sagan~\cite{sagan}, the thesis in this article is that \emph{we might well be anomalously made of starstuff!} !

 {\it  Affaire \`a suivre ...}

\section*{Acknowledgements}

\newnewtext{We would like to thank the anonymous referee for constructive and insightful comments, that helped in improving the presentation of our results. NEM wishes to acknowledge discussions with Diego Blas and Chris McCabe on axion DM models and their searches}.
SB acknowledges support from
the Research Center for Astronomy of the Academy of Athens in the
context of the program  ``{\it Tracing the Cosmic Acceleration}''.
The work  of NEM is supported in part by the UK Science and Technology Facilities  research Council (STFC) under the research grant
ST/P000258/1. The work of JS has
been partially supported by projects  FPA2016-76005-C2-1-P (MINECO), 2017-SGR-929 (Generalitat de Catalunya) and MDM-2014-0369 (ICCUB).
This work is also partially supported by the COST Association Action CA18108 ``{\it Quantum Gravity Phenomenology in the Multimessenger Approach (QG-MM)}''.
NEM acknowledges a scientific associateship (``\emph{Doctor Vinculado}'') at IFIC-CSIC-Valencia University, Valencia, Spain.

\appendix

\section{Arbitrary string mass scale. \label{sec:appA}}

\numberwithin{equation}{section}

\setcounter{equation}{0}

\nickcorr{In this Appendix, we demonstrate how in the general case, where the string scale $M_s \ne \MPl$, one can avoid transplanckian values for the UV cut-off $\mu$ by appropriately constraining the range of $M_s$. }

\nickcorr{Indeed, this 
follows from the fact that, in such a case, the condition \eqref{A=0} for an approximately constant ${\mathcal K}^0$ during inflation  is replaced by
\begin{align}\label{A2=0}
{\mathcal A}  = 1  -  1.95 \,  \times 10^{-5} \,  \Big(\frac{H}{M_{\rm Pl}}\Big)^2 \, \Big(\frac{\mu}{M_{s}}\Big)^4 
\simeq 0 \quad \Rightarrow  \quad
\frac{\mu}{M_s} \simeq 15 \, \Big(\frac{\MPl}{H}\Big)^{1/2}~,
 \end{align}
 where the $\simeq$ in the above relations are to be interpreted as within an error of order of at most a $\%$. Indeed, an approximately constant $\mathcal K^0$ in \eqref{k02} is guaranteed provided that at the end of the inflationary period its value is diminished no more than an order of magnitude, that is 
 \begin{align}\label{KN}
 {\mathcal K}^0_{\rm end} (t_{\rm end}) \simeq \Big(e^{-1} - e^{-2}\Big)\, {\mathcal K}^0_{\rm begin} (t(\eta=H^{-1})) 
 \end{align}
 Taking in to account that, in units of cosmic Robertson-Walker time $t$, the end of inflation occurs for $H \, t_{\rm end} \sim {\mathcal N}$, with ${\mathcal N}$, the number of e-foldings, which is expected from the data~\cite{planck} to be of order $\mathcal N= {\mathcal O}(60-70)$, we thus observe from \eqref{KN} that an 
 \begin{align}\label{aN}
0 \lesssim  {\mathcal A} \lesssim \xi \, (3{\mathcal N})^{-1} \sim  \xi (0.0048 - 0.0056 ), \quad \xi = 1-2,
 \end{align}
 suffices for our purposes, which leads to the aforementioned uncertainty of at most a $\%$ in the value of $\mu$ in \eqref{A2=0},
 \begin{align}\label{A=0N} 
 \frac{\mu}{M_s} \simeq 15 \Big( 1-\frac{\xi}{3\mathcal N} \Big)^{1/4}\, \Big(\frac{\MPl}{H}\Big)^{1/2} \simeq (0.998 - 0.999) \times 
 15 \, \Big(\frac{\MPl}{H}\Big)^{1/2}.
 \end{align}
 If one insists on phenomenologically acceptable ranges of $H \ll M_{\rm Pl}$, e.g. \eqref{Hinfl}, then one obtains: 
\begin{align}
\label{transpl2}
 \mu \sim 10^{3} \, M_{s}~,
 \end{align} 
 which replaces  \eqref{transpl}. Then, on combining \eqref{theta} and \eqref{slowroll}, we see that a {\it sufficient} condition to guarantee the smallness of $|\Theta \ll1 $ is 
 \begin{align}\label{HMs}
 H/M_s \ll 3.83, 
 \end{align}which, on account of  \eqref{transpl2} implies 
 \begin {align}\label{murange}
 \mu \gg 2.61 \times (10^{-3} - 10^{-2}) \, M_{\rm Pl}.
 \end{align} 
This, in turn, leads to the observation that the cutoff scale $\mu$ can be at least of order of $M_{\rm Pl}$. 
Thus, by allowing $M_s \ne \MPl$, we can in principle {\it avoid a transplanckian cutoff} $\mu$, since we may set $\mu \sim \MPl$
which is a quite natural order of magnitude for the UV completion of the low-energy effective theory. In such a case, \eqref{HMs} implies the following range of the minimal allowed order of magnitude of the string scale 
 $M_s \gtrsim 10^{-3} \, M_{\rm Pl}$. Saturating from above $M_s \lesssim \MPl$ we thus obtain the following range for the string scale
 \begin{align}\label{msr}
 \MPl \, \gtrsim  \, M_s \gtrsim 10^{-3} \, M_{\rm Pl}~,
 \end{align}
  in order to guarantee the Lorentz-violating solution \eqref{lv} for the KR background.}

\end{document}